\documentclass{aa}  

\usepackage{graphicx}
\usepackage{txfonts}
\usepackage{color}
\usepackage{longtable}
\usepackage{pdflscape}

\usepackage{hyperref}
\begin{document} 
   \title{The XXL Survey\thanks{Based on observations obtained with \textit{XMM-Newton}, an ESA science mission with instruments and contributions directly funded by ESA Member States and NASA.}}

   \subtitle{L. Active galactic nucleus contamination in galaxy clusters: Detection and cosmological impact}

    \author{Sunayana Bhargava
          \inst{1}, 
          Christian Garrel \inst{1,2},
          Elias Koulouridis \inst{3},
          Marguerite Pierre \inst{1},
          Ivan Valtchanov \inst{4},
          Nicolas Cerardi \inst{1},
          Ben J. Maughan \inst{5},
          Michel Aguena \inst{6},
          Christophe Benoist \inst{7},
          Cale Baguley \inst{5},
          Miriam E. Ramos-Ceja \inst{2},
          Christophe Adami \inst{8},
          Lucio Chiappetti \inst{9},
          Cristian Vignali \inst{10},
          \and
          Jon P. Willis.  \inst{11}
          }
   \institute{AIM, CEA, CNRS, Universit\'e Paris-Saclay, Universit\'e Paris Diderot, Sorbonne Paris Cite, F-91191 Gif-sur-Yvette, France\\
              \email{sunayana.bhargava@cea.fr}
         \and
             Max-Planck-Institut für extraterrestrische Physik (MPE), Giessenbachstrasse 1, D-85748 Garching bei München, Germany
         \and
             Institute for Astronomy \& Astrophysics, Space Applications \& Remote Sensing, National Observatory of Athens, GR-15236 Palaia Penteli, Greece
         \and 
             Telespazio UK for European Space Agency, ESAC, Camino Bajo del Castillo s/n, ES-28692 Villanueva de la Ca\~nada, Madrid, Spain
         \and 
             School of Physics, HH Wills Physics Laboratory, Tyndall Avenue, Bristol, BS8 1TL, UK
        \and
             Université Paris Cit\'e, CNRS(/IN2P3), Astroparticule et Cosmologie, F-75013 Paris, France   
         \and
             Laboratoire Lagrange, Universit\'e  C\^ote d’Azur, Observatoire de la C\^ote d’Azur, CNRS, Blvd de l’Observatoire, CS 34229, 06304 Nice cedex 4, France
         \and
             Aix Marseille Univ., CNRS, CNES, LAM, Marseille, F-13007 France
        \and
            INAF, IASF Milano, via Corti 12, I-20133 Milano, Italy
        \and
            INAF/IRA, Istituto di Radioastronomia, Via Piero Gobetti 101, I-40129 Bologna, Italy
        \and
            Department of Physics and Astronomy, University of Victoria, 3800 Finnerty Road, Victoria, BC, Canada
            }

   \date{Received XXX; accepted YYY}

 
  \abstract
   {X-ray observations of galaxy clusters are impacted by the presence of active galactic nuclei (AGNs) in a manner that is challenging to quantify, leading to biases in the detection and measurement of cluster properties for both astrophysics and cosmological applications.}
   {We detect and characterise clusters contaminated by central AGNs within the XXL survey footprint and provide a systematic assessment of the cosmological impact of such systems in X-ray cluster samples.}
   {We introduce a new automated class for AGN-contaminated (AC) clusters in the XXL source detection pipeline. The majority of these systems are otherwise missed by current X-ray cluster-detection methods. The AC selection is also effective in distinguishing AGN and cool-core presence using supplementary optical and infrared information.}
   {We present 33 AC objects, including 25 clusters in the redshift range, $0.14 \leq z \leq 1.03$, and eight other sources with significantly peaked central profiles based on X-ray observations. Six of these are new confirmed clusters. We computed the missed fraction of the XXL survey, which is defined as the fraction of genuine clusters that are undetected due to their centrally peaked X-ray profiles. We report seven undetected AC clusters above $z > 0.6$, in the range where X-ray cluster detection efficiency drops significantly. The missed fraction is estimated to be at the level of $5\%$ for the 50 square-degree XXL area. The impact on cosmological estimates from missed clusters is negligible for XXL, but it produces a tension of $\sim 3\sigma$ with the fiducial cosmology when considering larger survey areas.}
   {This work demonstrates the first systematic attempt to quantify the percentage of missed clusters in X-ray surveys as a result of central AGN contamination. Looking towards surveys such as eROSITA and \textit{Athena}, larger areas and increased sensitivity will significantly enhance cluster detection, and therefore robust methods for characterising AGN contamination will be crucial for precise cluster cosmology, particularly in the redshift $z > 1$ regime.}

\keywords{surveys -- X-rays : galaxies: clusters -- cosmological parameters -- X-rays: galaxies: clusters -- galaxies: clusters: general -- galaxies: active -- large-scale structure of Universe}

\titlerunning{The XXL Survey: AGN contamination in X-ray selected clusters}
\authorrunning{S. Bhargava et al.}
\maketitle
%

\section{Introduction}

The growth of galaxy clusters from the highest primordial density peaks makes them indispensable probes for the measurement of cosmological parameters. The number of clusters observed as a function of mass and redshift is extremely sensitive to the underlying matter and energy content of the Universe. However, since galaxy clusters are not detected according to their total mass, but rather via observable mass proxies, the precise modelling of selection effects is a crucial component to ensure an accurate sampling of clusters over cosmic time. One key advantage for X-ray cluster surveys, which detect diffuse emission from the intracluster medium (ICM), is that they are significantly less sensitive to projection effects, since the X-ray surface brightness is more centrally concentrated than the galaxy distribution of the cluster. This has allowed for the creation of many effective cluster catalogues \citep[e.g.][hereafter XXL Paper XX]{Ebeling1997,Bohringer2000,Mehrtens2012,Adami2018}, as well as more recent samples \cite[][]{Klein2019,Brunner2022}, including those that go to redshifts of $z > 1$ \citep[e.g.][]{Willis2013,Trudeau2020}.
Despite the efficacy of X-ray cluster searches, approximately 90 percent of sources in X-ray surveys are point-like objects, of which the majority are active galactic nuclei (AGNs). Sufficient angular resolution can allow one to distinguish between clusters and AGNs, but this is harder at intermediate to high redshifts, where the extent of cluster emission becomes comparable to the point spread function (PSF) of most X-ray missions (e.g. FWHM $6''$ on-axis for \textit{XMM-Newton}). As a consequence, AGNs may be misclassified as clusters and vice versa \citep{Donahue2020,Bulbul2021}. Moreover, galaxy clusters may be contaminated by X-ray emission from an unresolved AGN within or along the line of sight. Famously, the Phoenix cluster at $z=0.597$ was first misclassified as an X-ray point source in the ROSAT Bright Source Catalogue \citep{Voges1999} due to the presence of a bright AGN embedded in the cluster centre. In more recent work by \citet[][hereafter XXL Paper XXXIII]{Logan2018}, an XXL sample of cluster candidates ($z > 1$) with associated \textit{Chandra} observations revealed the presence of significant contamination from previously unresolved AGNs in approximately one third of the sample. It is also difficult to distinguish between AGNs and cool-cores in such systems due to the similarity in their X-ray surface brightness profiles \cite[particularly in the inner 10-30 kpc region, see][]{Fabian1994}. While consequences are less drastic for nearby clusters where the \textit{XMM} PSF is compensated by the low redshift, this illustrates the importance of the subject well. 

Modelling the impact of AGN contamination is important in the context of the XXL survey \citep[][hereafter XXL Paper I]{Pierre2016}. This is the largest \textit{XMM} programme totaling $\sim$ 7Ms. It covers two extragalactic areas of 25 deg$^{2}$ each at a point-source sensitivity of $\sim 6 \times 10^{-15}$ erg s$^{-1}$ cm$^{-2}$ in the [0.5--2] keV band (completeness limit). Given that one of the survey's key goals is to serve as a pathfinder for future wide-area X-ray missions such as \textit{Athena} for the next decade, the accurate selection of galaxy clusters is a necessary aspect, especially given that AGN density in clusters increases with redshift \citep[e.g.][]{Martini13,Bufanda17,krishnan17,Koulouridis18b}. AGN contamination within X-ray cluster surveys is typically addressed statistically by using realistic models of clusters, field AGNs, and AGNs embedded or projected onto clusters to calibrate the selection function \citep[e.g.][]{Kafer2020}, but future surveys will likely need to employ cosmological hydrodynamical simulations in which AGNs and cluster evolution are treated self-consistently \citep[see][]{Biffi2018, Koulouridis18a,Zhang2020}. Unfortunately, there remains a lack of observational data on which to base such models, motivating the work presented in this paper. The final XXL data release aims to have approximately 400 cluster candidates of which AGN contamination may constitute a significant fraction. Looking forwards, the eROSITA all-sky X-ray survey will likely detect $10^5$ clusters \citep{Merloni2012} together with more than three million X-ray AGNs. Therefore, to obtain large X-ray cluster samples with sufficient purity, an automated method is required to select clusters with point source contamination. 

This work presents a systematic search for the presence of AGN contamination within or projected onto X-ray-selected clusters. We applied a pipeline-driven classification blindly to all significant detected objects within the full XXL survey footprint; therefore, this work also delivers the first estimate of the level of AGN contamination over the redshift range of the XXL cluster sample. The outline of the paper is as follows. In Section \ref{sec:sims} we describe the simulations used to model AGN-contaminated (AC) clusters in X-ray data. In Sections \ref{sec:accrit} and \ref{sec:acselfunc} we state the selection criteria for AC objects and their selection function. Section \ref{sec:sample} describes the properties of the AC sample on the latest XXL dataset, including redshift estimates and multi-wavelength methods of confirmation. Section \ref{sec:xrayproperties} details the X-ray properties of the AC sample. 
In Section \ref{sec:accosmology}, we estimate the missed fraction within XXL and its consequences for the final cosmological analysis of XXL and other X-ray surveys. We summarise our results in Section \ref{sec:summary}. Throughout the paper, unless otherwise stated, we assume a WMAP9 cosmology with $\Omega_{\rm M} = 0.28$, $\Omega_\Lambda = 0.72$, and $H_0 = 70$ km s$^{-1}$Mpc${^{-1}}$. 

\section{Modelling AGN-contaminated clusters using simulations}\label{sec:sims}

\subsection{Injections into simulated \textit{XMM} observations}

\begin{figure*}
    \centering
    \includegraphics[width=0.46\textwidth]{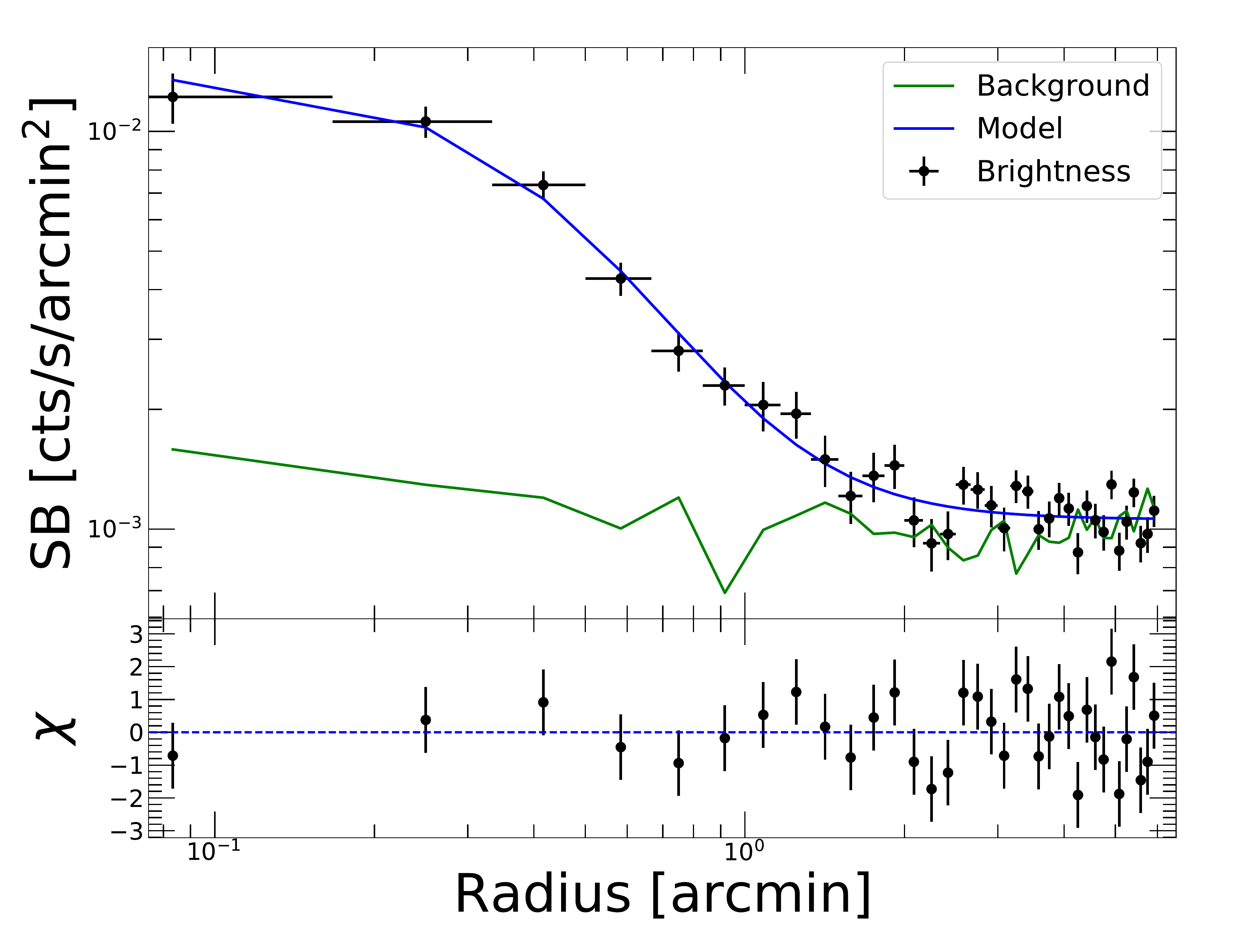}
    \includegraphics[width=0.46\textwidth]{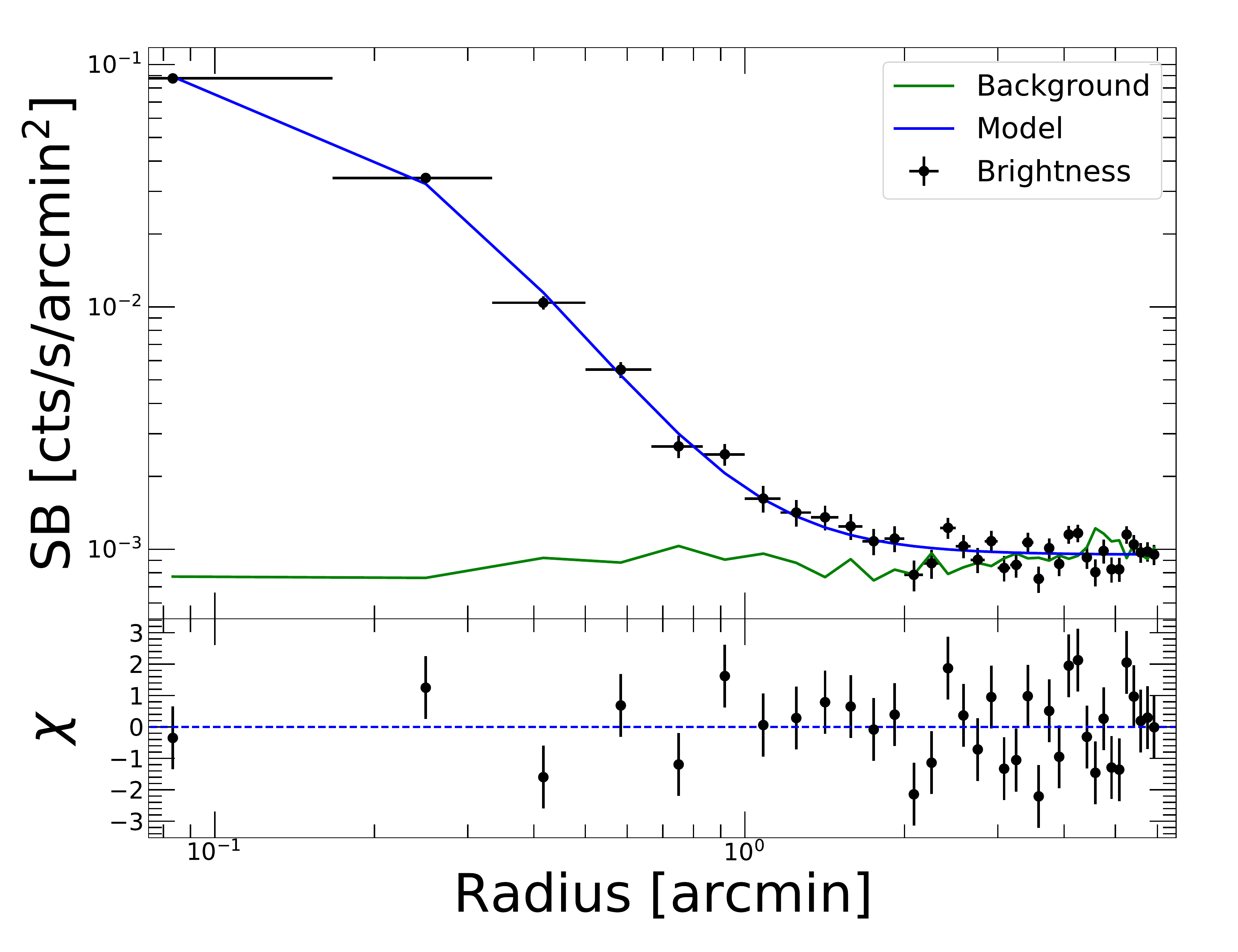}
    \includegraphics[width=0.9\textwidth]{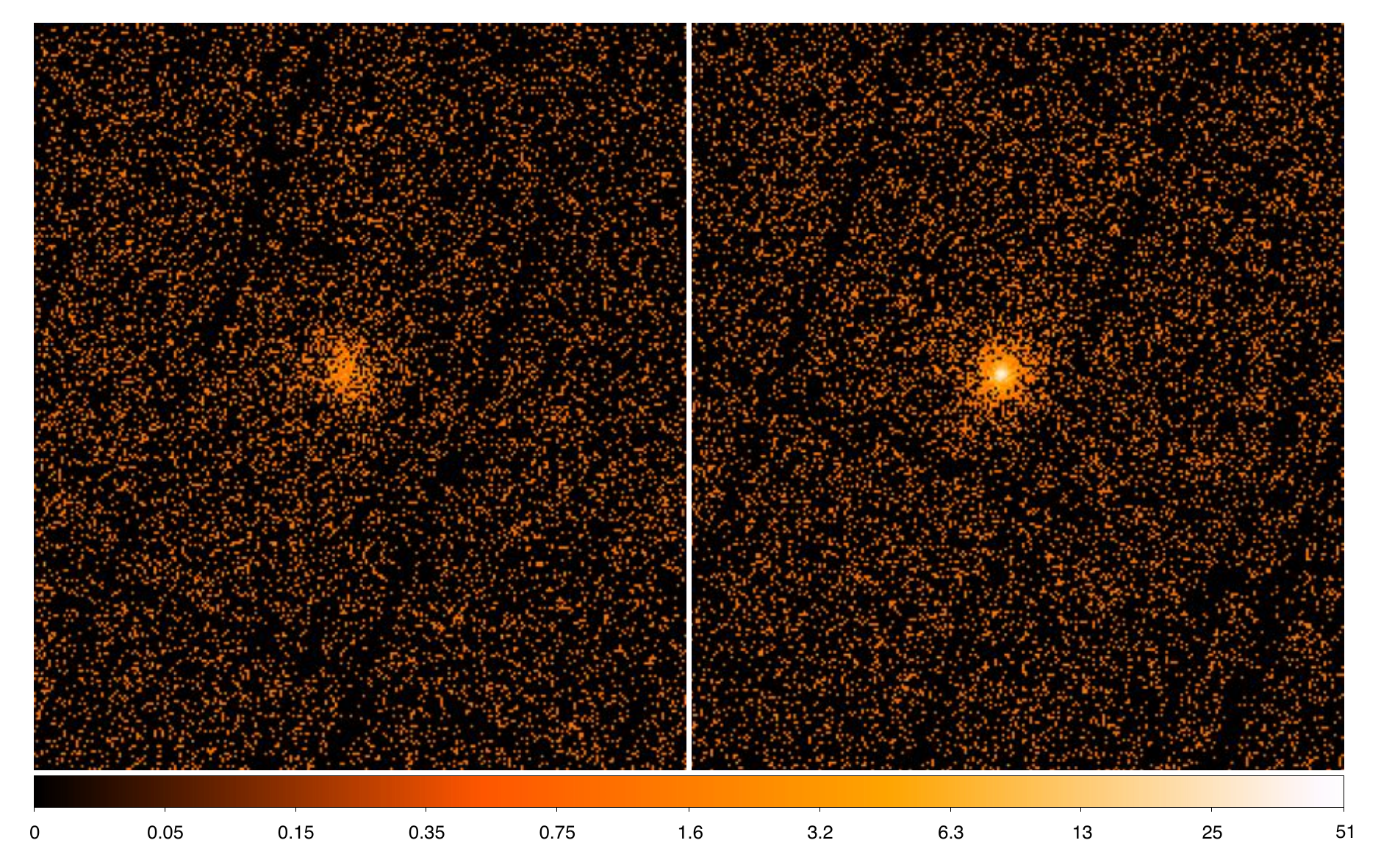}
    \caption{Comparison of simulated uncontaminated (left) and point-source contaminated (right) clusters. \textbf{Top panel:} surface brightness (SB) distribution showing the extracted cluster profiles (black crosses) are plotted against the blue line corresponding to the fitted $\beta$-model; the green line displays the particle background level extracted from each image. \textbf{Bottom panel:} Simulated 10ks \textit{XMM} pointings showing a cluster with a core radius $r_c=20''$ and count rate ${\rm CR}=0.1$ cts/s. For the contaminated cluster, 50$\%$ of the overall source counts lie within the central point source. The colour bar shows the number of photons within each pixel ($2.5''$ per pixel scale).}
    \label{fig:acsbcomparisons}
\end{figure*}

We performed realistic Monte Carlo image simulations of \textit{XMM}-like observations (hereafter pointings) to assess the detection threshold of clusters with central point source contamination using the \textsc{InstSimulation} software \citep{Valtchanov2001}. Soft-band \textit{XMM} pointings were produced from scratch using a combined exposure time of 10 ks (see Figure \ref{fig:acsbcomparisons}). Two background components - the non-resolved vignetted AGN photon background and the unvignetted, uniform particle background - were added according to \cite{ReadPonman2003}. These simulations faithfully reproduce the characteristics of the three EPIC detectors and have been used to characterise the cluster selection function of the XXL and X-CLASS surveys \citep[see][hereafter XXL Paper XLVI]{Pacaud2006,Koulouridis2021,Garrel2021}. 

In order to model the detection differences for AGN-contaminated clusters, we first modelled "pure" uncontaminated clusters according to a single-beta profile 
\begin{equation}\label{eq:betamodel}
    S_X(r) = S_0\left[1+\left(\frac{r}{r_c}\right)^2\right]^{-3\beta+0.5}
\end{equation}
where the core radius $r_c$ is measured in arcseconds, and a fixed value of $\beta=2/3$ is used throughout. The total count rate in the soft [0.5-2] keV band and core radius were varied as shown in Table \ref{tab:ACsimsproperties}. The clusters were populated in random positions within three off-axis shells (0-$5'$, 5-$10'$, 10-$13'$) measured from the \textit{XMM} aimpoint - the number of clusters per shell was adjusted according to its core radius to minimise the occurrence of overlaps between sources. Altogether 850 simulations of pure clusters were rendered, with 5900 clusters simulated (based on the breakdown of clusters according to Table \ref{tab:ACsimsproperties}).

The simulations for clusters with point-source contamination were produced identically, with the addition of a point source placed in the centre of the cluster. We used three flux ratios for the contamination level: 0.25, 0.5 and 1, i.e. where the central point source had one quarter, half, or the same count rate of the cluster in the soft band. The total rate in this instance is the sum of both the central point and cluster count rates. The point source was always positioned in the centre of the cluster in all cases. In total, 2550 simulations were rendered for the contaminated clusters (850 corresponding to each level of contamination). Finally, a set of 180 simulations were produced for field AGNs modelled based on the soft-band logN-logS distribution from \cite{Moretti2003} down to the flux limit of $5 \times 10^{-16}$ erg s$^{-1}$cm$^{-2}$ at 10ks, yielding approximately 830 randomly distributed point sources per pointing. Both point-source and extended-source profiles were convolved using the latest ELLBETA PSF model from \cite{Read2011}, available from the \textit{XMM} calibration data, which takes into account the strong distortions of the PSF at large off-axis angles.

\begin{table}
     \caption{Input configuration for \textit{XMM} simulations of cluster profiles. The numbers in brackets indicate the number of clusters simulated per pointing as a function of core radius and total count rate. The run no. is the number of realisations per total count rate and core radius.}
         \begin{tabular}{cccc}
            \hline
            \noalign{\smallskip}
            Core radius [arcsec] & Count rate [cts/s] &  Run no. \\
            \noalign{\smallskip}
            \hline
            \noalign{\smallskip}
            3 (11)& 0.005, 0.01, 0.02, 0.05, 0.1  & 20\\
            5 (11) & 0.005, 0.01, 0.02, 0.05, 0.1  & 20\\
            10 (8) & 0.005, 0.01, 0.02, 0.05, 0.1 & 20\\
            20 (8) & 0.005, 0.01, 0.02, 0.05, 0.1 & 20\\
            50 (6) & 0.005, 0.01, 0.02, 0.05, 0.1 & 30\\
            100 (4) & 0.005, 0.01, 0.02, 0.05, 0.1 & 60\\
            \noalign{\smallskip}
            \hline
         \end{tabular}
         \label{tab:ACsimsproperties}
\end{table}

\subsection{XMM pipeline processing}\label{subsec:xmmpipeline}

All simulated sources were then processed through the latest version of the XXL source detection pipeline, which consists of a three-step process. Soft X-ray band observations were created and subsequently filtered using the wavelet decomposition method described in \cite{StarckPierre1998}. This technique is considered to be optimal for filtering X-ray images that contain few photon counts and Poisson noise, and has proven effective for cluster detection in the regime of short exposure times. Secondly, \textsc{SExtractor} \citep{SExtractor} was used to detect sources within the inner $13'$ of the field to avoid border effects. The background level was iteratively estimated using $3\sigma$ clipping, and a full background map was constructed by bicubic spline interpolation. While the simulated background is known, we performed this step to match the processing of the real XXL data. An isophotal analysis was then performed to determine the X-ray centroid position, brightness and shape within a flexible elliptical aperture. Finally, these parameters were inputted into the \textsc{Xamin} maximum likelihood fitting routine that applies several source models on the soft band photon image. For a detailed description of the individual model fits, we refer the reader to \citet[][hereafter XXL Paper XXIV]{Faccioli2018}, however, we provide a self-contained description of the relevant models below. The \textsc{pnt} model is a precise point spread function (PSF) model for point-like sources. The \textsc{ext} model is a spherically symmetric $\beta$ model for pure extended sources \citep{CavaliereFuscoFemiano1976}. Finally, the \textsc{epn} model is a $\beta$ model superposed to a central PSF for extended sources containing a central point source.

\subsection{The extended and central point (\textsc{epn}) model}

The \textsc{epn} model is introduced to recover clusters with central AGN contamination. This is required in addition to the \textsc{ext} model which can miss clusters that are too peaked in the core region. In the \textsc{epn} fit, the candidates are fitted using a superposition of the convolved $\beta$ profile and ELLBETA PSF model. 

We defined two parameters to quantify the likelihood of the \textsc{epn} fit with respect to a) the point-like \textsc{pnt} and b) the simple extended \textsc{ext} fit. Both the \textsc{epn\_stat\_pnt} and \textsc{epn\_stat\_ext} values are defined as the difference in the best-fitted values ($E_{BF}$) of the Cash ($C$)-statistic \citep{Cash1979} for each model. The third key parameter in the \textsc{epn} model is the \textsc{epn\_ratio}, which is the ratio of the count rate estimated from the \textsc{pnt} and \textsc{ext} models. The three key properties are therefore
\begin{equation}
    \begin{split}
    &\textsc{epn\_stat\_pnt} = E_{BF}|_\textsc{pnt} - E_{BF}|_\textsc{epn}, 
    \\
    &\textsc{epn\_stat\_ext} = E_{BF}|_\textsc{ext} - E_{BF}|_\textsc{epn},
    \\
    &\textsc{epn\_ratio} = {\rm CR}_\textsc{pnt}/{\rm CR}_\textsc{ext},
    \end{split}
\end{equation}
where $E_{BF}$ is the best fitted value of the Cash ($C$)-statistic \citep{Cash1979} for each model. The higher the value of the \textsc{epn\_stat\_pnt} or \textsc{epn\_stat\_ext}, the better the fit from the \textsc{epn} model compared to either the \textsc{pnt} and \textsc{ext} models alone. In more physical terms, the \textsc{epn\_stat\_ext} value determines that the contaminated cluster is sufficiently peaked while remaining extended, while the second (\textsc{epn\_stat\_pnt}) distinguishes the contaminated cluster from a point source. The \textsc{epn\_ratio} is analogous to a flux ratio between the central point source and cluster. 

\section{Defining the AC parameter space}
\label{sec:accrit}

Since AGN-contaminated clusters are a particular class of objects, they must be distinguishable from existing XXL source criteria. We recap these categories below \cite[for a more detailed description, the reader is referred to][]{Pacaud2006}.The C1 class refers to cluster candidates where the level of purity is above 90\% and contamination from point sources is deemed negligible. The C2 class refers to cluster candidates with an assigned purity of 50\%, and hence this class also includes misclassified AGNs, image artefacts, and spurious detections. The XXL pipeline criteria for the C1 and C2 classes is outlined in Table \ref{tab:ACcritcomp}. 

\begin{figure}
    \centering
    \includegraphics[width=0.5\textwidth]{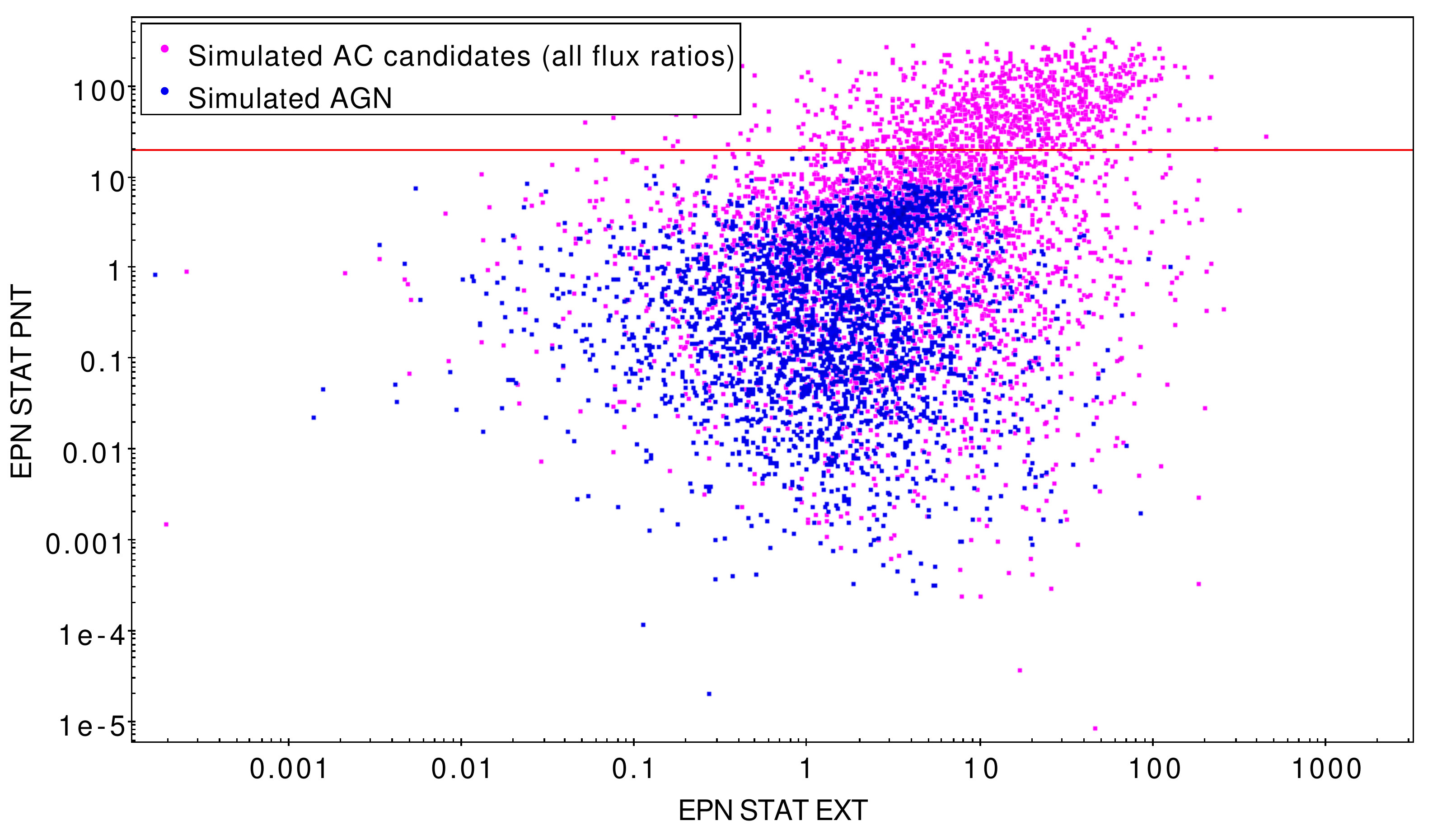}
    \caption{Simulated AC versus field AGN comparison for the simulated dataset. The red line denotes the cut for the AC class, where a largely pure fraction of AC objects are expected.}
    \label{fig:ACvsAGNonly}
\end{figure}

We define a new class for the \textbf{A}GN-contaminated \textbf{c}lusters, hereafter the AC class. We first distinguish these sources from field AGNs, and, subsequently, from the uncontaminated cluster population. From the set of simulations described in Section \ref{sec:sims}, we correlated the input and \textsc{Xamin} output sources with a maximum radius of 37.5" for clusters (both contaminated and uncontaminated) following the prescription outlined in \cite{Pacaud2006}. Point sources were correlated within 12.5" of an input source. Figure \ref{fig:ACvsAGNonly} shows the distribution of the simulated field AGN in the \textsc{epn\_stat\_ext} versus \textsc{epn\_stat\_pnt} parameter space. As expected, the point sources do not produce sufficiently high likelihood values for the \textsc{epn} model, as the $E_{BF}$ for these objects is highest for the \textsc{pnt} model alone. We applied a cut at $\textsc{epn\_stat\_pnt} \geq 20$ to separate the AC and AGN populations.

\begin{figure*}
    \centering
    \includegraphics[width=0.6\textwidth]{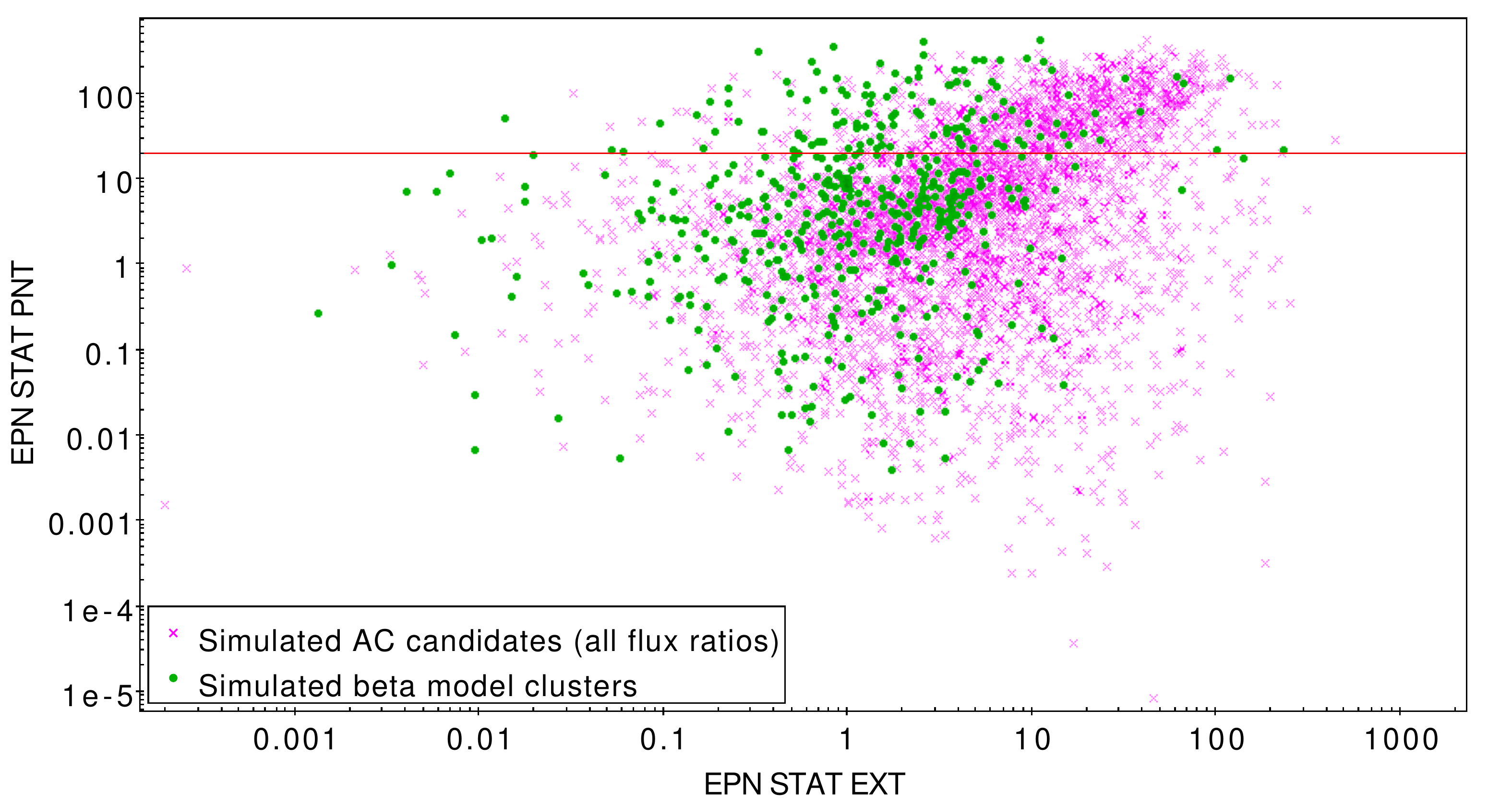}
    \includegraphics[width=0.65\textwidth]{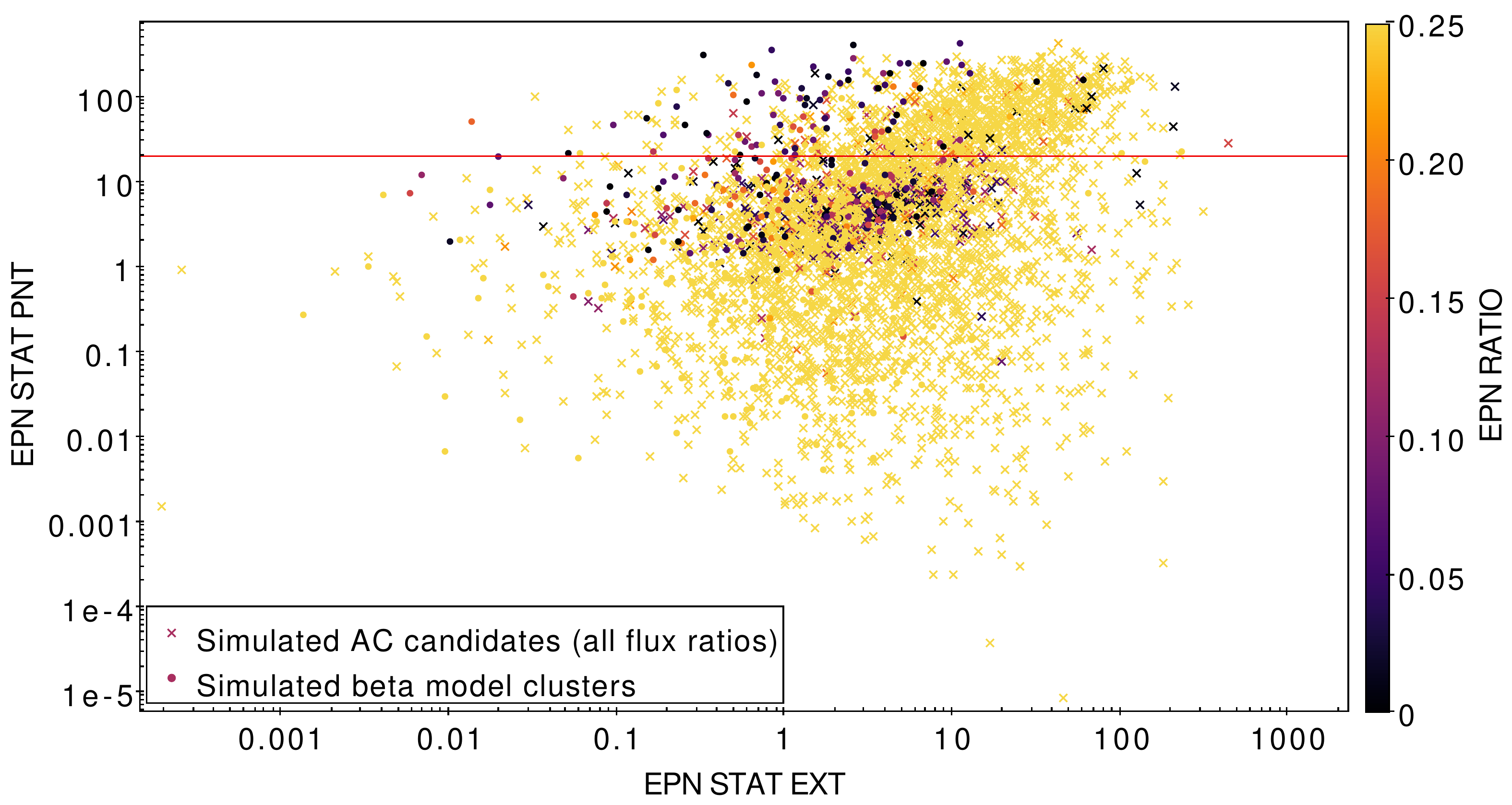}
    \caption{Population of simulated AC and uncontaminated clusters in the output \textsc{XAmin} parameter space. \textbf{Top panel:} Distribution of AC sources (pink crosses) compared with uncontaminated clusters (green circles). \textbf{Bottom panel:} Distribution of AC sources and uncontaminated clusters shaded according to the \textsc{epn\_ratio} parameter. The red line highlights the $\textsc{epn\_stat\_pnt} > 20$ cut. Dark shaded circles above the red line indicate pure clusters that are separable from the AC population due to their having an \textsc{epn\_ratio} $\leq 0.2$.}
    \label{fig:ACvsC1}
\end{figure*}

Next we segregated AC candidates from the population of pure uncontaminated clusters. The top panel of Figure \ref{fig:ACvsC1} shows that both pure and contaminated clusters, in green and pink respectively, exist above \textsc{epn\_stat\_pnt} > 20, since both are types of extended objects. We used the \textsc{epn\_ratio} to separate the `peakiness' of the two classes, selecting a threshold of $\textsc{epn ratio} \geq 0.2$. After applying this cut, the majority of pure clusters have lower $\textsc{epn\_ratio}$ values compared to the AC class (Figure \ref{fig:ACvsC1}, bottom panel). We also imposed a cut on the core radius, $\textsc{epn\_ext} \geq 5$", similarly to the C1 and C2 criteria. We selected a slightly higher value of 5" rather than 3" for the pure clusters to compensate for the fact that the AC sources are, by defintion, more peaked. The final criteria for the AC selection is summarised in Table \ref{tab:ACcritcomp}. We emphasise that given the use of X-ray image simulations, no light cone information is provided, and therefore the density of points in Figures \ref{fig:ACvsAGNonly} and \ref{fig:ACvsC1} is not physically relatable to the real ratio of C1 and AC clusters. Nevertheless, we estimated the misclassification rate of C1 to AC clusters for the simulated dataset. Out of 5900 clusters in total (see Section \ref{sec:sims}), 3521 are recovered as pure C1 by the \textsc{Xamin} detection algorithm. 149 are classed as AC (less than 3\% of the total set). Among the 149 misclassified C1, over 90\% have an input core radius of $r_{c} \leq 5''$, highlighting that the highest misclassification rate occurs at smaller radii; i.e. clusters that appear more peaked are more likely to be classed as AC rather than C1. Overall, the number of predicted C1 to AC misclassifications is much smaller than the number of AC sources presented in Section \ref{sec:sample}. Finally, we re-simulated the detection process using a cosmological simulation - not including AGN contamination - over a 25 deg$^2$ area (Bhargava et al. in preparation). We processed the field following the tile system detailed in Section \ref{subsec:dataprocessing}, taking into account pointing overlaps. The results show the fraction of C1 clusters misclassified as AC is similar to that obtained with the single-pointing simulations.

We define two sub-classes within the AC category: 1) the pure AC class, which consists of objects that meet only the selection criteria from the \textsc{epn} model, and 2) the C1/C2-AC class, comprising sources that satisfy both criteria. Both of these classes have a concerted impact for X-ray surveys. The pure AC class serves as an indicator of cluster candidates that are not recovered by the latest XXL pipeline due to a highly peaked emission profile. We used this classification to assess the missed fraction of clusters in Section \ref{sec:accosmology}. The second C1/C2-AC class refers to known clusters, but with some unmodelled AGN contribution or cool-core signature. The impact of such sources is more astrophysical; while they do not contribute to the missed cluster fraction, the peaked morphology of C1/C2-AC clusters, in particular if originating from AGN contamination, means their use in scaling relations can be challenging and requires special attention \citep{Eckert2016, Sereno2020, LovisariMaughan2022}.

\section{The AC selection function}\label{sec:acselfunc}
\begin{figure}
    \includegraphics[width=0.5\textwidth]{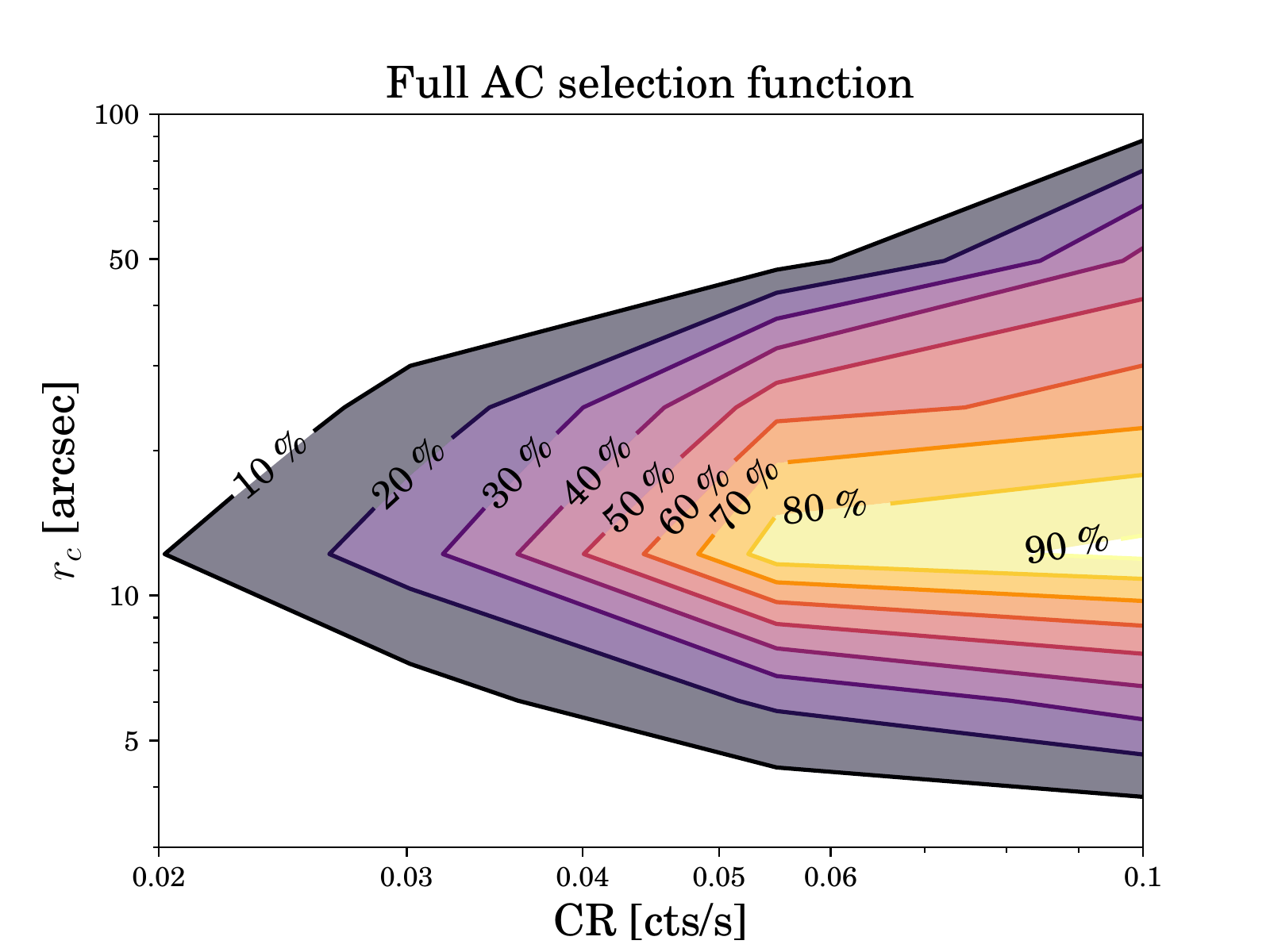}
    \caption{Contours of AC detection probability as a function of the total count rate (CR) in the 0.5--2 keV band and the input core radius ($r_c$) from the $\beta$ model. An exposure time of 10 ks was used along with nominal background value of $b=1$. The flux ratio chosen was 0.5, i.e. half the count rate of the cluster is contained within the central AGN.}
    \label{fig:pure_ac_sel_func}
\end{figure}

To determine the AC selection function, the detection probability of point-source contaminated clusters is computed for each combination of core radius and count rate described in Section \ref{sec:sims}. This is for all sources in the output catalogue that fulfil the AC criteria. The resulting selection function is shown in Figure \ref{fig:pure_ac_sel_func}. The selection function is plotted in the $r_c$-CR observable plane. While the overall shape is consistent with the one derived for the pure C1 case (left panel of Figure \ref{fig:c1selfuncwithcontam}), the most notable difference between the two cases is the more peaked shape of the AC detection probability. Owing to the centrally concentrated X-ray emission within the AC objects, the detection rate falls off more sharply compared to that of C1 clusters as a function of CR, while the range of core radii is narrower. The C1 selection function is illustrated as part of a more detailed assessment of the impact of AC clusters for cosmological applications (Section \ref{sec:accosmology}). 

We emphasise that in order to define the class of AGN-contaminated clusters, we only used the \textsc{Xamin} pipeline parameters. From this classification alone, it is not possible to determine the exact nature of the AC object - simply that it is an extended source with a peaked central emission profile that is better fit by the \textsc{epn} model than either the \textsc{ext} and \textsc{pnt} fits alone. We identify three principal reasons for this: an X-ray point source located at the cluster position (either physically associated or as a result of a foreground/background projection), a cluster with a prominent cool-core, or an X-ray-bright nearby extended object, such as a galaxy with an active nucleus. In principle, X-ray cluster samples are biased by the occurrence of any of these particular features. We aim to characterise the number of AC objects in the XXL survey that come under each of these categories, using complementary, multi-wavelength methods of confirmation.

\begin{table*}
    \centering
    \caption{Summary of the various types of source and their selection criteria. If more than one condition is specified, all conditions must be used unless explicitly stated otherwise.}
    \begin{tabular}{llll}
            \hline
            \noalign{\smallskip}
            Classification & Source type & XXL Paper XXIV & This work \\
            \noalign{\smallskip}
            \hline
            \noalign{\smallskip}
            C1 & Extended & \textsc{ext > 5"} & \textsc{ext > 3"} \\
            & & \textsc{ext\_stat > 33} and \textsc{ext\_det\_like > 32} & Unchanged \\
            \hline
            \noalign{\smallskip}
            C2 & Extended & \textsc{ext > 5"} & \textsc{ext > 3"}  \\
            & & \textsc{ext\_stat > 15} & Unchanged \\
            \hline
            \noalign{\smallskip}
            AC & Extended + central point & \textsc{epn\_ext > 5"} & \textsc{epn\_ext > 5"} \\
            & & \textsc{epn\_stat\_ext > 20} and \textsc{epn\_stat\_pnt > 20} &
            \textsc{epn\_stat\_pnt > 20} \\
            & & or \textsc{epn\_stat\_pnt > 100} & \textsc{epn\_ ratio > 0.2} \\
            \hline
    \end{tabular}
    \label{tab:ACcritcomp}
\end{table*}

\section{The catalogue}\label{sec:sample}

\subsection{Data processing and sample selection}\label{subsec:dataprocessing}
We implemented the AC criteria within the latest version (hereafter V4.3) of the XXL pipeline to undertake a systematic search for AC clusters within the survey footprint. Details of the most recent XXL pipeline are given in XXL Paper XXIV; below we summarise the salient aspects. First, event lists were created from raw observation data files (ODFs) using the SAS software \citep{XMMSAS} tasks \texttt{emchain} and \texttt{epchain}, filtered for solar soft photon flares. The cleaned event lists were then used to produce images of $2.5''$ per pixel to correctly sample the \textit{XMM} PSF ($\simeq 6''$ on-axis) using \texttt{evselect}. Three images were produced - one for each EPIC detector (MOS1, MOS2, and PN) - for three energy bands: [0.3-0.5], [0.5-2.0], and [2.0-10.0] keV. In what follows we predominantly focus on [0.5-2.0] keV images as this is most relevant for cluster detection and characterisation. Departing from the earlier use of approximately 700 single \textit{XMM} pointings spread over 50.9 deg$^2$ of the extragalactic sky, the most up-to-date version features images that are mosaicked into $68' \times 68'$ 'tiles' (the term 'mosaic' is reserved for images consisting of more than one EPIC detector). One tile was created per EPIC instrument, pixelised at $2.5''$ using the SAS tasks \texttt{attcalc} and \texttt{evselect}. The tiling layout is designed such that there is a $4'$ overlap between tiles, with approximately 20-25 pointings per tile. The three individual tile images were coadded into a single mosaic prior to running the XXL source detection pipeline (described in Section \ref{subsec:xmmpipeline}). We detect 27 AC candidates in the northern field, 23 of which are 'pure' AC objects, three are C2-AC, and one is C1-AC. In the southern field we recovered 20 such candidates, 18 of which are pure AC, one is C1-AC, and the other is C2-AC. In XXL Paper XX, a third C3 class was also defined corresponding to optically confirmed clusters selected as C1/C2 by a previous pipeline version, but not by the present one. Typically these clusters exhibit an X-ray emission that is weak enough to be at the detection limit of the pipeline. In this study, we recover one C3 cluster known from the literature using the new AC class. The system, XLSSC 063, is a cluster with a spectroscopic redshift $z = 0.276$ (see Figure \ref{fig:XLSSC63}), with [NII], [OI] and [OII] emission lines in the spectrum of the BCG, indicating the presence of ionised gas in the central galaxy. 

\subsection{Visual screening}

Given that this is the first instance of applying the purely pipeline-driven AC classification to X-ray data, a visual screening process was conducted to confirm the final AC sample. The screening procedure is based on X-ray and optical images. Optical imaging data was taken from Hyper Suprime-Cam \citep[][hereafter HSC]{HSC} in the \textit{gri} bands, the Canada France Hawaii Telescope Lensing Survey\footnote{https://www.cfht.hawaii.edu/Science/CFHTLS/} (CFHTLS, \textit{i}-band images) for the northern XXL field, and from the Blanco Cosmology Survey \citep[BCS,][\textit{i}-band images]{Desai2012} and the Dark Energy Survey Data Release 1 \citep{DESDR1} (\textit{gri} band) for the southern field. Visual inspection is relatively rapid to perform and provides useful information on the broad nature of each X-ray source, as, for example, bright clustered galaxies consistent with a low-redshift cluster, nearby galaxies, clusters with background/foreground/member AGNs, QSOs, two blended point-like sources, stars, or a significant extended X-ray source with a grouping of faint galaxies consistent with a high-redshift cluster. Altogether, 14/47 objects were discarded and 33 were retained. 

Among the 14 discarded objects, five of these were lone QSOs without any visible optical overdensity of galaxies, one was an X-ray detector artefact caused by a nearby star, three were nearby bright stars in the optical images, and five were removed due to their extended X-ray profile appearing as a result of two blended QSOs in the optical image. For all the discarded sources, we do not observe any systematic trend in their \textsc{Xamin}-derived properties compared with genuine contaminated clusters or active galaxies.

\begin{figure*}
    \centering
    \includegraphics[width=0.41\textwidth]{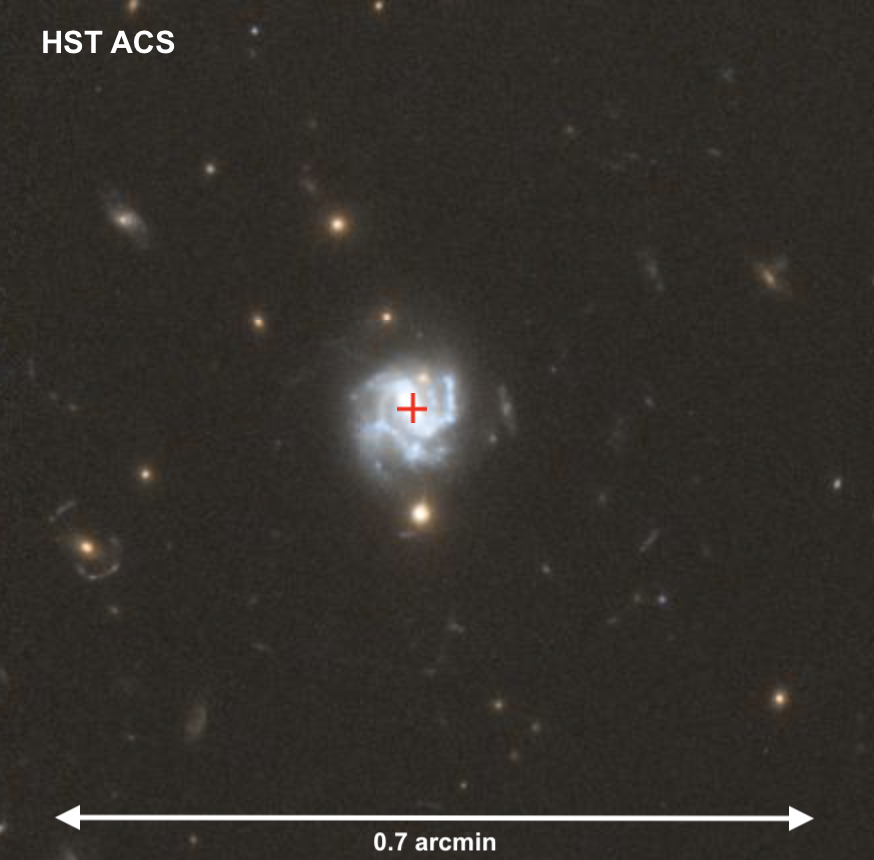}
    \includegraphics[width=0.45\textwidth]{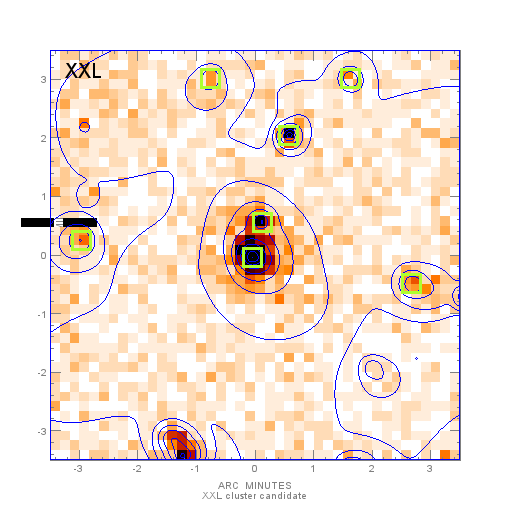}
    \caption{Zoomed-in view of a new AC cluster candidate, XLSSU J022055.4-033332 ($z_{phot}=1.03$). \textbf{Left:} HST ACS F850LP/F775W composite image around the cluster position (red cross), highlighting the distribution of red sequence cluster members behind the star forming spiral galaxy ($z_{spec}=0.15$). \textbf{Right:} Raw $7 \times 7$ arcminute \textit{XMM} image centred on the same source, showing the peaked X-ray profile due to the superposition of point-like (galaxy) and extended (cluster) emission. The X-ray contours are shown in blue. Green squares indicate X-ray-pipeline-detected objects.}
    \label{fig:HSTimageforNT005}
\end{figure*}

The remaining 33 sources (20 in the north, 13 in the south) are considered to be genuine extended sources based on X-ray and optical information, with some level of point-source contamination or cool core. The resulting AC catalogue, presented in Table \ref{tab:acsample}, is a heterogeneous sample, shedding light on the fact that similarity in X-ray profiles can nevertheless be obtained by different types of objects. While we are principally interested in the case of AGN contamination in clusters, we find that the AC classification is effective at detecting active fossil and galaxy groups, as well as single active galaxies. In two cases (XLSSU J022129.1-040531 and XLSSU J021830.7-050126), the AC object is centred on a QSO that is located very close to a distant cluster - XLSSC 034 ($z=1.036$) and XLSSC 064 ($z=0.874$). In these instances, the detection of the AC object is due to the blended X-ray emission from the cluster and the point source. In one further example, we detect a high-redshift ($z_{phot} = 1.03$) cluster, first discovered as RCS J0220.9-0333 in \cite{Jee2011} due to a strong red sequence among cluster members, and later confirmed via the SZ signal in \cite{Hilton2018}. We detected this cluster for the first time due to a 'boost' in the X-ray emission from a low-redshift foreground galaxy in alignment (Figure \ref{fig:HSTimageforNT005}), allowing us to quantify the occurrences of cluster-galaxy projections along the line of sight in the AC sample. 

\subsection{Redshift confirmation}
Out of the 33 candidates, 11 have spectroscopic redshift confirmations published in XXL Paper XX. For six additional objects, we derived spectroscopic redshifts from public (e.g. SDSS, GAMA, AAT) or XXL private data stored in the CEntre de don\'eeS Astrophysiques de Marseille\footnote{https://www.lam.fr/cesam/} (hereafter CESAM). Two objects have spectroscopic confirmation from  the New Technology Telescope (NTT) operated by the European Southern Observatory (ESO). In total, 19 objects in the full catalogue are spectroscopically confirmed. For the objects that possess no spectroscopic confirmation, we report a photometric redshift estimate within 120 arcseconds of the X-ray cluster centre where available. 

Photometric redshifts of the clusters were measured using the Wavelet Z Photometric (WaZP) cluster finder, which uses wavelet-based density maps of galaxies selected in photometric redshift space, removing any assumptions on the cluster galaxy population \citep{Aguena2021}. Details on the method used to compute the individual galaxy redshifts is detailed in \cite{Gschwend2018}. Where referenced, we refer only to the WaZP-based cluster photometric redshifts. All sources with a confirmed WaZP redshift estimate have a S/N $\simeq 3.0$, above which the occurrence of false detections is considered to be negligible. We prioritised the use of a wavelet-based cluster finder rather than one based on the red sequence to confirm the AC sources for the main reason that we are searching for clusters with central AGN contamination, which may appear more 'blue' \citep{Klesman2014}, posing issues for colour-based cluster finders. In the absence of WaZP estimates, we used photometric information from the XMM-BCS survey \citep{Suhada2012} or those which were publicly available via the NASA/IPAC Extragalactic Database (NED). 

We confirm the cluster nature of an AC object if there are at least three concordant spectroscopic redshifts within the extent of the X-ray emission, or if an obvious brightest cluster galaxy (BCG hereafter), close to the X-ray centroid, has a spectroscopic redshift (mirroring the criteria used in XXL Paper XX). Cluster names with the prefix ‘XLSSC’ pertain to spectroscopically confirmed clusters. It constitutes a cumulative cluster catalogue, in the sense that objects are published in subsequent independent papers. In particular six new confirmed clusters are published afresh in this paper (210, 211, 648-651). They are tagged by the last footnote in Table \ref{tab:acsample}. The term ‘cluster candidate’ is used to refer to clusters with insufficient information to be spectroscopically confirmed, and which nevertheless have either a photometric redshift estimate and/or a clear visual overdensity of galaxies. Such objects are provisionally indicated with the ‘XLSSU' acroynm. The source coordinates of these objects may be updated when the final XXL source catalogue is published (Bhargava et al., in preparation) if a \textsc{Xamin} version later than 4.3 is used.

\subsection{Indication of AGN presence}\label{subsec:agnpresence}
We performed two diagnostic checks to indicate the presence of an AGN within the AC objects. We did this for all 33 sources - clusters and individual galaxies - as we aim to quantify to what extent the peaked X-ray profile is indicative of an AGN, irrespective of the object morphology. We began by searching for any publicly available optical spectra for QSOs at or near the object position, using SDSS, NED, and CESAM databases. If no QSO spectrum was available, we searched for the presence of emission lines in the BCG spectrum as an indicator of ionised gas, which may suggest the presence of AGN activity. Four cluster candidates were followed up with the MISTRAL spectrograph\footnote{http://www.obs-hp.fr/guide/mistral/MISTRAL\_spectrograph\_camera.shtml} to search for emission lines that could confirm AGN presence (see details in Section \ref{app:acatlas}). 

\begin{figure*}
    \centering
    \includegraphics[scale=0.95]{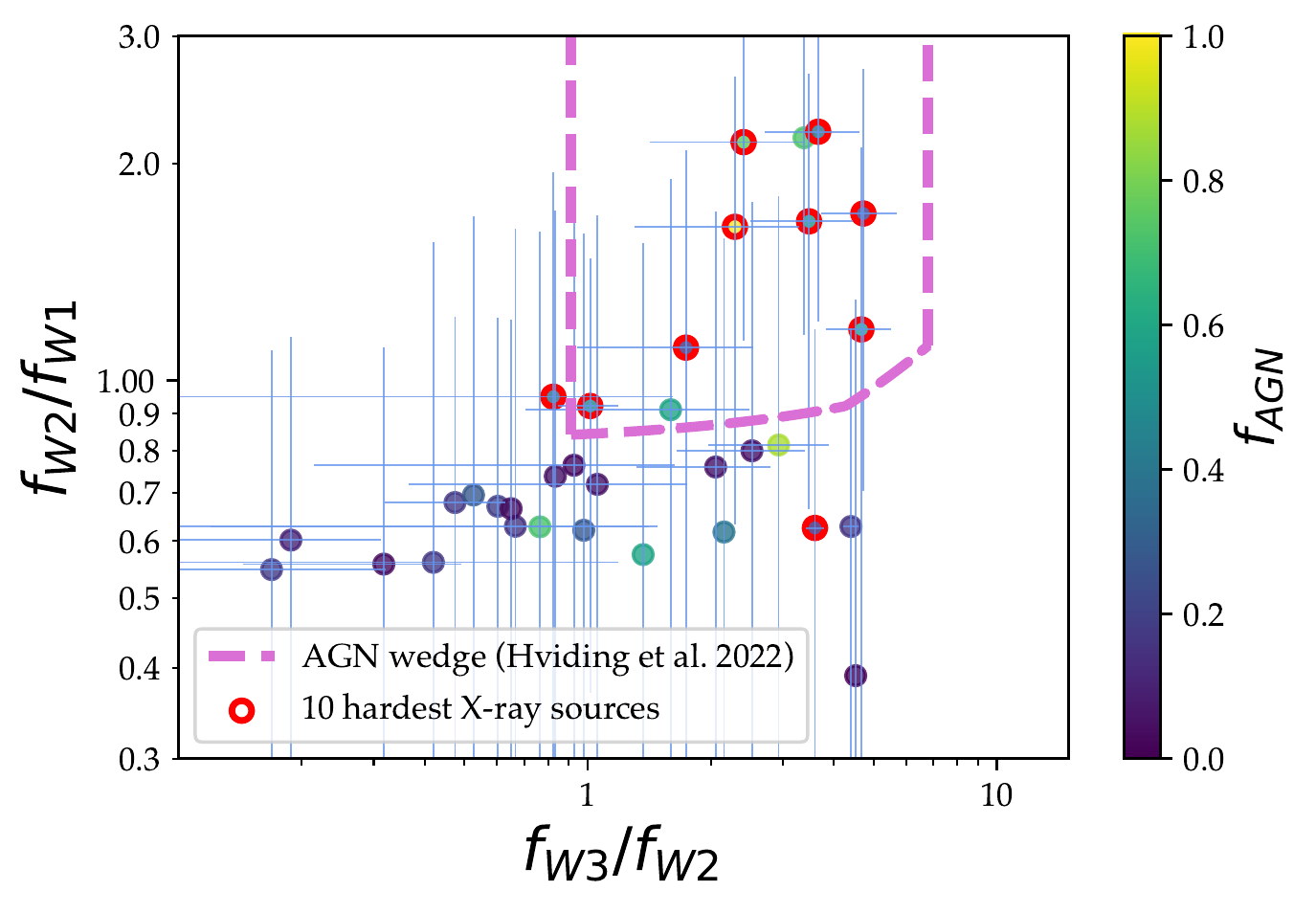}
    \caption{Distribution of 33 XXL AC objects in the mid-infrared colour space. The dashed magenta line encloses the spectroscopic type I and II AGN criteria. The red markers denote AC objects that have the ten largest hardness ratios (an independent X-ray indicator of AGN presence) described in Section \ref{subsec:coolcorepresence}. All objects are coloured according to the level of point-source contamination in the soft X-ray band, which is defined in Section \ref{sec:xrayproperties}.} 
    \label{fig:wiseACAGN}
\end{figure*}

Secondly, we used observations from the publicly available All-Sky Wide-field Infrared Survey Explorer (WISE) data release in four photometric bands, centred at 3.4, 4.6, 12, and $22\mu$m and referred to as W1, W2, W3, and W4, respectively. We probed the infrared power law of AGNs by measuring the flux $f$ of the AC sources in adjacent bands, namely $f_{W1}$, $f_{W2}$, and $f_{W3}$. Hot dust emission from the torus heated by AGN activity is expected to result in high flux ratios, hence allowing us to confirm AGN presence. We searched for all mid-infrared counterparts within an angular radius of $60''$ from the AC position, resulting in 33 mid-IR matches (the full AC catalogue) all within <$11''$ of the object centre. We computed the mid-infrared colour properties of each source in order to assess how many fall within the bounds of type I and II optical spectroscopic AGNs, based on a new selection criterion described in \cite{Hviding2022}. The results are shown in Figure \ref{fig:wiseACAGN}. We report that out of the 33 candidates, ten are revealed to be within the designated 'AGN wedge' (the objects are marked within Table \ref{tab:acsample}). The majority of these sources have independently confirmed redshifts for a QSO at the cluster centre, reinforcing the AGN contamination hypothesis for these cases. The information related to the AGN for each source is listed in Appendix \ref{app:acatlas}. 

\subsection{Indication of cool-core presence}\label{subsec:coolcorepresence}
We elaborate on the AC class further by classifying the fraction of AC clusters where the peaked X-ray profile may be due, fully or at least partially, to the presence of a cool core. In order to do this, we performed a simple hardness ratio test. The hardness ratio (HR) is defined as ${\rm HR = (H-S)/(H+S)}$, where H is the hard ($2-10$~keV) band and S is the soft ($0.5-2$~keV) band count rate measured in the same sized aperture. Out of the full sample, we report nine clusters that have a hardness ratio of -1, referring to clusters with no measurable X-ray emission in the hard band. If the peaked X-ray emission is visible only within the soft band, this can indicate that cooling gas within the cluster core is contributing to the peaked surface brightness profile, rather than clear AGN activity (which is correlated strongly with a non-zero hardness ratio). These nine clusters with a hardness ratio of -1 are marked accordingly within the table. The hardness ratio allows us to examine the overall spectral shape of the AC sources without a dedicated analysis. Given that we are limited by the number of photon counts, spectroscopic confirmation of point sources within the cluster emission is not feasible for all of the AC sources. However we find the X-ray hardness ratio is consistent with the WISE AGN diagnostic described in Section \ref{subsec:agnpresence}, as shown in Figure \ref{fig:wiseACAGN}, suggesting that this can assist in determining the presence of AGNs or cooling flows in each of the sources. We acknowledge there are some AC candidates that do not correspond to clear optical overdensities of galaxies; the most common reasons for this are high ($z$ > 0.6) redshift clusters (see Figure \ref{fig:redshift_comparison_ac_vs_c1}), or foreground AGNs that dominate the optical image. 

\begin{table*}
    \centering
    \caption{Full sample of 33 AC sources. Col. 1 displays the XLSSC name or the new source tag for the cluster based on the latest version of the XXL pipeline. Cols. 2 and 3 give the cluster position; col. 4 is the estimated redshift (see footnote); col 5. is the automated \textsc{Xamin} pipeline classification; col 6. gives the object type. Col 7. provides the spectroscopic redshift of the QSO counterpart where available. Col. 8 provides the AGN contamination fraction measured in the soft X-ray band. The horizontal line divides the 25 clusters and cluster candidates from eight non-cluster AC objects.}
    \begin{tabular}{lcccccccl}
    \hline
    \noalign{\smallskip}
        Object name & R.A. & Dec & $z$ & Class & Type & QSO $z_{spec}$ & $f_{\rm AGN}$ & Notes \\
        \noalign{\smallskip}
        \hline
        \noalign{\smallskip}
        XLSSC 045 & 36.368 & -4.261 & $0.56$ & a & Cluster & 0.56 & 0.20 & $^{a\ddagger}$ \\
        XLSSC 063 & 34.655 & -5.673 & $0.28$ & a & Cluster & - & 0.32 & $^a$ \\
        XLSSC 090 & 37.122 & -4.857 & $0.14$ & a & Cluster & - & 0.28 & $^{a\ddagger}$ \\
        XLSSC 095 & 31.962 & -5.206 & $0.14$ & a & Cluster & - & 0.25 & $^{a\ddagger}$ \\
        XLSSC 115 & 32.680 & -6.58 & $0.04$ & C2-A & Cluster & - & 0.19 & $^{a\ddagger}$ \\
        XLSSC 150 & 37.659 & -4.992 & $0.29$ & C2-A & Cluster & - & 0.20 & $^a$ \\
        XLSSC 210 & 37.624 & -5.225 & $0.19$ & C2-A & Cluster & - & 0.27 & $^{d\S}$ \\
        XLSSC 211 & 33.370 & -5.195 & $0.44$ & a & Cluster & 0.44 & 0.56 & $^{d\S}$ \\
        XLSSC 518 & 349.822 & -55.325 & $0.18$ & C1-A & Cluster & - & 0.19 & $^{a\ddagger}$ \\
        XLSSC 519 & 353.020 & -55.211 & $0.27$ & a & Cluster & - & 0.20 & $^{a\ddagger}$ \\
        XLSSC 595 & 351.690 & -53.811 & $0.21$ & a & Cluster & - & 0.18 & $^{a\ddagger}$ \\
        XLSSC 648 & 356.773 & -53.848 & $0.64$  & a & Cluster & - & 0.27 & $^{d\S}$ \\
        XLSSC 649 & 355.119 & -53.730 & $0.19$ & a & Cluster & - & 0.42 & $^{f\dagger\S}$ \\
        XLSSC 650 & 355.102 & -55.164 & $0.29$ & a & Cluster & - & 0.25 & $^{f\dagger\S}$ \\
        XLSSC 651 & 356.504 & -56.185 & $0.10$ & a & Cluster & - & 0.23 & $^{d\ddagger\S}$ \\
        XLSSU J020435.7-061922 & 31.149 & -6.324 & $0.90$ & a & Cluster candidate & 0.91 & 0.45 & $^{c\dagger}$ \\
        XLSSU J020514.7-045638 & 31.311 & -4.944 & $0.29$ & a & Cluster candidate & 0.36 & 0.49 & $^{c\dagger}$ \\
        XLSSU J022055.4-033332 & 35.230 & -3.558 & $1.03$ & a & Cluster candidate & - & 0.65 & $^c$ \\
        XLSSU J023322.1-045506 & 38.343 & -4.918 & $0.78$ & a & Cluster candidate & 0.78 & 0.71 & $^{b\dagger}$ \\
        XLSSU J232212.6-553259 & 350.553 & -55.550 & - & a & Cluster candidate & 0.82 & 0.49 & - \\
        XLSSU J232713.5-560337 & 351.806 & -56.061 & $0.97$ & a & Cluster candidate & - & 0.20 & $^{e\ddagger*}$\\
        XLSSU J232936.7-555349 & 352.400 & -55.897 & $0.31$ & a & Cluster candidate & 2.03 & 0.36 & $^d$ \\
        XLSSU J233006.5-545553 & 352.526 & -54.932 & - & a & Cluster candidate & - & 0.37 & $^{\dagger}$ \\
        XLSSU J233809.3-555350 & 354.537 & -55.896 & $0.60$ & a & Cluster candidate & - & 0.17 & $^e$ \\
        XLSSU J234705.7-535653 & 356.773 & -53.948 & $0.65$ & a & Cluster candidate & - & 0.39 & $^b$ \\
        \hline
        \noalign{\smallskip}
        XLSSU J020218.3-065958 & 30.576 & -6.999 & $0.05$ & a & Active galaxy & - & 0.27 & $113^a$ \\
        XLSSU J020519.5-062702 & 31.331 & -6.451 & $0.01$ & a & Active galaxy & - & 0.26 & $^{d*}$ \\
        XLSSU J020802.9-050302 & 32.011 & -5.051 & - & a & Active galaxy & 1.86 & 0.27 & $^{g\dagger}$ \\
        XLSSU J021830.7-050126 & 34.628 & -5.023 & $0.87$ & a & QSO & 3.00 & 0.48 & $64^{a\dagger}$\\
        XLSSU J021905.5-051038 & 34.774 & -5.178 & $1.65$ & a & Active galaxy & 1.66 & 0.57 & $^{g\dagger}$ \\
        XLSSU J022129.1-040531 & 35.372 & -4.092 & $1.04$ & a & QSO & 1.23 & 0.32 & $34^{a}$ \\
        XLSSU J022402.5-044134 & 36.011 & -4.693 & $0.04$ & a & Active galaxy & - & 0.23 & $^g$ \\
        XLSSU J022445.6-030224 & 36.190 & -3.040 & - & a & Active galaxy & 1.23 & 0.58 & $^\dagger$ \\
    \hline
    \end{tabular}
    \tablefoot{The `a' class refers to objects which satisfy only the AC criteria, while `C1/C2-A' refers to those which satisfy both the C1/C2 and AC criteria. \newline $^a$ denotes spectroscopic redshift estimates reported in XXL Paper XX. 
    \newline
    $^b$ denotes WaZP photometric redshift estimates.\newline
    $^c$ denotes publicly available photometric redshift estimates. 
    \newline
    $^d$ denotes spectroscopic redshifts stored in CESAM. \newline
    $^e$ denotes photometric redshifts from the XMM-BCS survey \citep{Suhada2012,Bleem2015}. \newline
    $^f$ denotes spectroscopic redshifts from NTT. \newline
    $^g$ denotes publicly available spectroscopic redshifts.\newline
    * denotes objects that were classified as C1 in XXL Paper XX but without redshift estimates. \newline 
    $^\dagger$ denotes AC objects which fall within the WISE type I/II AGN wedge (Figure \ref{fig:wiseACAGN}) \newline
    $^\ddagger$ denotes AC clusters which display an indication of a cool core (Section \ref{subsec:coolcorepresence}) \newline
    $^\S$ denotes new XLSSC clusters, first published in this work. \newline
    Numbers denote the known XLSSC object that is blended with the AC source. Redshifts are quoted for the object in these cases, with additional redshifts provided in the relevant column where applicable. 
    }
    \label{tab:acsample}
\end{table*}    

\section{X-ray properties of the AC sample}\label{sec:xrayproperties}

\subsection{Count rate measurements}

\begin{figure}
    \centering
    \includegraphics[width=0.45\textwidth]{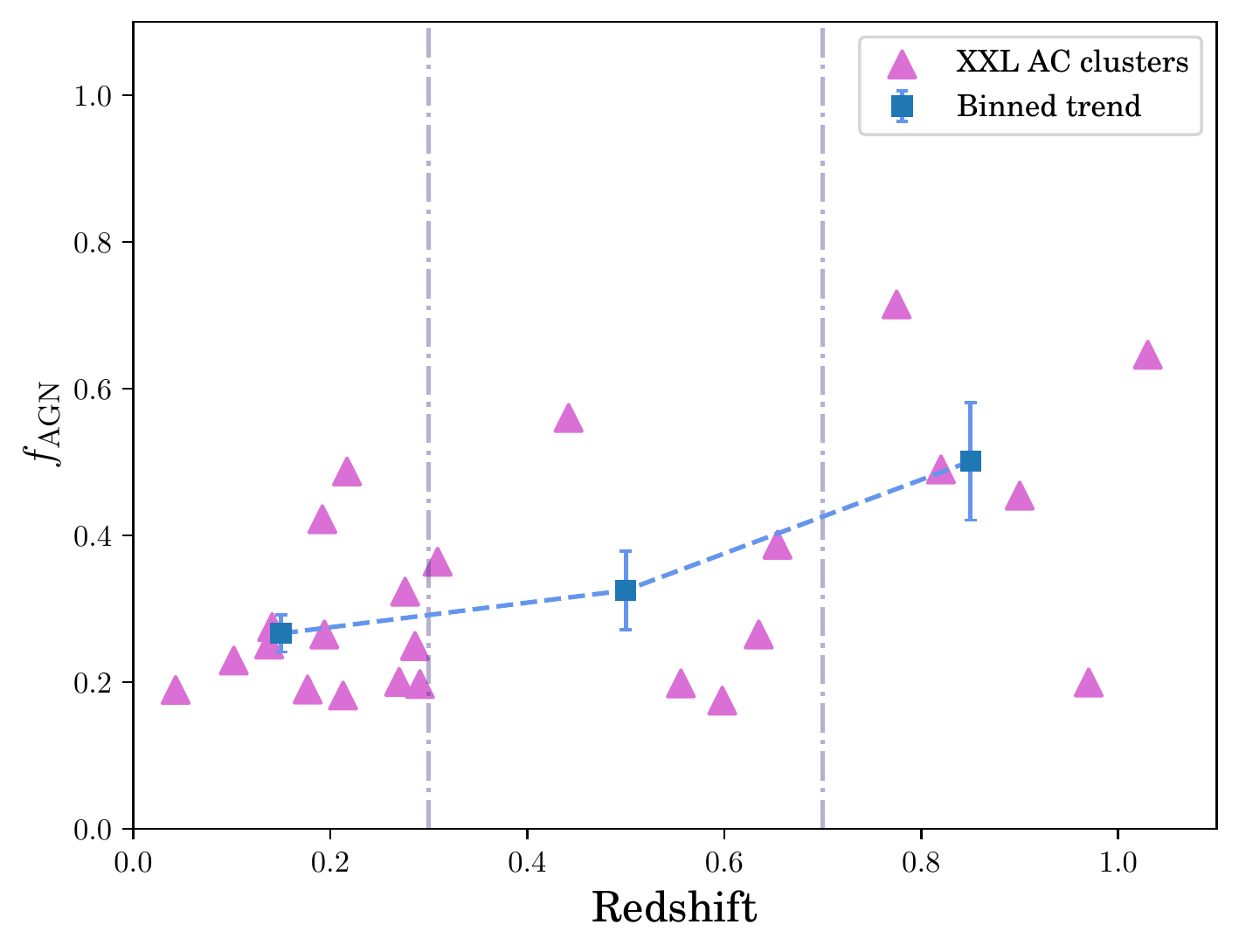}
    \caption{Trend of $f_{\rm AGN}$ as a function of redshift for the AC clusters. Pink triangles illustrate the individual systems, with the binned trend and standard error displayed in blue. The lilac dashed vertical lines denote the binned redshift boundaries: $z < 0.3$, $0.3 < z < 0.7$, $z > 0.7$.}
    \label{fig:contam_z_trend}
\end{figure}

For the AC candidates, the \textsc{Xamin} pipeline provides both an angular core radius ($r_c$) and flux estimate from the $\beta=2/3$ surface-brightness profile. This allows us to derive an approximate count rate for the source within a given radius. The \textsc{epn} model fit also provides the ratio of the fluxes measured for the point source and extended model, from which the individual count rates corresponding to the cluster and AGN can be inferred. The total count rate for the source is determined by summing the individual PN and MOS detectors as follows
\begin{equation}
    {\rm CR} = {\rm CR_{PN}} + 2 \times {\rm CR_{MOS}}
\end{equation}
and the individual rates for the cluster and AGN can be computed via
\begin{equation}
    \begin{split}
    &{\rm CR}_{\textrm{cluster}} = {\rm CR}/(1+\textsc{epn\_ ratio})
    \\
    &{\rm CR}_{\textrm{AGN}} = \textsc{epn\_ratio} \times {\rm CR}/(1+\textsc{epn\_ratio})
    \end{split}
\end{equation}
The fraction of AGN contamination, $f_{\rm AGN}$, of the AC sources can hence be defined as the ratio of the point source contribution to the total flux, which is given as
\begin{equation}
    f_{\rm AGN}={\rm CR}_{\rm AGN}/{\rm CR}
\end{equation}
We plot the distribution of $f_{\rm AGN}$ as a function of the AC redshift in Figure \ref{fig:contam_z_trend}. We find the trend is indicative of a positive correlation between contamination level and the redshift of the cluster. It is important to note, however, this might not necessarily indicate a stronger AGN presence, but rather a decrease of angular resolution of the instrument. This may result in a larger scatter in the $f_{\rm AGN}$ for clusters above a given redshift. Deeper observations are required to more precisely measure the level of point source contribution in these objects. We do not perform a direct comparison with other studies of point-source contamination in high-redshift clusters \cite[e.g.][XXL Paper XXXIII]{Willis2013}, owing to the considerably different methods of selection. It is not useful to compare the current AC sample to clusters detected using the C1 criteria as these are necessarily more diffuse and less peaked in their emission profiles. In particular, high-redshift C1 systems with point source contamination occur predominantly in the XMM-SERVS area – an approximately 4 deg$^2$ region in the northern field, where the sensitivity is up to four times the nominal value for the full XXL area. In contrast, the nature of the AC selection allows for the detection of high redshift AGN-contaminated candidates with considerably lower exposure times on average. The release of the final XXL cluster catalogue will allow us to make a more pointed comparison between the differences of the C1 and AC selection, to better assess AGN population statistics in high redshift X-ray clusters.

\section{Assessment of AGN contamination in the XXL survey}\label{sec:accosmology}

\subsection{Missed fraction of clusters due to AGN}

\begin{figure*}
    \centering
    \includegraphics[width=0.7\textwidth]{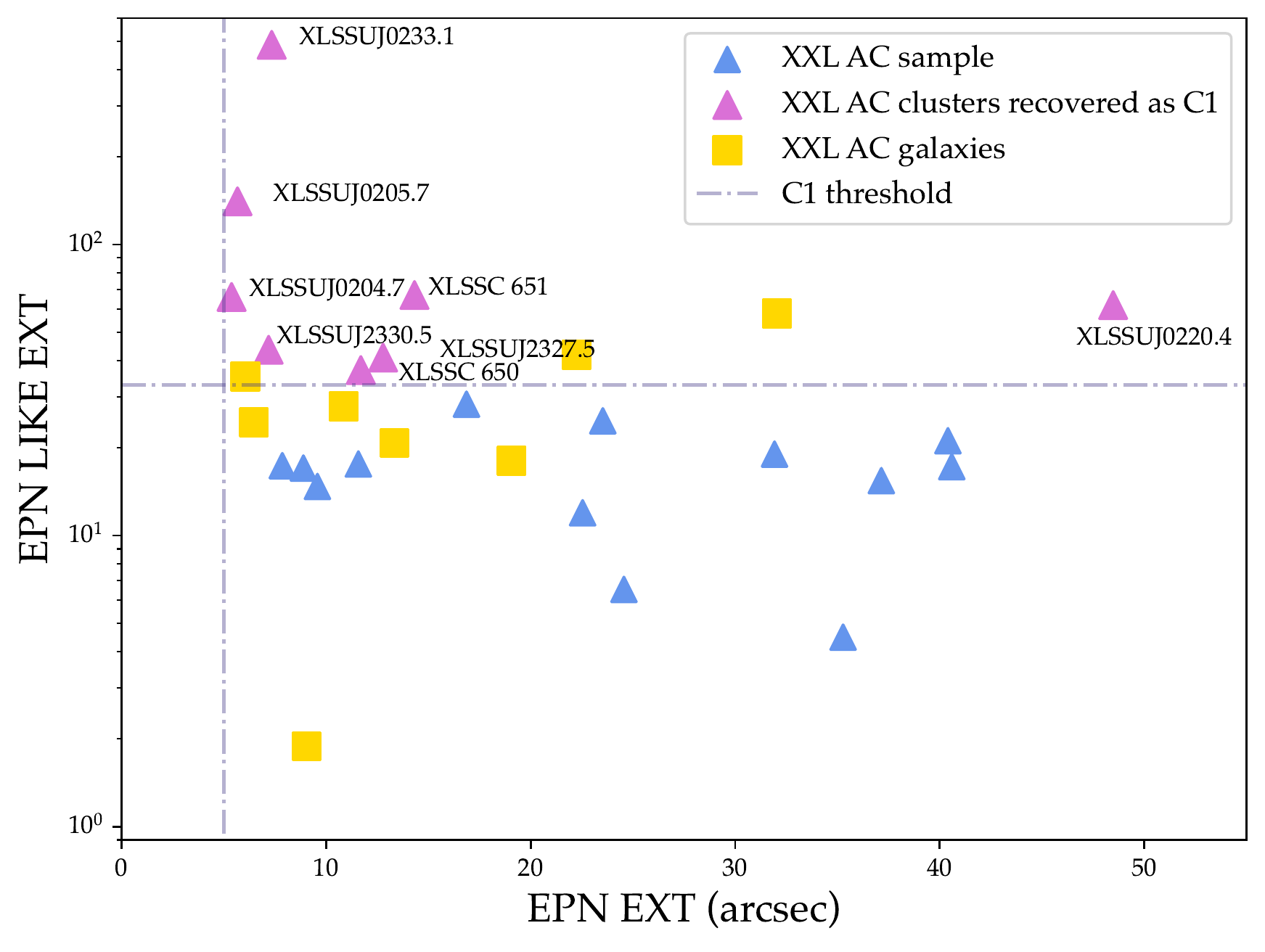}
    \caption{AC clusters recovered as C1 by detection pipeline after the point-source contribution from the central region of the cluster is removed. The recovered clusters are given by the pink triangles, while the total AC population is displayed in blue. Yellow squares denote AC galaxies that would not appear in the final sample following visual screening. The lilac dot-dashed line demarcates the C1 threshold. We note that four AC clusters do not appear in this plot as their \textsc{epn like ext} value is exactly 0, indicating that their profiles were too peaked to be fitted adequately by the extended model.}
    \label{fig:acrecoveredasc1}
\end{figure*}

In this work, we investigated two related but nevertheless distinct concepts - cluster contamination and sample contamination. Cluster contamination refers to the level of point source contamination within the individual system, while sample contamination describes the impact of such objects on the overall purity and completeness of a cluster sample for cosmological use. Previous work by \cite{Bohringer2013} defined the contamination fraction within X-ray surveys to be the number of non-cluster sources within a flux-limited cluster sample. However, since the C1 class defined within XXL is calibrated to be above 90\% pure, we instead focused on quantifying some of the C1 incompleteness. This is done by analysing the 'missed' fraction of clusters - those that are missed from the final sample due to the presence of AGNs. To do this, we computed the fraction of AC sources that would be classed as C1 if the emission from the central AGN were removed. The \textsc{epn} model is a superposition of the \textsc{ext} and \textsc{pnt} models, so we analysed the \textsc{epn\_ext} and \textsc{epn\_ext\_like} parameters that were analogous - but strictly speaking, not identical - to the C1 selection criteria presented in Table \ref{tab:ACcritcomp}. This set of criteria corresponds directly to the extended $\beta$-model component of the \textsc{epn} fit, and is used to mirror the selection for the C1 class shown in Table \ref{tab:ACcritcomp}. Out of the 33 objects, we find that 11 fulfil the criteria, and eight of these are clusters. The distribution of these clusters is displayed in Figure \ref{fig:acrecoveredasc1}. If we consider the C1 sample used in the latest XXL cosmological analysis (XXL Paper XLVI), this corresponds to a missed fraction of 5\%. In other words, 5\% of genuine clusters are excluded from the cosmological dataset due to contamination from a central AGN.

\begin{figure}
    \centering
    \includegraphics[width=0.45\textwidth]{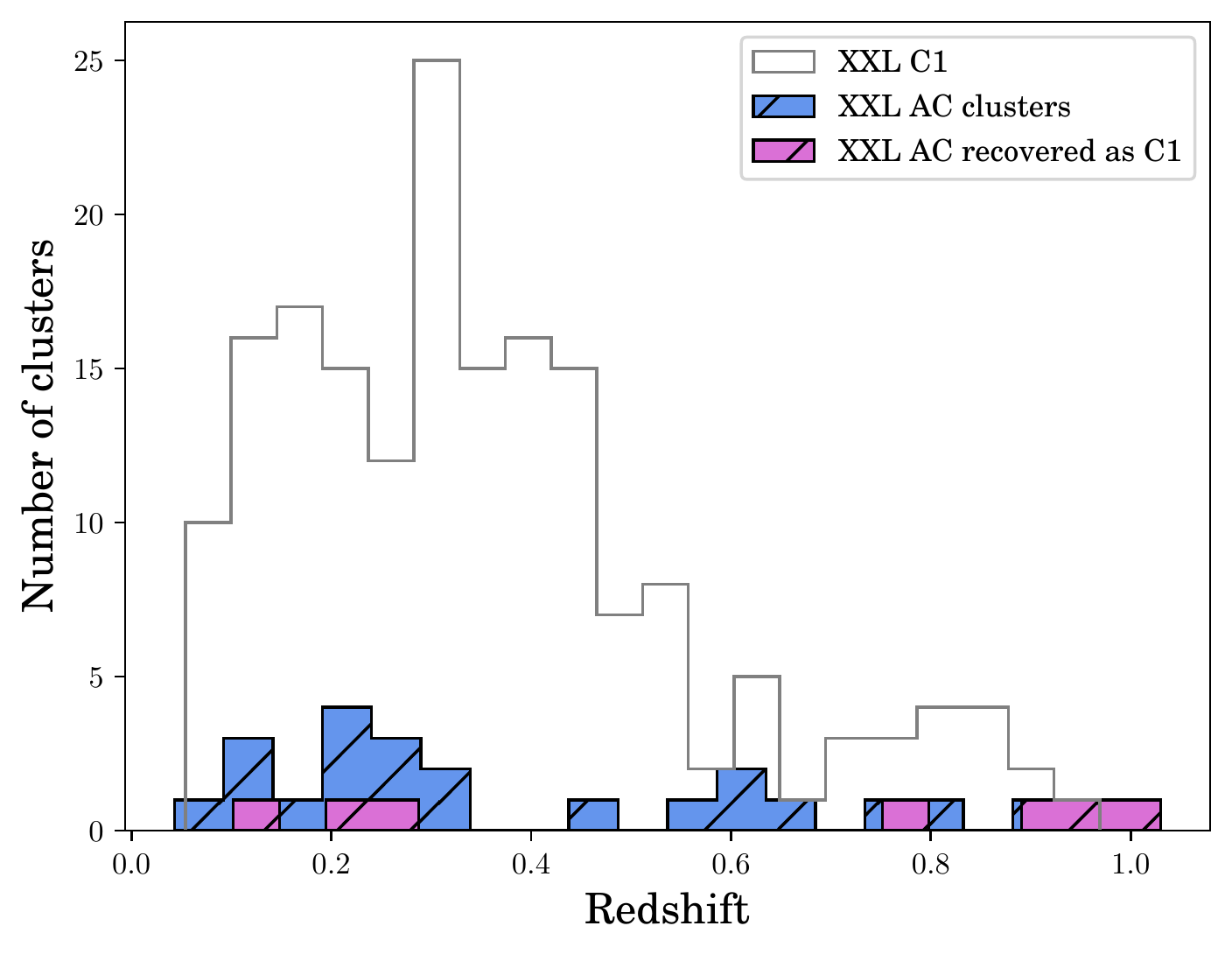}
    \caption{Redshift distribution of the AC clusters in this study compared to the C1 sample used in the previous cosmological analysis of XXL Paper XLVI. The ratio of AC to C1 clusters is approximately 50 percent at $z \sim 0.6$, indicating that the population of AC clusters may increase as the number of detected C1 objects decrease.}
    \label{fig:redshift_comparison_ac_vs_c1}
\end{figure}

Strikingly, after removing the point-source contribution from the AC sources, clusters can be recovered in the $0.8 \leq z \leq 1$ range, suggesting that AC and cool-core clusters may help to explain the deficit of detected X-ray clusters above $z > 0.6$ (Figure \ref{fig:redshift_comparison_ac_vs_c1}). Such a deficit has been reported in X-ray cluster samples based on the predicted number density of clusters using the Planck CMB cosmological model \cite[][]{Planck2014}. This deficit has been observed within both the northern and southern XXL fields, yet its origin remains unclear \citep{Clerc2014,Pacaud2018}. Independent X-ray samples such as \cite{McDonald2013} similarly report a deficit of high-redshift ($z \geq 0.75$), cuspy, cool-core clusters. While the AC sample of objects is considerably smaller in size compared to the C1, Figure \ref{fig:redshift_comparison_ac_vs_c1} shows that is more homogeneous across the overall redshift range. Since the X-ray luminosity depends on the gas density squared, it is expected that X-ray clusters at high redshift are more likely to be detected if they have more peaked profiles, hence the AC classification is a critical tool to recover clusters that are otherwise missed by the C1 classification alone. Finally, we note that all the C1/C2-AC objects, after removing the central point-like emission, are no longer classed as C1 objects. This is not unexpected since we are removing the majority of the flux from objects that are already classed as C1/C2 by the pipeline. While these objects do not impact the cosmological dataset, since they are known by definition in the C1 selection function, the number of C1/C2-AC may inform as to the cool-core fraction of the C1 sample. Interestingly, these objects also displayed no clear signature from an AGN from the criteria outlined in Section \ref{subsec:agnpresence}, suggesting that their AC classification may support a cool-core morphology rather than clear point-source contamination. 

The missed cluster fraction from other X-ray surveys such as eROSITA and \textit{Athena} is likely to be determined by five main factors. Three of these are instrumental (the sensitivity, PSF size, and background level) and two are survey-dependent (exposure time and survey area). Given the flux limit in XXL is $\sim 80$ photons for C1 clusters, we are able to detect a cluster of luminosity $L = 10^{44}$ erg s$^{-1}$ at this limit up to a redshift $z \sim 0.8$. Assuming the same background level and exposure time, the increased sensitivity of the \textit{Athena} Wide Field Imager (WFI) will reach the equivalent SNR limit $\geq 5$ for such a cluster at a redshift $z \sim 1.9$. This will result in the detection of many more systems, therefore also increasing the number of the clusters missed due to AGN presence. Given that the peak of cosmic AGN activity occurs at $z \sim 2$ \citep{Aird2015}, we can infer that the missed fraction for \textit{Athena} is likely to be larger than for the XXL survey. 

\subsection{Impact on cosmological parameters}

As described in Section \ref{sec:acselfunc}, the selection function for AC clusters was computed in the CR--$r_c$ parameter space using three flux ratios. We then applied the XXL pipeline to select AC objects in the \textsc{Xamin} output parameter space (\textsc{epn pnt stat}, \textsc{epn ext} and \textsc{epn ratio}). This is subsequently mapped back into the input CR--$r_c$ parameter space; this quantifies the probability of detecting clusters at each CR--$r_c$ combination. We emphasise that it is not the distribution of \textsc{Xamin} output CR and $r_c$ values but rather the input values, following the prescription first described in \cite{Pacaud2006}.

While the number of missed clusters within the XXL survey is limited by the small survey area and relatively high exposure on average for extended sources (resulting in fewer misclassifications between clusters and AGNs), consequences may be more drastic for larger X-ray surveys with smaller exposure times, where the detection of extended sources may be more impacted by point-source contamination \citep[see e.g.][]{Bulbul2021}.

\begin{figure*}
    \centering
    \includegraphics[width=0.45\textwidth]{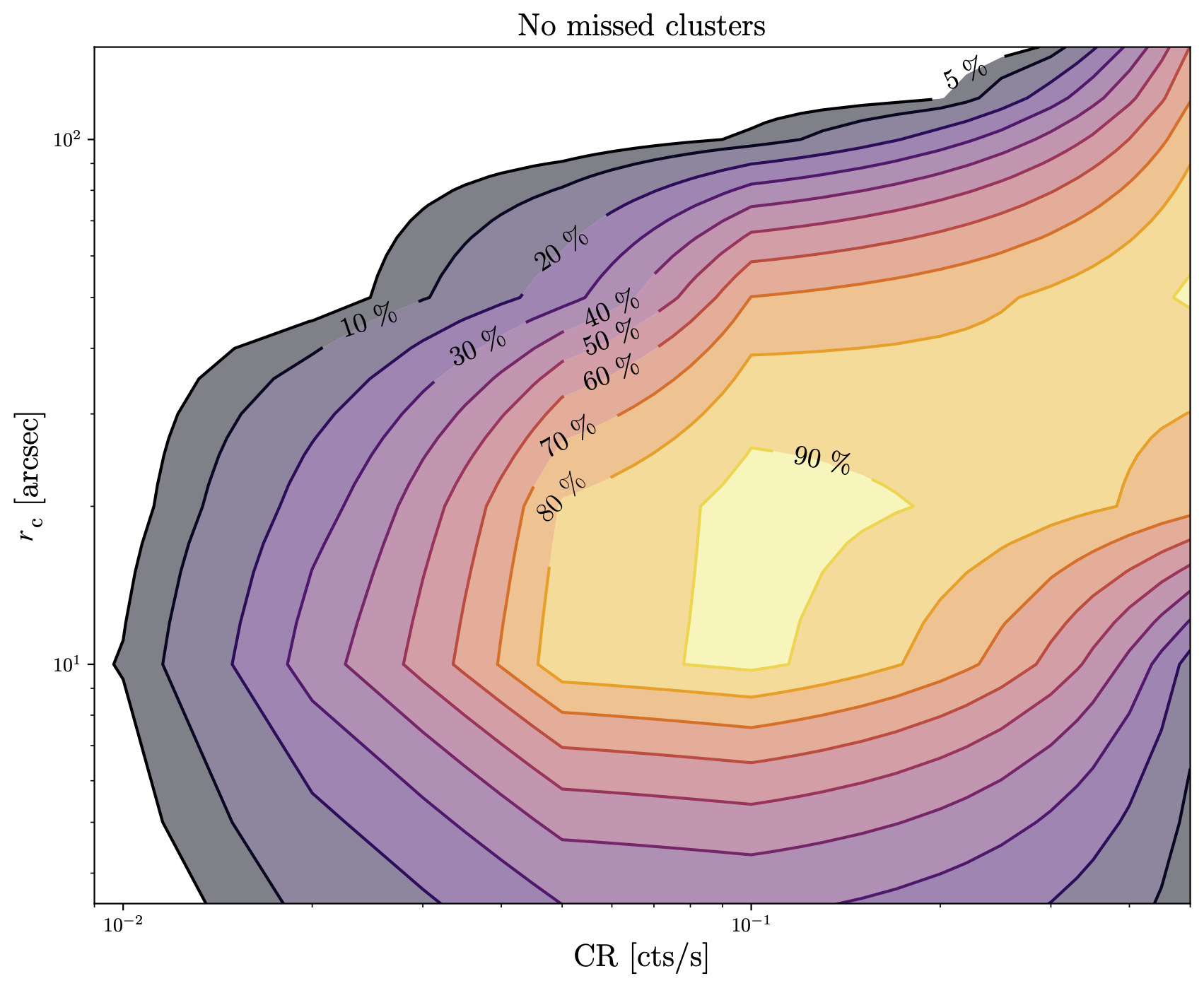}
    \includegraphics[width=0.45\textwidth]{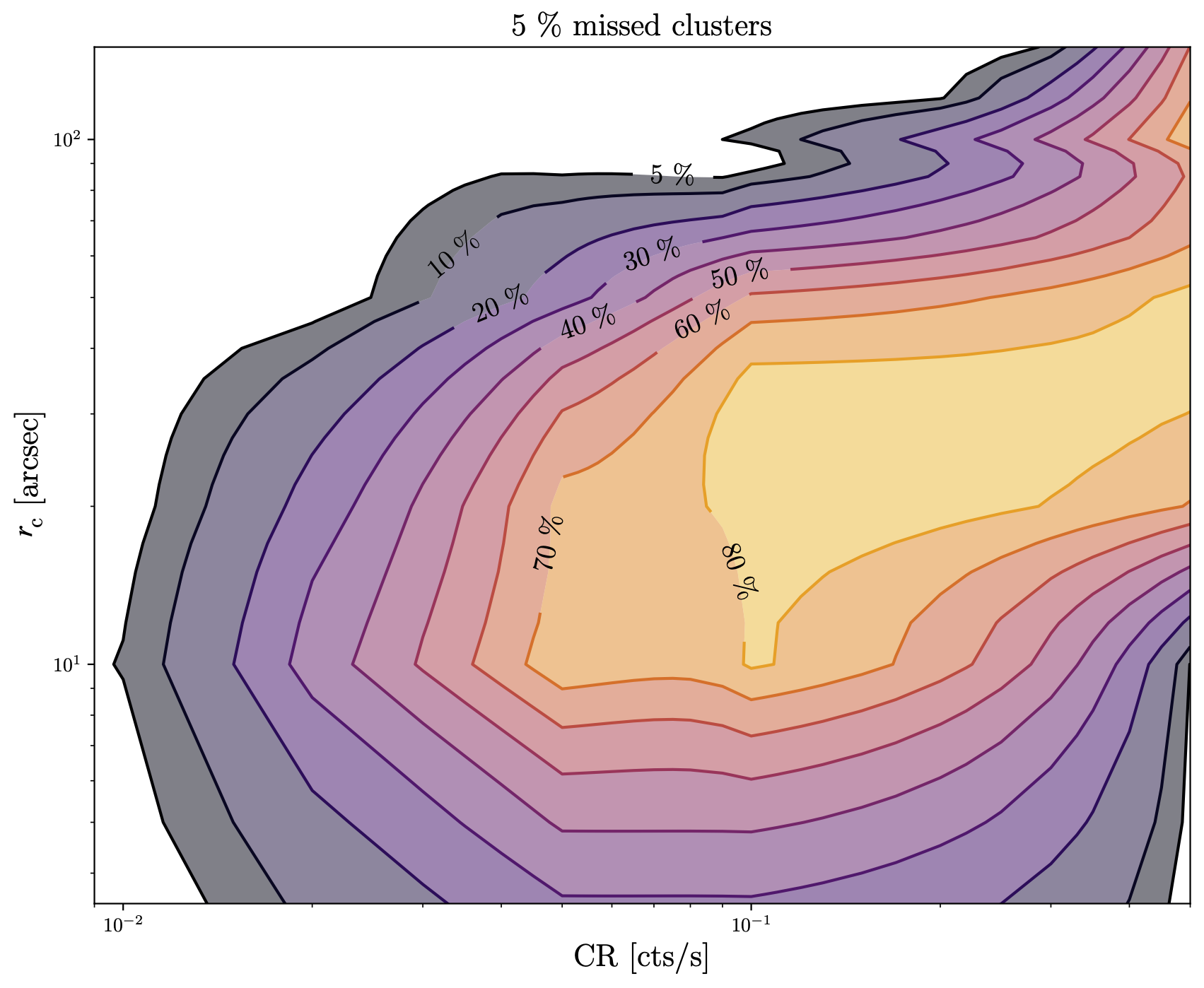}
    \caption{Impact of AGN contamination on XXL selection function in the CR-$r_c$ parameter space. We considered two cases: a pure case where no clusters are missed due to AGN contamination (left), and the measured missed fraction from XXL, constituting 5\% of the cluster population. The detection probability is reduced in the case of 5\% missed clusters, particularly in the CR > 0.1 region, reinforcing the hypothesis that very peaked clusters are excluded by the C1 selection alone.}
    \label{fig:c1selfuncwithcontam}
\end{figure*}

We therefore estimated the impact of missed clusters by modelling the C1 in tandem with the pure AC selection function, to take into account the lost fraction of clusters

\begin{equation}\label{eq:selfunccontam}
    {\rm C1}_{\rm selfunc, \rm contam} = {\rm C1}_{\rm selfunc} - d \left< \frac{{\rm C1}_{\rm selfunc}}{ {\rm AC}_{\rm selfunc}}\right>  {\rm AC}_{\rm selfunc} 
\end{equation}
where $d$ is chosen to denote the percentage of clusters missed from the C1 sample due to their pure AC classification. We compared diagrams for the selection function for two $d$ values: 0 (no missed clusters) and 0.05 (5\% contamination), shown in Figure \ref{fig:c1selfuncwithcontam}. The 5\% value is chosen based on the eight clusters that are 'missed' out of 178 in the latest cosmological sample. The difference between the two cases reveals the change in overall shape of the detection probability based on the fraction of missed clusters from the final sample. Both the C1 and AC selection functions in this case were computed using the simulations described in Section \ref{sec:sims}.

To quantify the cosmological impact of mis-modelling the selection function due to the presence of AC clusters, we study the $\Omega_m$--$\sigma_8$ parameter space for the case of 5\% missed clusters for two levels of sky coverage: a) $47.36\ \mathrm{deg}^2$ (XXL-like) and b) $1000\ \mathrm{deg}^2$. 

We used the ASpiX \citep[][XXL Paper XLVI]{Clerc2012b} method to perform the cosmological analysis with the following method. In each case, we generated a predicted diagram for a fiducial cosmology with a selection function corresponding to the percentage of clusters missed due to AGN contamination (5\%). We then rescaled it to the chosen survey area and applied Poisson noise (the same seed is used in all cases). Finally, we applied a Markov Chain Monte-Carlo (MCMC) approach to estimate posterior distributions and the log-likelihood is chosen to only account for Poisson noise,
\begin{equation}\label{eq:log_like_cosmo}
    {\cal L} = \sum_{z_i} \left[ \bar n - \sum_j \hat N_j \ln(\bar n_j) \right]_{z_i}
\end{equation}
where $\bar n$ is the total number of predicted clusters in the redshift bin, $i$, and $\hat N_j$, and $\bar n_j$, respectively, refer to the observed and predicted number of clusters in the CR--HR bin $j$ for a redshift $z_i$. We ran two different analyses: with the selection function computed taking into account that 5\% of sources are 'missed' by the C1 class ($d=0.05$ in Equation \ref{eq:selfunccontam}), and one with the pure C1 selection function ($d=0$). The fiducial parameters are chosen to be the ones measured from XXL Paper XLVI (XXL-HSC ASpiX + XXL cluster clustering + BAO), namely $\Omega_m = 0.364$ and $\sigma_8 = 0.793$.

\begin{figure*}
    \centering
    \includegraphics[width=0.45\textwidth]{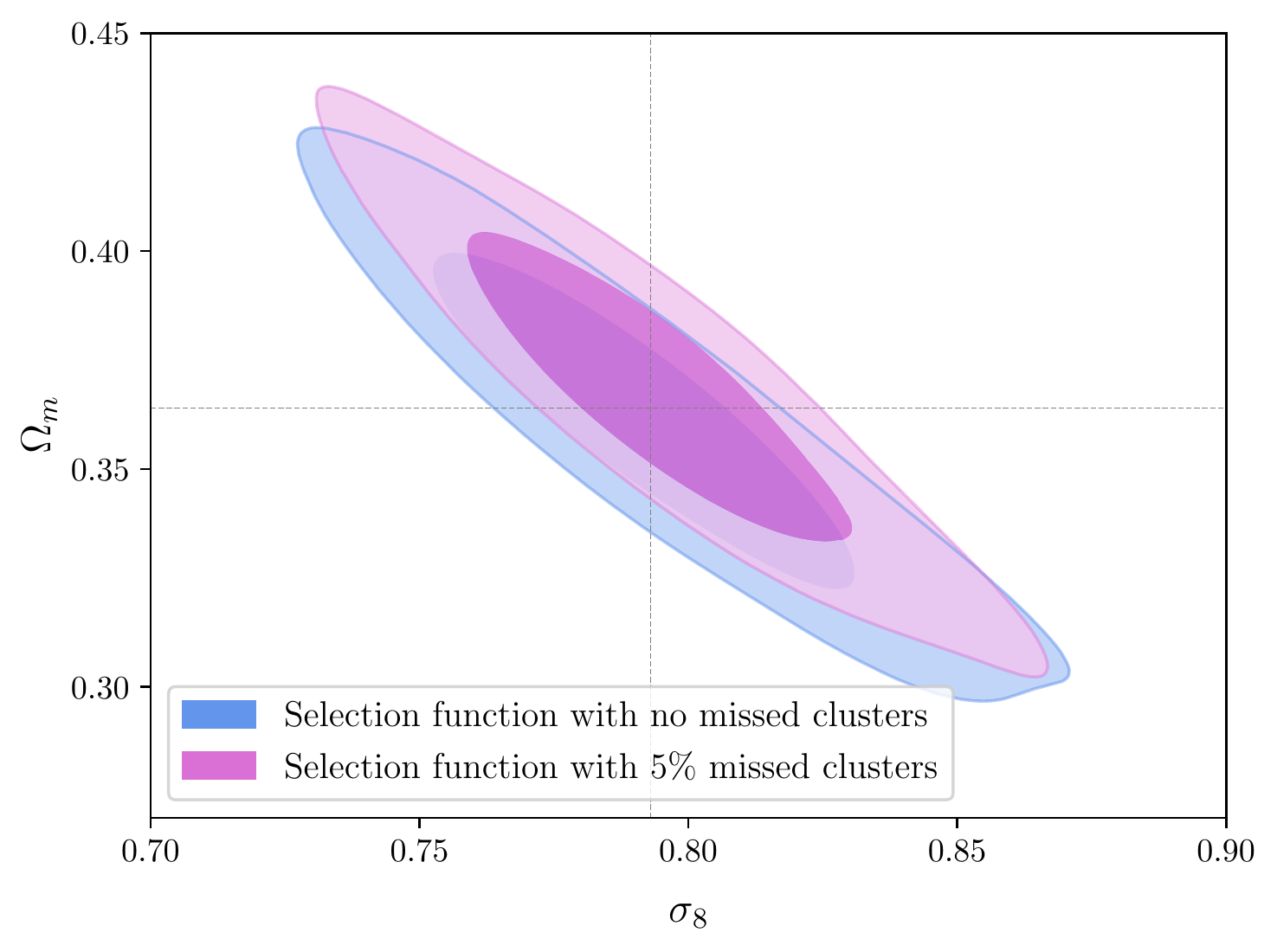}
    \includegraphics[width=0.45\textwidth]{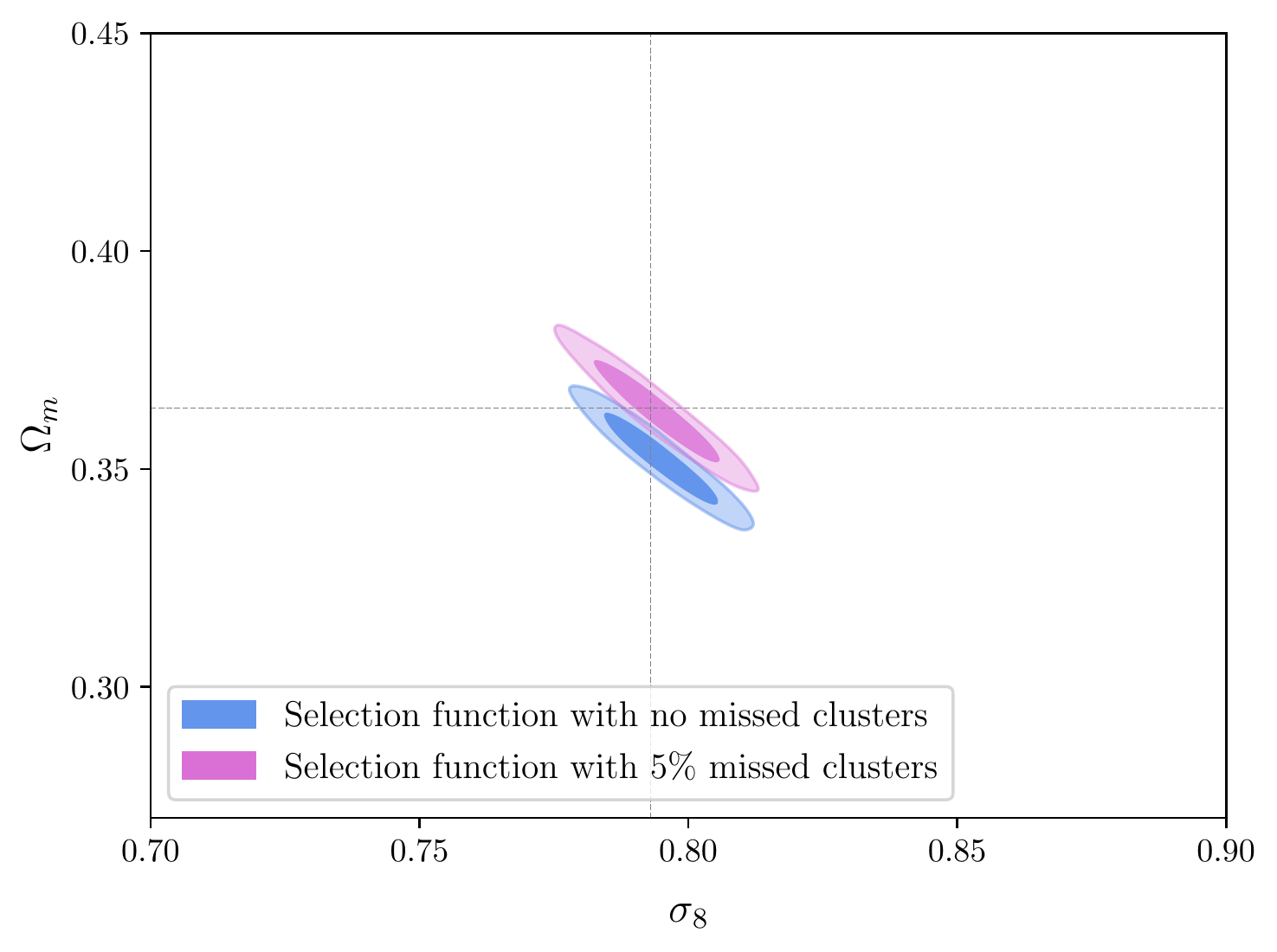}\\
    \caption{Bias introduced in $\Omega_m$--$\sigma_8$ plane without accounting for missed clusters in the cluster selection function. All other parameters, including scaling relation coefficients, are fixed. The pink contours show the recovered constraints from a complete selection function, while the blue contours reflect the results obtained by fitting the cosmological parameters assuming only pure, extended sources. The contours represent the 68\% and 95\% confidence intervals respectively. The dashed lines indicate the fiducial input values used for the mock number density of clusters (see XXL Paper XLVI for further details). The left panel shows the results for an XXL-like survey area and the right one shows results for a survey of $1000\ \mathrm{deg}^2$.}
    \label{fig:cosmo_results}
\end{figure*}

The cosmological posterior estimates for the two surveys are shown in Fig.~\ref{fig:cosmo_results}. As expected, for an XXL-like survey, parameter uncertainties are dominated by Poisson noise. We found that a correct modelling of the selection function results in a $\Omega_m$--$\sigma_8$ posterior distribution that is consistent with the fiducial values, while a selection function accounting for only extended sources in its estimation also remains in good agreement within 1$\sigma$. However, going to a $1000\ \mathrm{deg}^2$ survey area, the discrepancy is significantly more pronounced and shows how AGN contamination could be problematic for X-ray surveys to come. While the more accurate modelling of the selection function, shown by the pink contour, encompasses the fiducial value within 1$\sigma$ well, the selection function that does not account for AGN contamination shows an $\sim 3\sigma$ tension with the fiducial values. We emphasise that increasing the survey size is used as a proxy for increasing the number of clusters in the sample. While the 1000 $\mathrm{deg^2}$ realisation predicts $\sim$ 4000 clusters based on the XXL selection function, we reiterate that eROSITA will likely detect $\sim 10^5$ clusters, and hence this tension may be larger.

Finally, the results obtained in XXL Paper XLVI are not significantly impacted by the contamination level from AGNs. As outlined in Figure \ref{fig:cosmo_results}, the impact of contamination at the 5\% level for an XXL survey area is negligible, though we stress that our model comparison fixes all relevant parameters aside from $\Omega_m$ and $\sigma_8$. However, for future surveys such \textit{Athena}, where clusters will be detected out to a redshift of $z \sim 2$, we anticipate a considerably larger contamination rate from AGNs within clusters. In a similar vein, all-sky missions such as eROSITA may plausibly have a contamination fraction that is larger than $5\%$. Due to the larger survey area, we suggest that without proper modelling of the selection function, a $5\%$ missed fraction of genuine clusters may constitute a lower rather than upper limit for such surveys.

\section{Summary and conclusions}\label{sec:summary}

The AC sample is the first pipeline-derived catalogue of clusters with measurable AGN contamination within the XXL survey. In particular, the characterisation of the 25 AC clusters, using X-ray and multi-wavelength diagnostics, forms a valuable dataset to better understand the evolution of AGN with clusters and their impact on cosmology. We used extensive \textit{XMM}-like image simulations to define a parameter space to capture AC clusters by modelling the point-like and extended X-ray emission simultaneously. We then applied this criteria within the XXL pipeline to generate a sample of clusters impacted by AGN presence or with cool core signatures. Our work revealed that AGN contamination in clusters is present well into the intermediate redshift range ($0.5 < z < 1$), consistently with other studies \cite[e.g.][]{Logan2018,Maughan2019}. We found that removing the point-source flux contribution from these objects allows for the recovery of genuine clusters in parts of the mass-redshift plane currently excluded by the canonical cluster selection function, implying that clusters are 'missed' by current X-ray detection methods. We estimated the impact of these missed clusters to be of the order 5\% in the most recent XXL dataset. Finally, we quantified the impact on the $\Omega_m$--$\sigma_8$ parameter space as a result of improperly accounting for these missed objects within the selection function. Consequences are not drastic for small XXL-like areas, but likely to be more substantial for other X-ray surveys. Our future work will involve finding additional, complementary methods to determine AGN presence within clusters, exploiting both the spectral and image properties of these objects; for example, machine learning methods to denoise and increase the spatial resolution of \textit{XMM-Newton} images \citep{Sweere2022} may lead to better classification and identification of AC systems. One limitation of our study is that unlike cool cores, contaminating AGN are not necessarily located in the centre of the X-ray cluster emission, and hence the \textsc{epn} model developed so far is limited in identifying only missed clusters where AGNs are sufficiently close to the X-ray centre. Future work will include placing the point sources in different locations relative to the cluster emission and assess the resulting detection probability. We will also aim to compare the evolution of AGN contamination in clusters using hydrodynamical simulations. A natural extension of this work will be to subsequently subtract the AGN emission from the cluster flux in order to render contaminated clusters usable for scaling laws and cosmological studies. With larger samples it will further be possible to quantify the co-evolving fraction of AGNs within clusters as a function of redshift. Overall, the class of AC clusters is rich for both astrophysical and cosmological uses, with a potentially significant impact on future X-ray studies. Larger and deeper datasets will allow for a more precise determination of the properties of these objects to maximise the cosmological potential of clusters.  

\section*{Acknowledgements}

XXL is an international project based around an XMM Very Large Programme surveying two 25 ${\rm deg^2}$ extragalactic fields at a depth of ${\rm \sim 6\ \times\ 10^{-15}\ erg\ cm^{-2}\ s^{-1}}$ in the [0.5-2] keV band for point-like sources. The XXL website is \url{http://irfu.cea.fr/xxl}. The authors would like to thank the anonymous referee for instructive comments that helped improve the manuscript considerably. The Saclay team (SB, MP, NC) acknowledges long term support from the Centre National d'Etudes Spatiales (CNES). SB acknowledges a CNES postdoc and support from CNRS, support from the ESA Archival Research Visitor Programme, and would like to thank P. A. Giles, A. Pellissier, J. B. Melin, and R. T. Duffy for fruitful comments. This work was supported by the Programme National Cosmology et Galaxies (PNCG) of CNRS/INSU with INP and IN2P3, co-funded by CEA and CNES. BJM acknowledges support from STFC grant ST/V000454/1. This work was based in part on observations made at Observatoire de Haute Provence (CNRS), France, with the MISTRAL instrument. This research has made use of the MISTRAL database, based on observations made at Observatoire de Haute Provence (CNRS), France, with the MISTRAL spectro-imager, and operated at CeSAM (LAM), Marseille, France.

This research made use of {\sc Astropy}\footnote{http://www.astropy.org}, a community-developed core Python package for astronomy \citep{astropy:2013, astropy:2018}. This work also made use of the \textsc{pyproffit} package \citep{Eckert2020}, as well as {\sc numpy}, {\sc scipy} and {\sc matplotlib}. The data underlying this study are available in the article. 

\bibliographystyle{aa}
\bibliography{ACPAPER}
\begin{appendix} 
\section{Atlas of final AC objects in this study}\label{app:acatlas}

We present a visual catalogue of all sources presented in Table \ref{tab:acsample}. The layout of the atlas is as follows. The left panel shows a $4 \times 4$ arcminute optical (HSC for the nothern field candidates, DES for southern) \textit{gri} cutout centred on the AC object. The orange contours represent the X-ray emission, centred on the X-ray peak value. The pink cross denotes the location of the weighted X-ray barycentre. The middle panel shows the $7 \times 7$ arcminute $i$-band image (CFHTLS for XXLn, BCS for XXLs) centred on the AC object (X-ray contours are displayed in blue). The right panel shows the $7 \times 7$ arcminute raw X-ray photon image. The green symbols correspond to XAmin pipeline detections. Green squares correspond to AC objects (and point sources), while green circles correspond to pure extended sources.

\begin{figure*}
    \centering
    \includegraphics[width=0.3\textwidth]{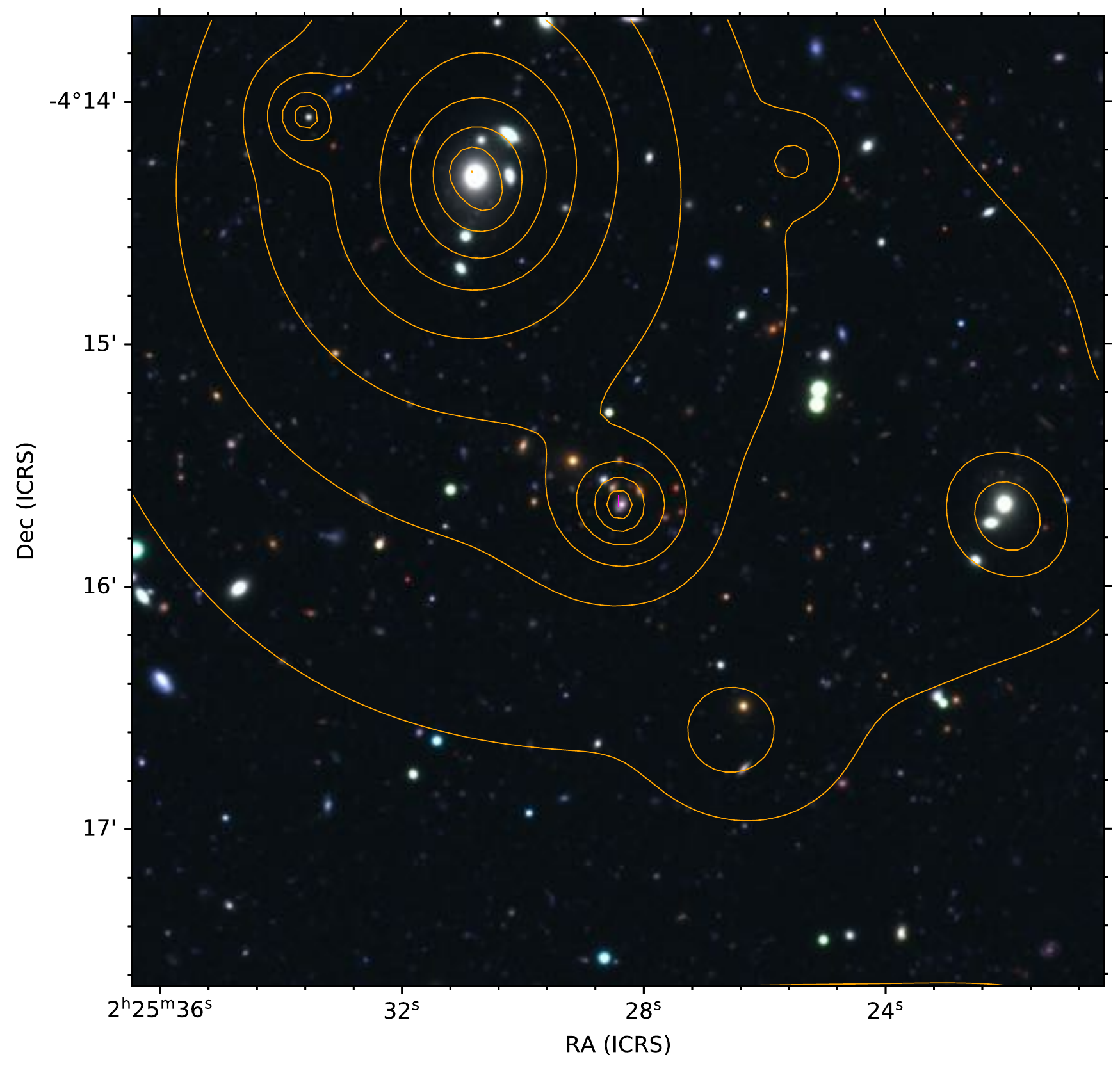}    
    \includegraphics[width=0.3\textwidth]{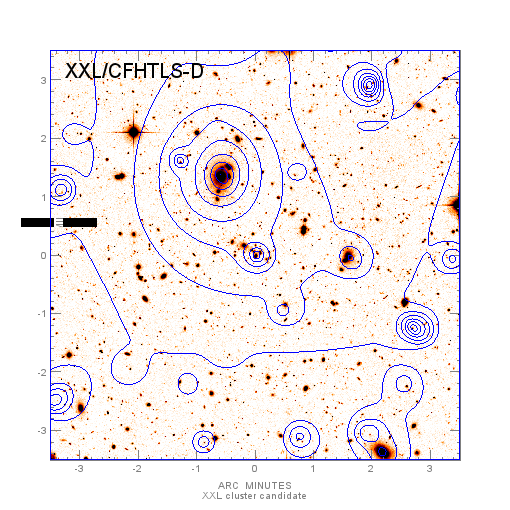}
    \includegraphics[width=0.3\textwidth]{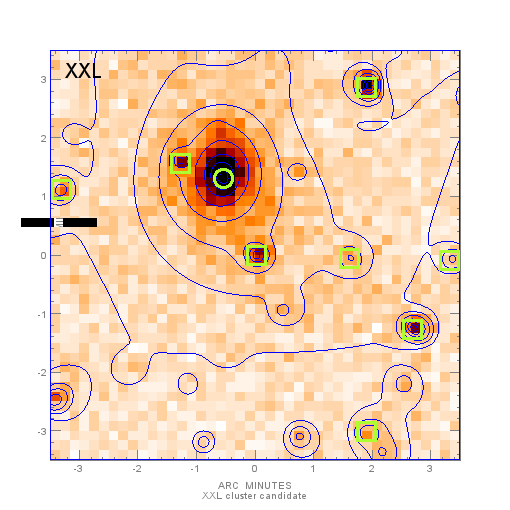}
    \caption{Cluster XLSSC 045 at $z_{spec}=0.556$. At the centre of the X-ray emission we find an AGN (not the BCG) at $z_{spec}=0.563$.}
    \label{fig:XLSSC45}
\end{figure*}

\begin{figure*}
    \centering
    \includegraphics[width=0.3\textwidth]{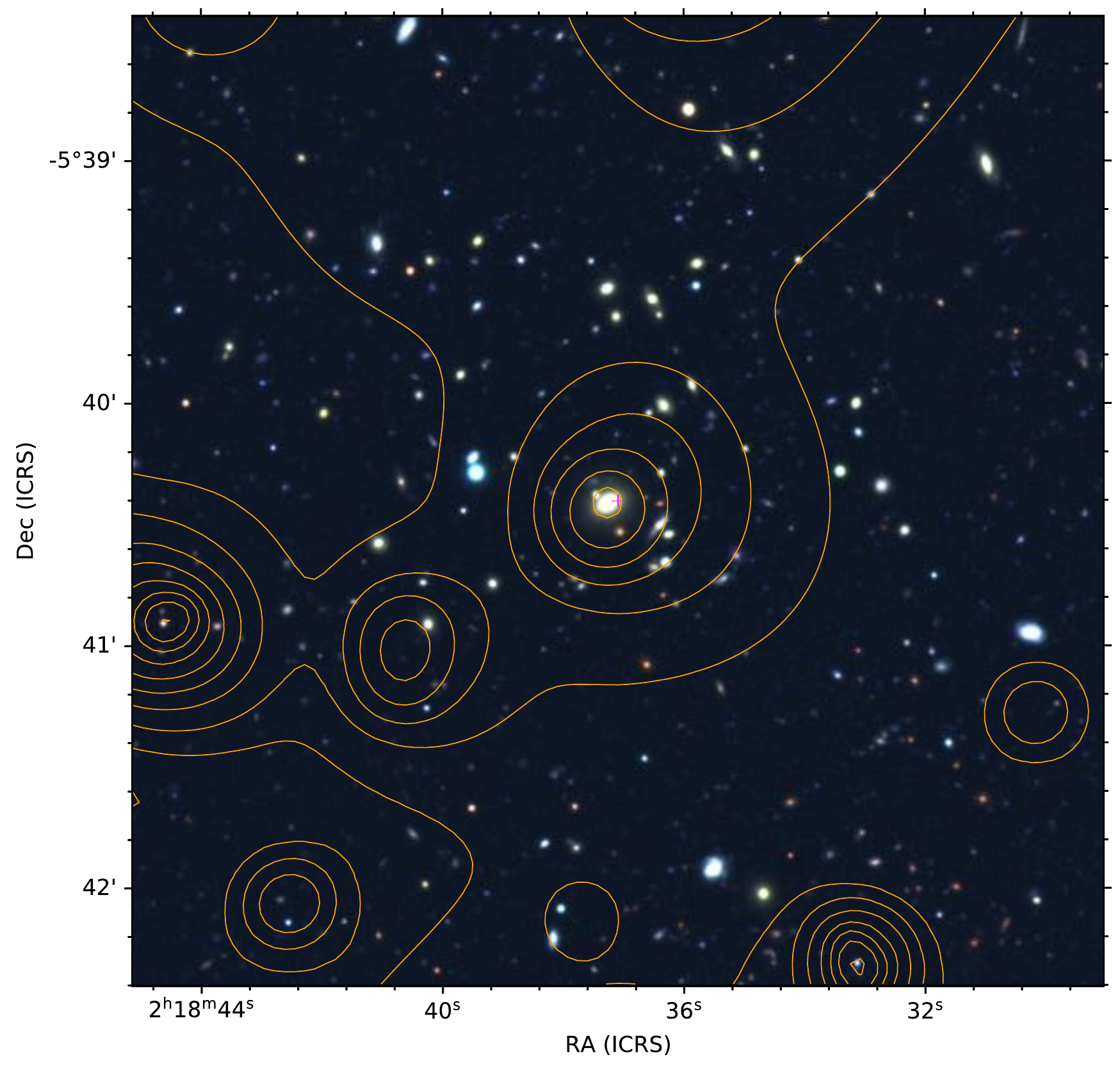}
    \includegraphics[width=0.3\textwidth]{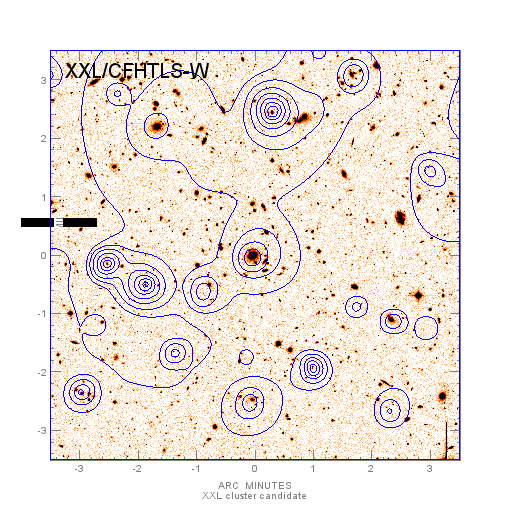}    
    \includegraphics[width=0.3\textwidth]{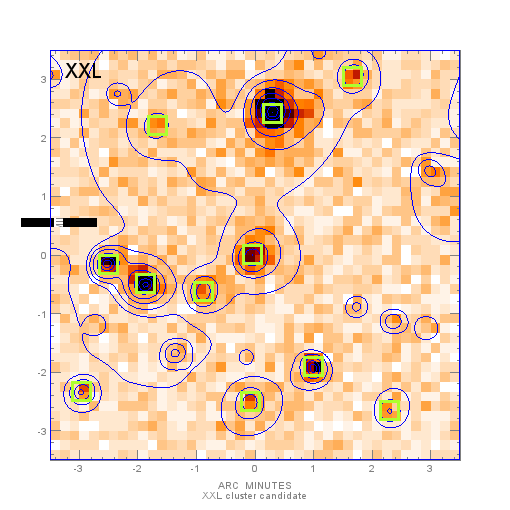}
    \caption{Cluster XLSSC 063 recovered as AC in this study. The optical spectrum of the BCG presents several emission lines, indicating the presence of ionised gas in the elliptical galaxy.}
    \label{fig:XLSSC63}
\end{figure*}

\begin{figure*}
    \centering
    \includegraphics[width=0.3\textwidth]{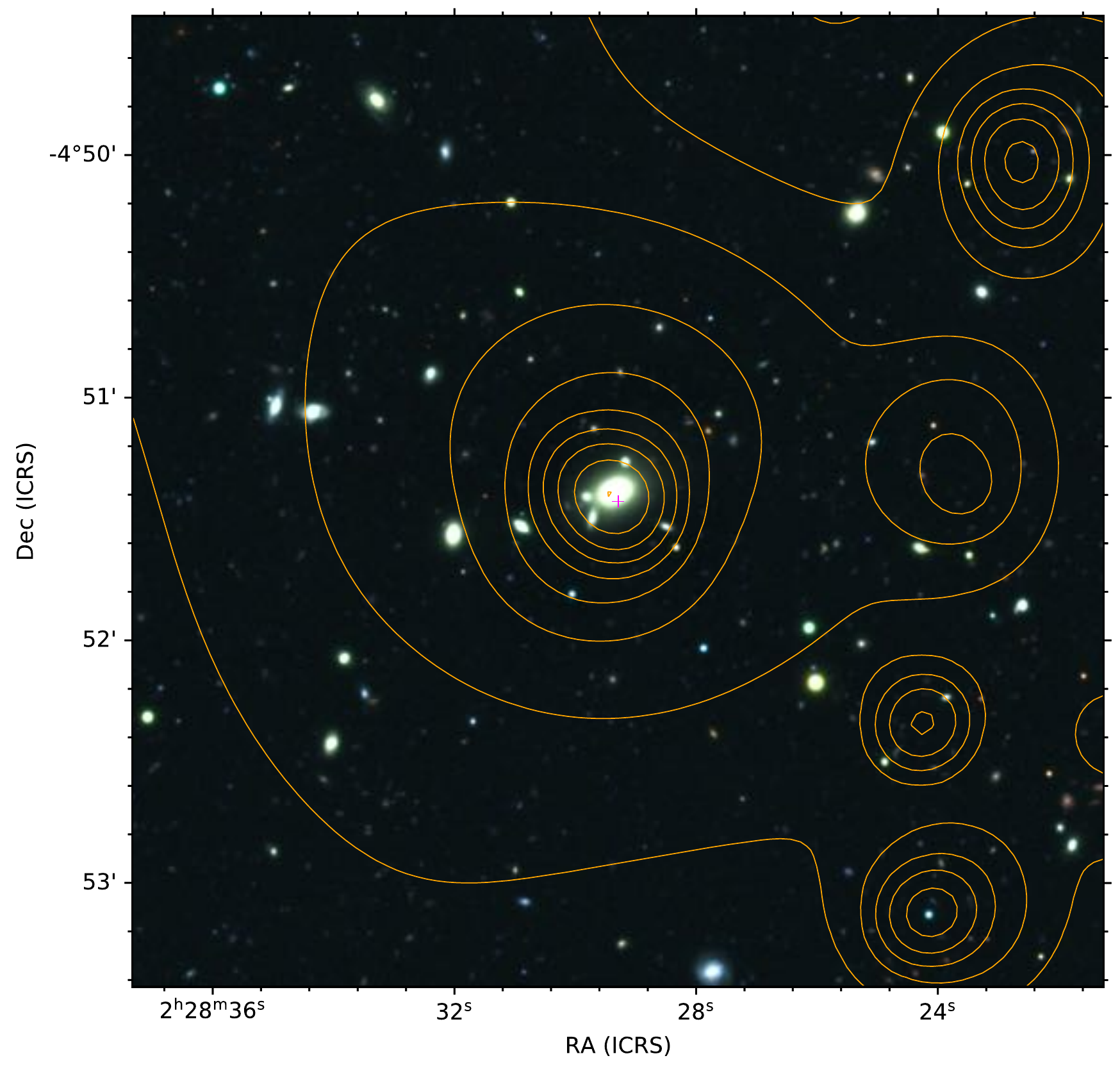}
    \includegraphics[width=0.3\textwidth]{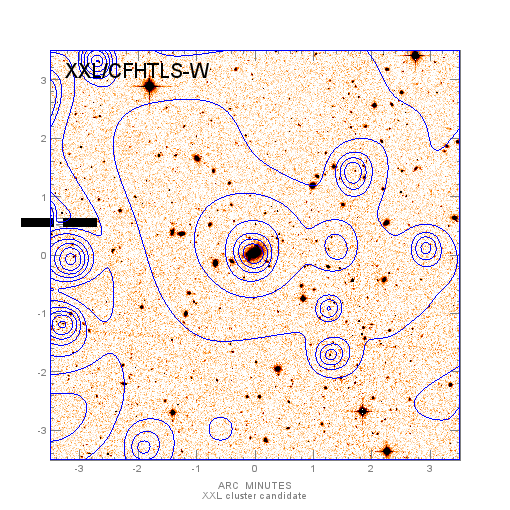}    
    \includegraphics[width=0.3\textwidth]{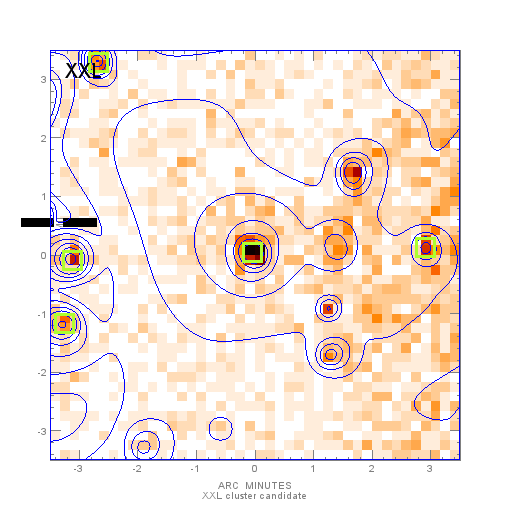}
    \caption{Fossil group XLSSC 090 with $z_{spec}=0.141$. The NII (6586\,$\AA$) line is clearly visible in the optical spectrum of the BCG, again indicating the presence of ionised gas in the elliptical galaxy.}
    \label{fig:XLSSC90}
\end{figure*}

\begin{figure*}
    \centering
    \includegraphics[width=0.3\textwidth]{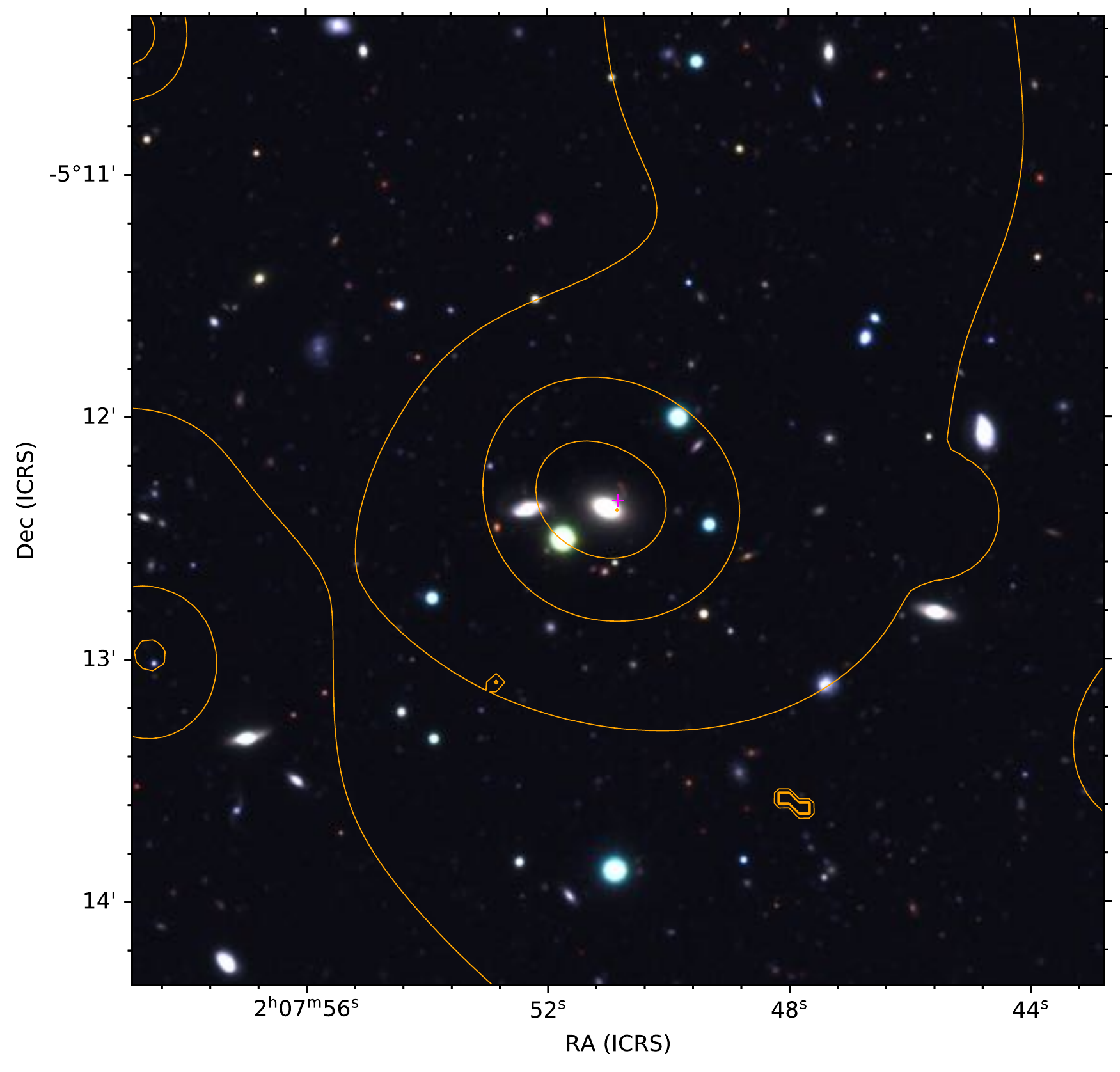}
    \includegraphics[width=0.3\textwidth]{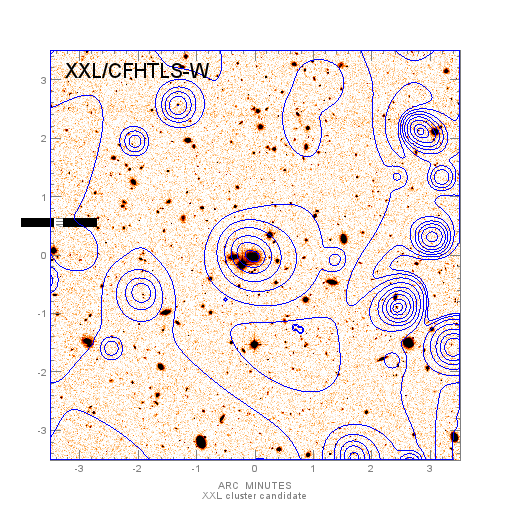}    
    \includegraphics[width=0.3\textwidth]{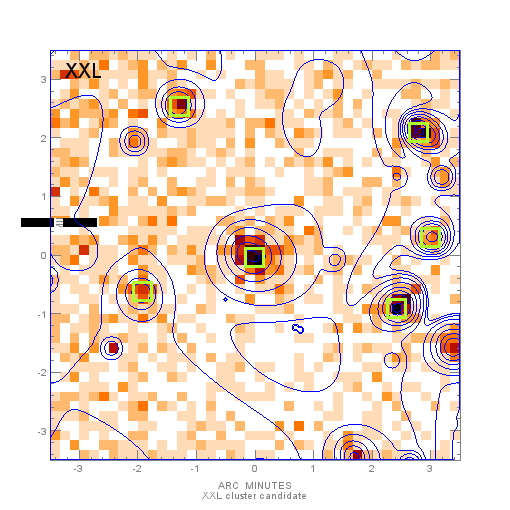}
    \caption{Cluster XLSSC 095. The NII (6586\,$\AA$) line is clearly visible in the optical spectrum of the BCG of this system.}
    \label{fig:XLSSC95}
\end{figure*}

\begin{figure*}
    \centering
    \includegraphics[width=0.3\textwidth]{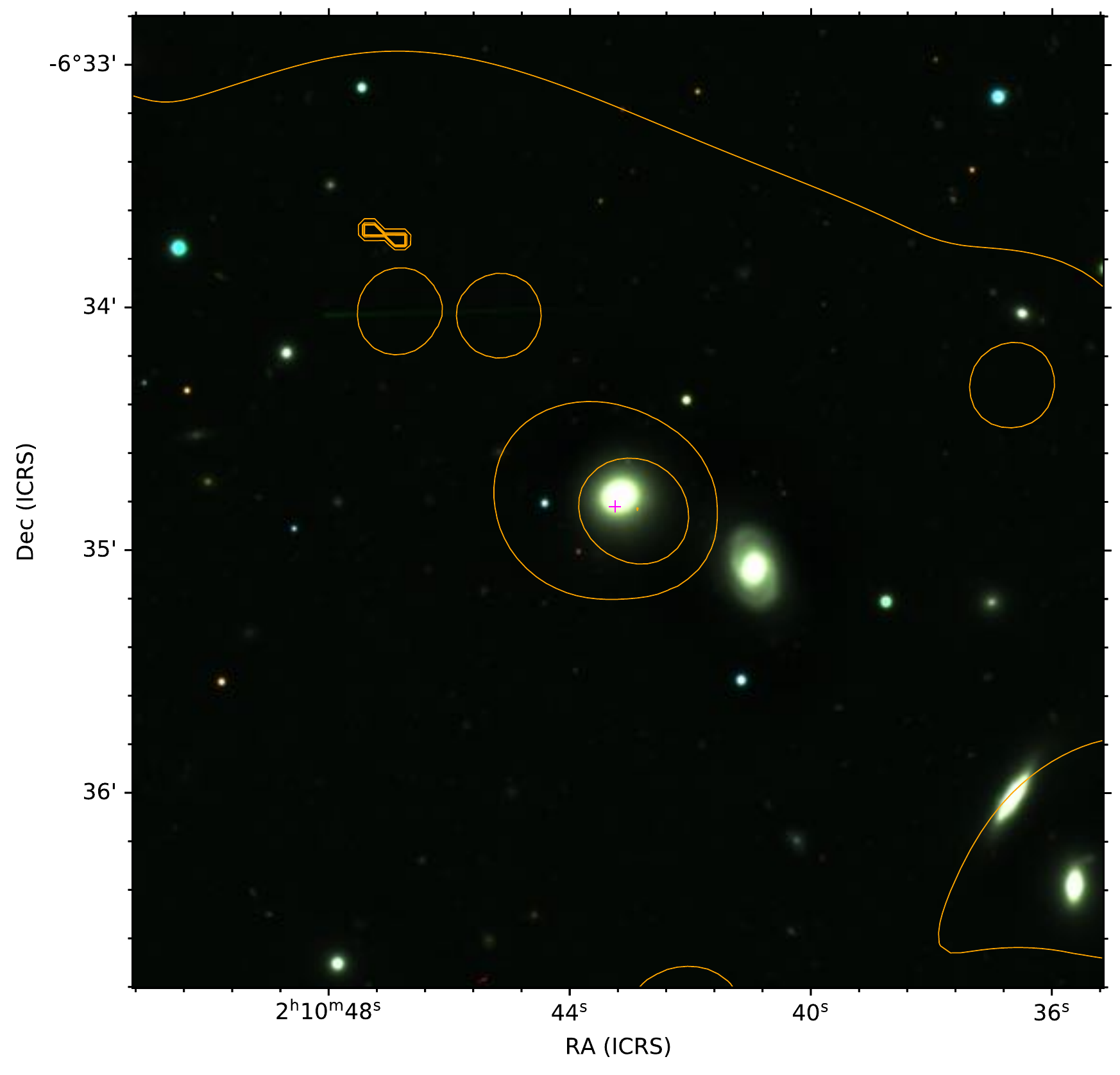}
    \includegraphics[width=0.3\textwidth]{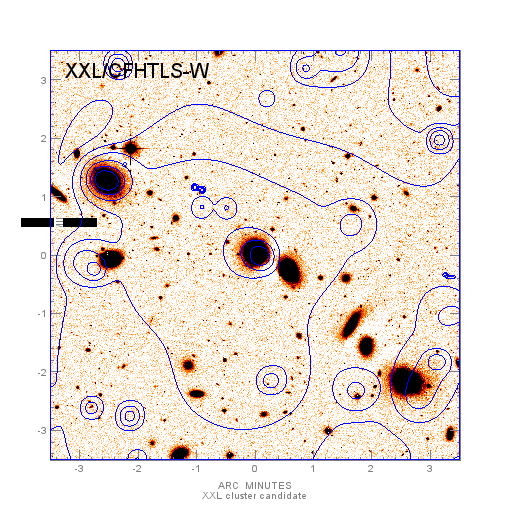}    
    \includegraphics[width=0.3\textwidth]{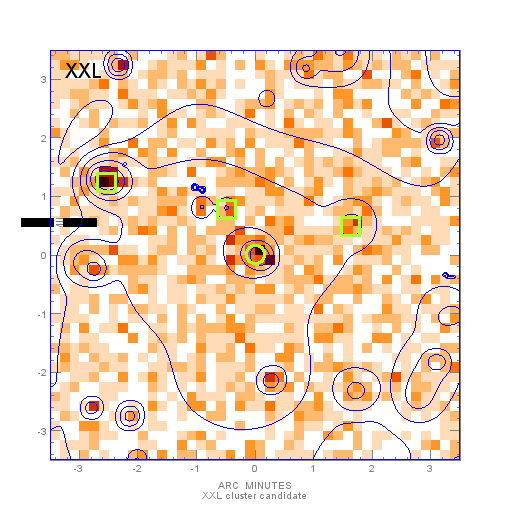}
    \caption{Galaxy group XLSSC 115.}
    \label{fig:XLSSC115}
\end{figure*}

\begin{figure*}
    \centering
    \includegraphics[width=0.3\textwidth]{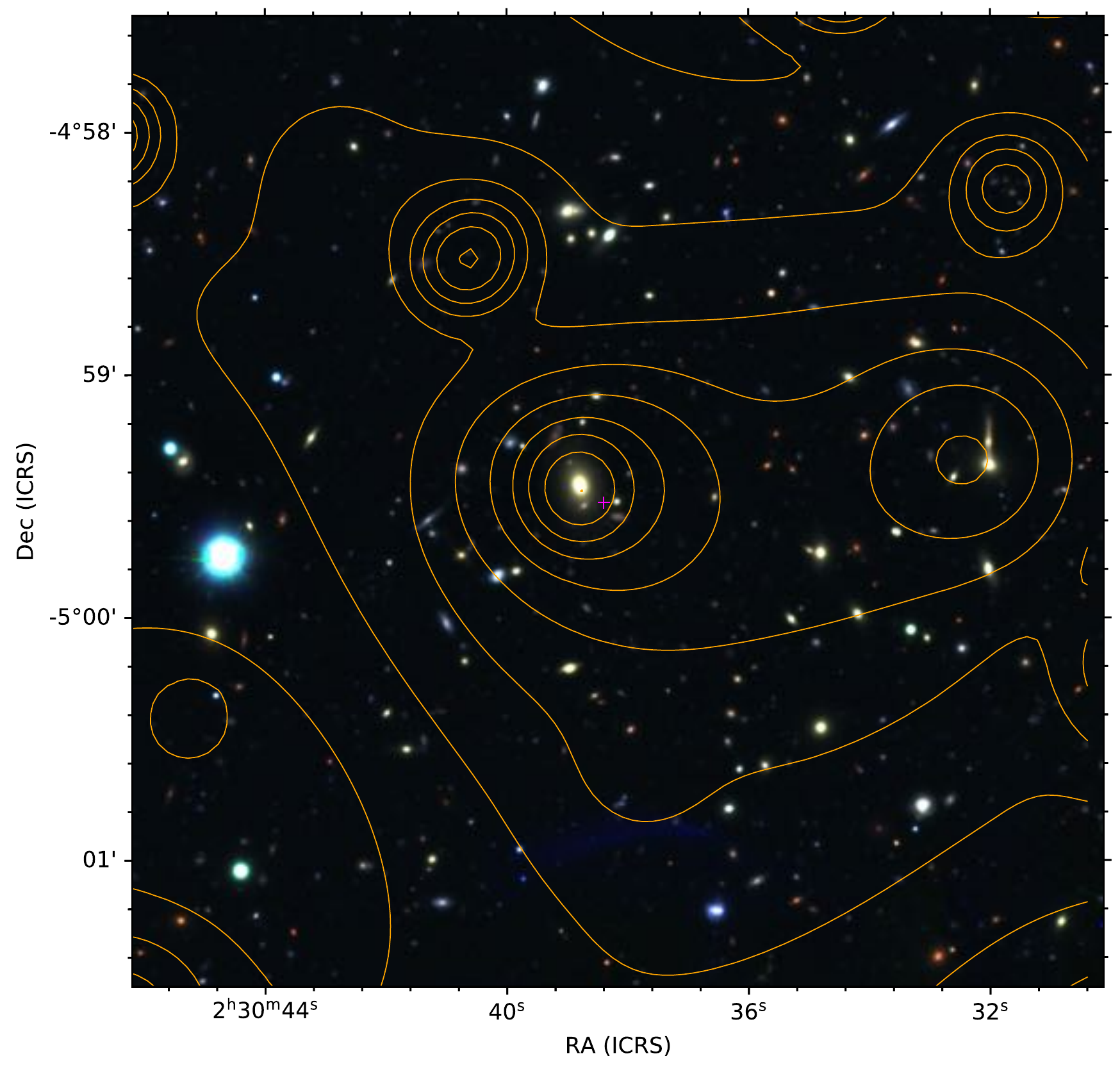}
    \includegraphics[width=0.3\textwidth]{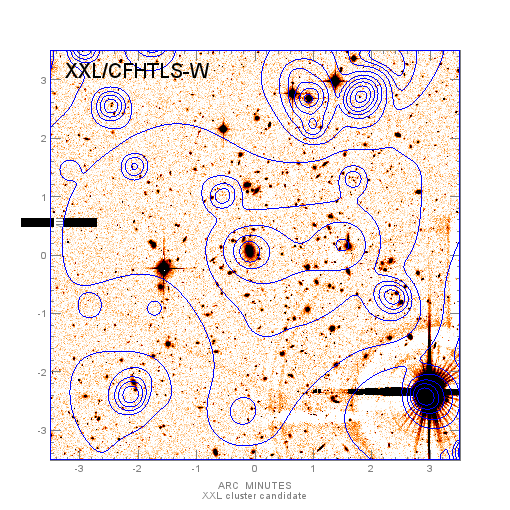}   
    \includegraphics[width=0.3\textwidth]{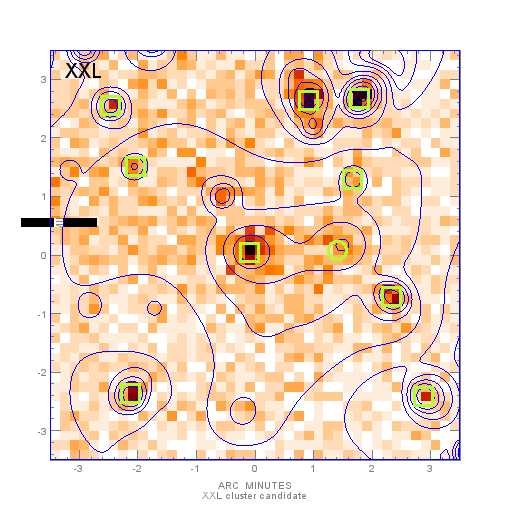}
    \caption{Cluster XLSSC 150. Two member galaxies observed with the MISTRAL instrument revealed passive objects.}
    \label{fig:XLSSC150}
\end{figure*}

\begin{figure*}
    \centering
    \includegraphics[width=0.3\textwidth]{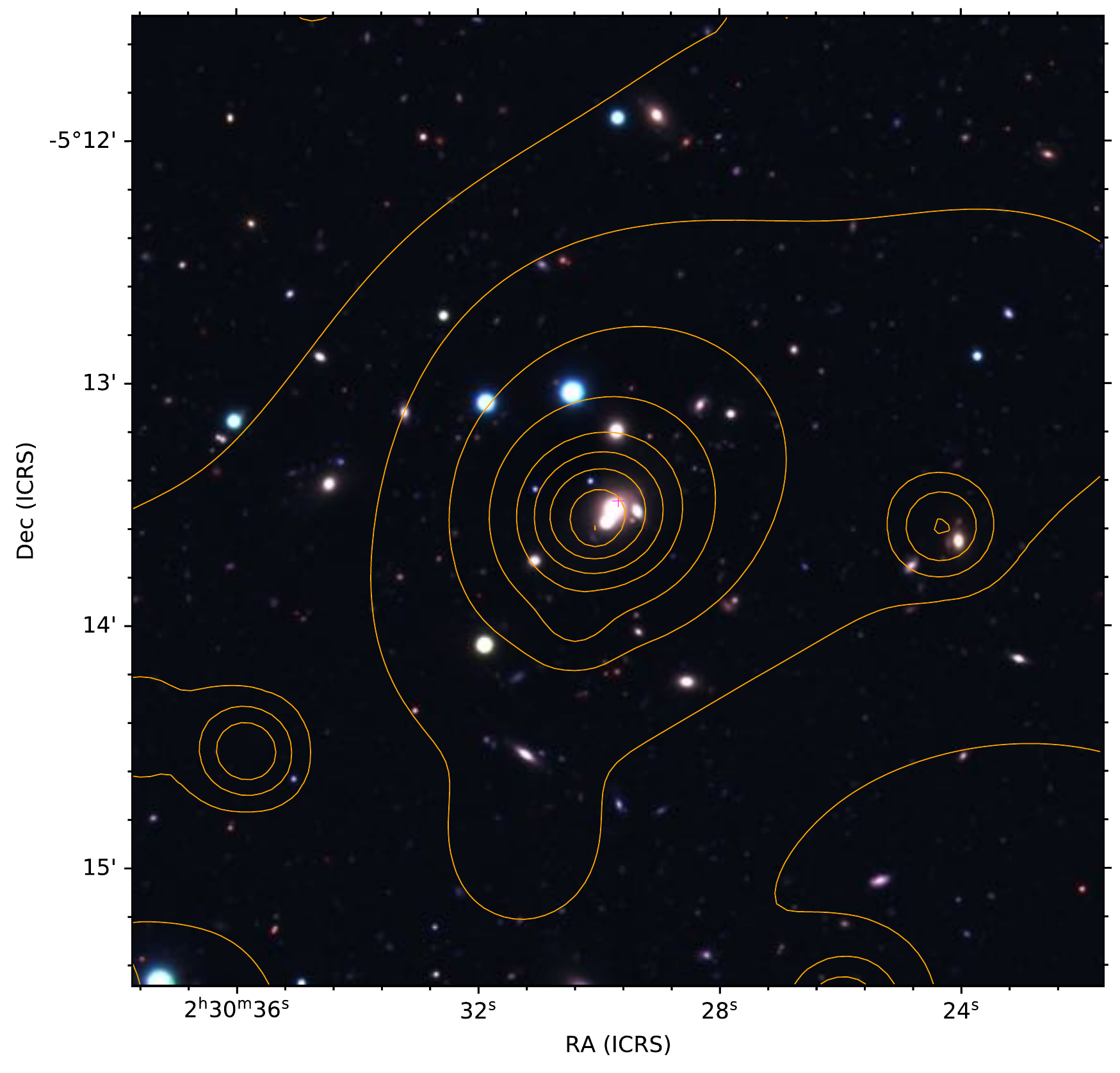}
    \includegraphics[width=0.3\textwidth]{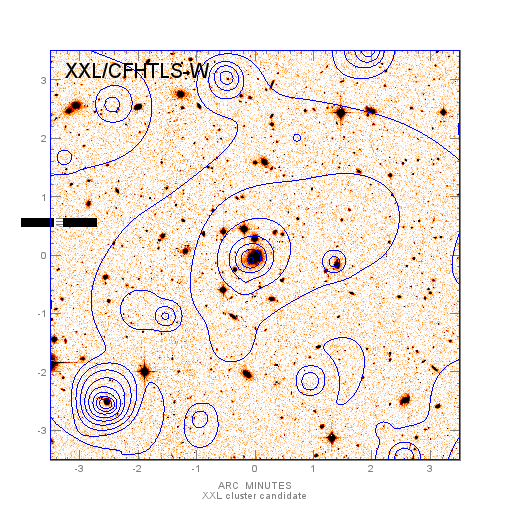}    
    \includegraphics[width=0.3\textwidth]{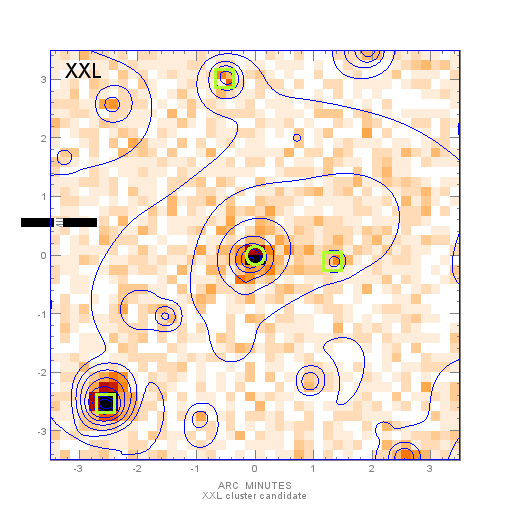}
    \caption{Cluster XLSSC 210. Many GAMA redshifts are available within the field including the BCG at $z = 0.19$ (CESAM). While there is no clear evidence of AGN activity, it may be hidden in the three merging elliptical galaxies in the centre. Two member galaxies observed by MISTRAL indicate passive objects.}
    \label{fig:XLSSC210}
\end{figure*}

\begin{figure*}
    \centering
    \includegraphics[width=0.3\textwidth]{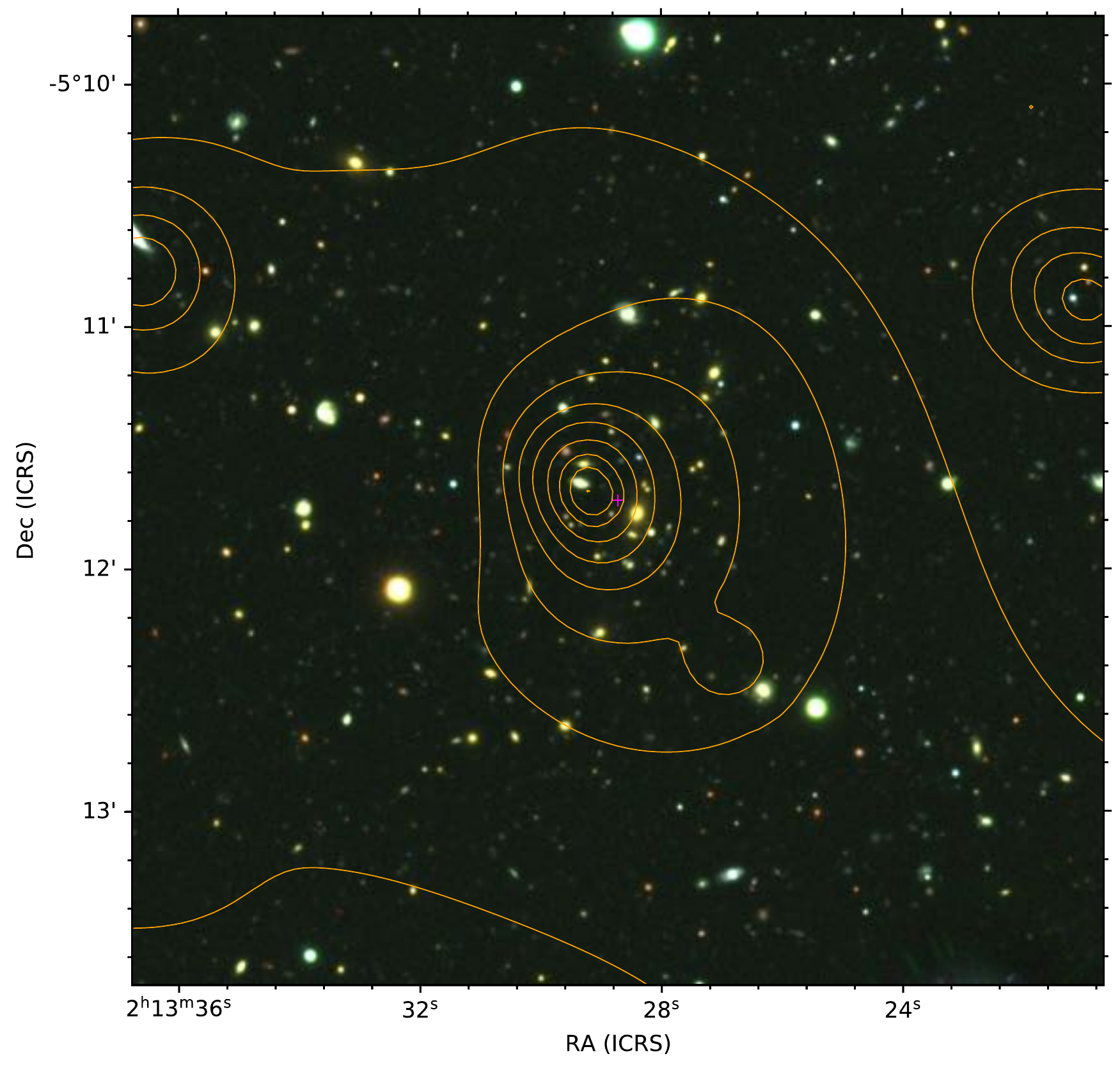}
    \includegraphics[width=0.3\textwidth]{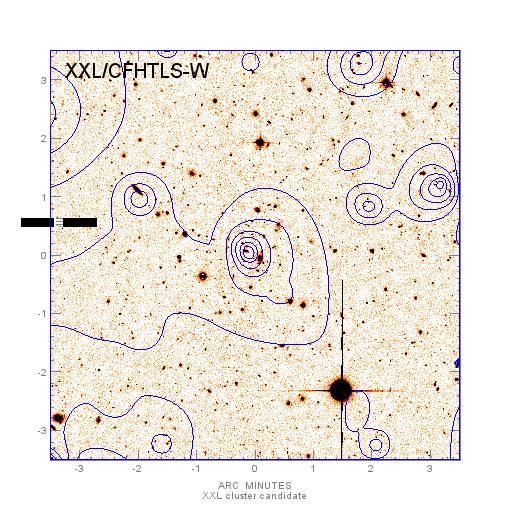}    
    \includegraphics[width=0.3\textwidth]{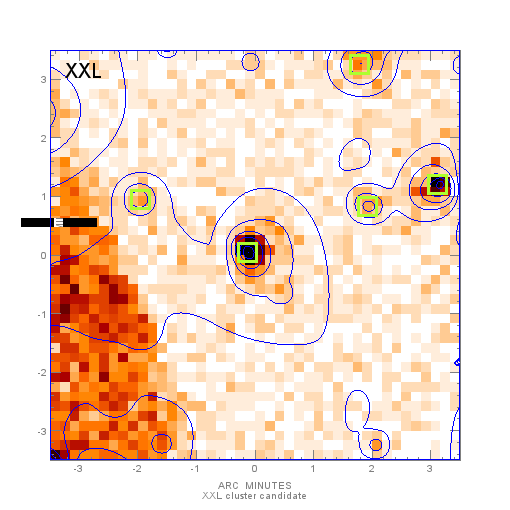}
    \caption{Cluster XLSSC 211. The X-ray emission is centred on a cluster member galaxy that hosts a broad-line AGN (SDSS), though it is not the BCG of the system.}
    \label{fig:XLSSC211}
\end{figure*}

\begin{figure*}
    \centering
    \includegraphics[width=0.32\textwidth]{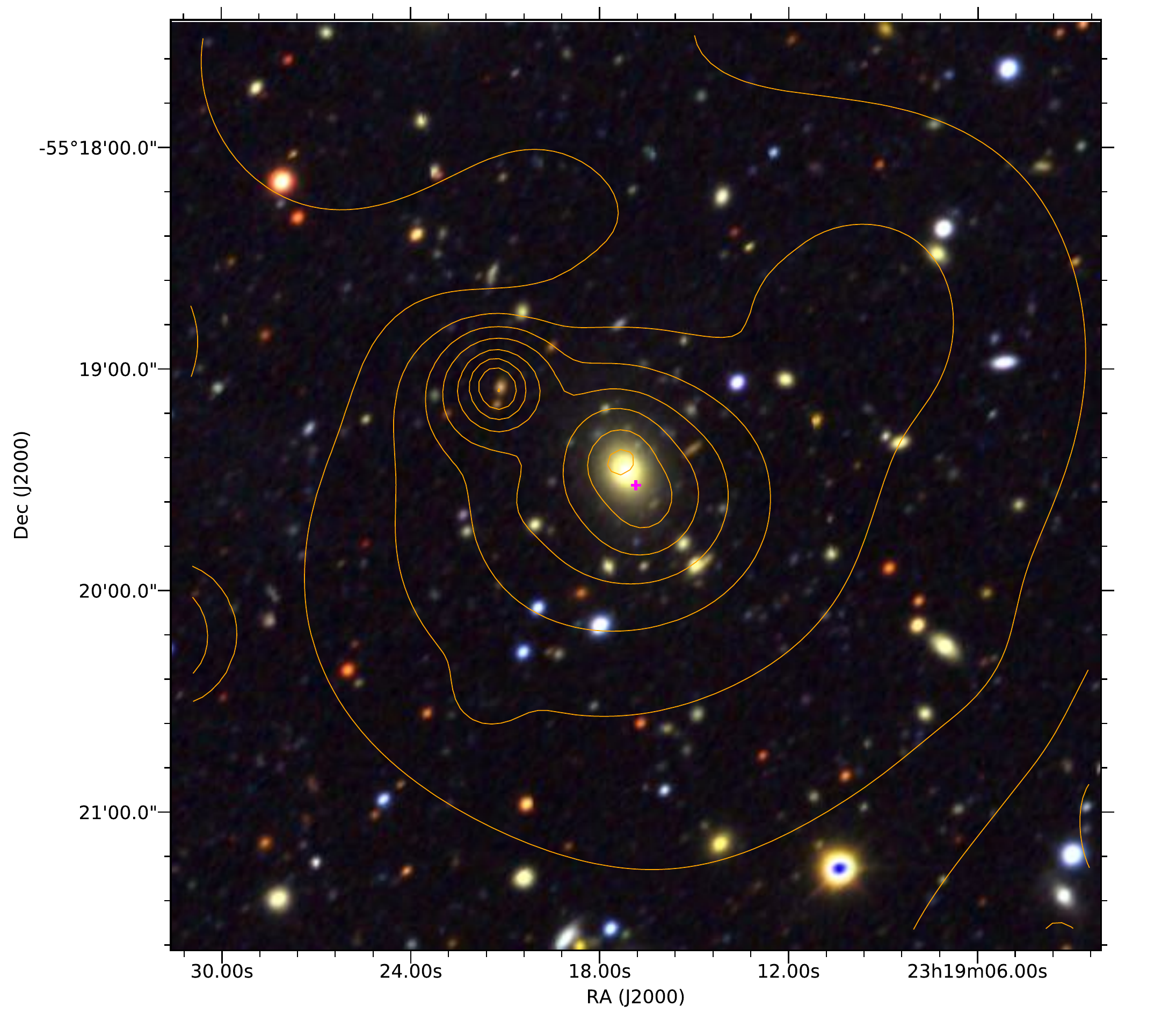}     
    \includegraphics[width=0.3\textwidth]{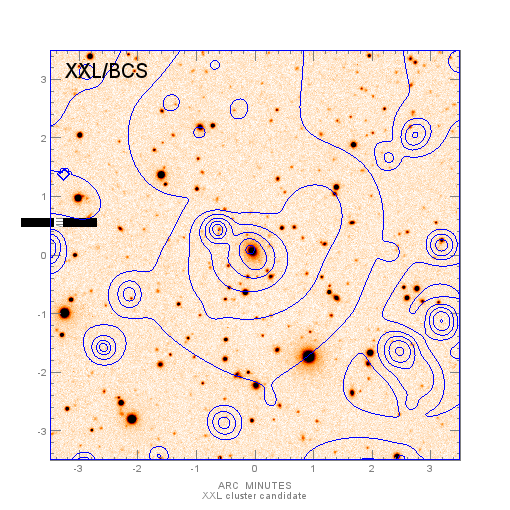}    
    \includegraphics[width=0.3\textwidth]{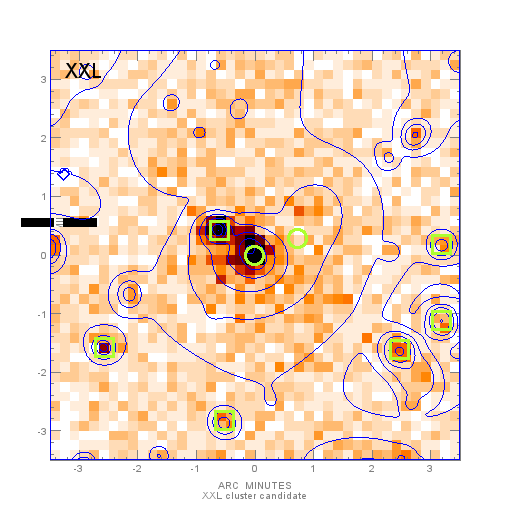}
    \caption{Cluster XLSSC 518, shown to have a possible cool-core signature and no obvious AGN presence.}
    \label{fig:XLSSC518}
\end{figure*}

\begin{figure*}
    \centering
    \includegraphics[width=0.32\textwidth]{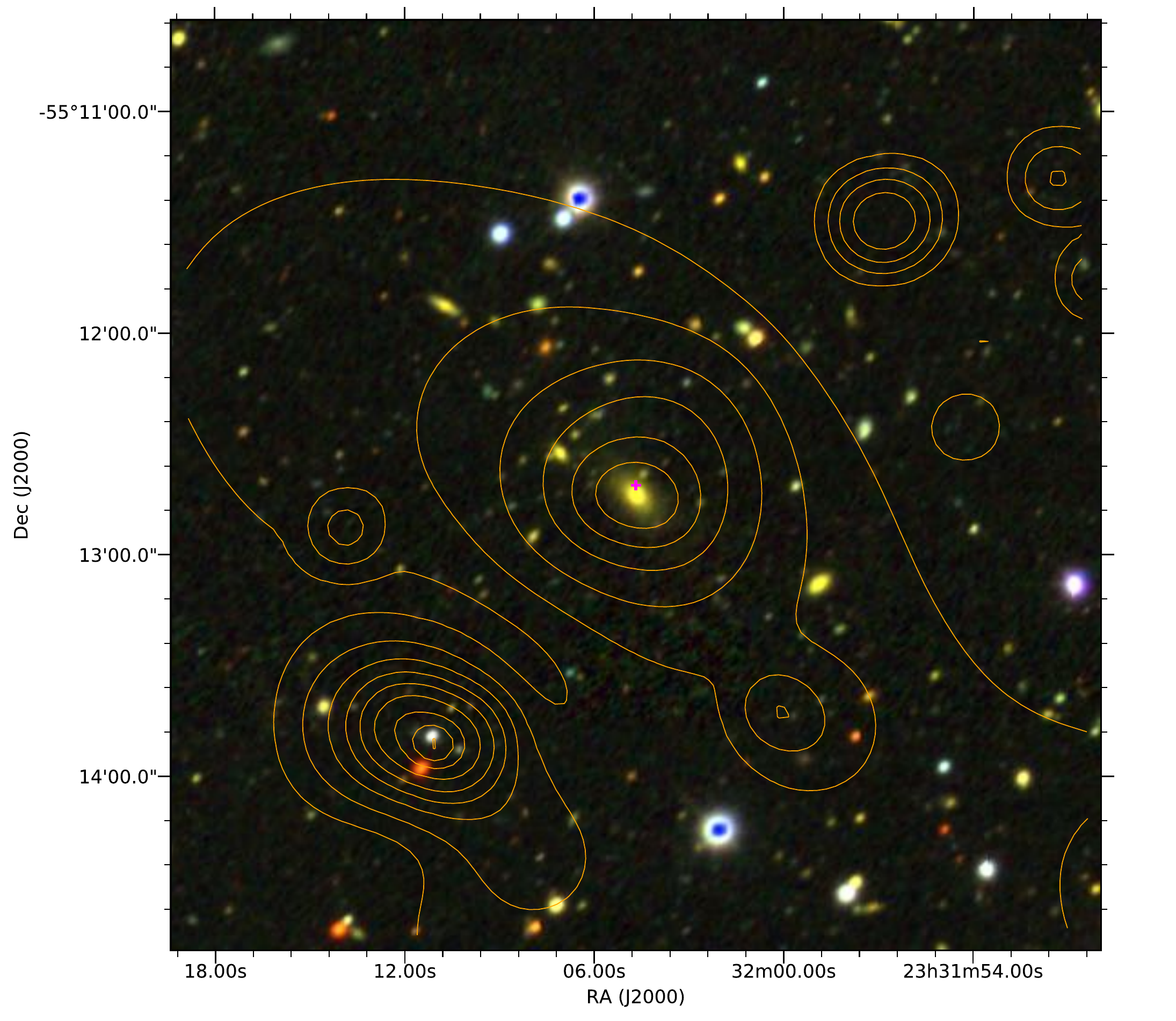} 
    \includegraphics[width=0.3\textwidth]{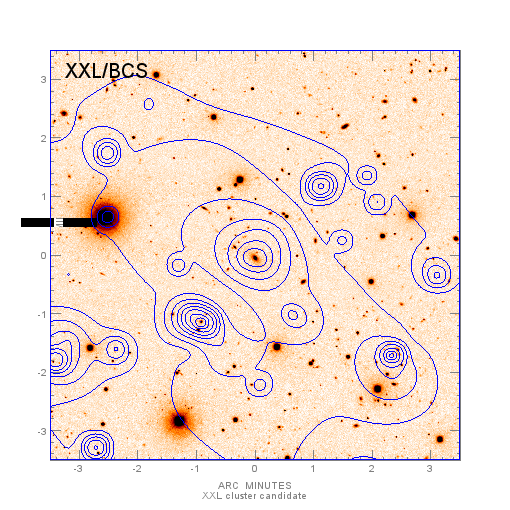}    
    \includegraphics[width=0.3\textwidth]{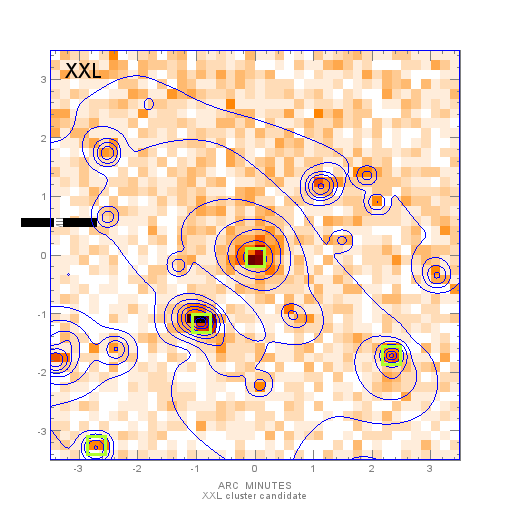}
    \caption{Cluster XLSSC 519. Emission lines in the optical spectrum of the BCG indicate the presence of ionised gas in the elliptical galaxy.}
    \label{fig:XLSSC519}
\end{figure*}

\begin{figure*}
    \centering
    \includegraphics[width=0.32\textwidth]{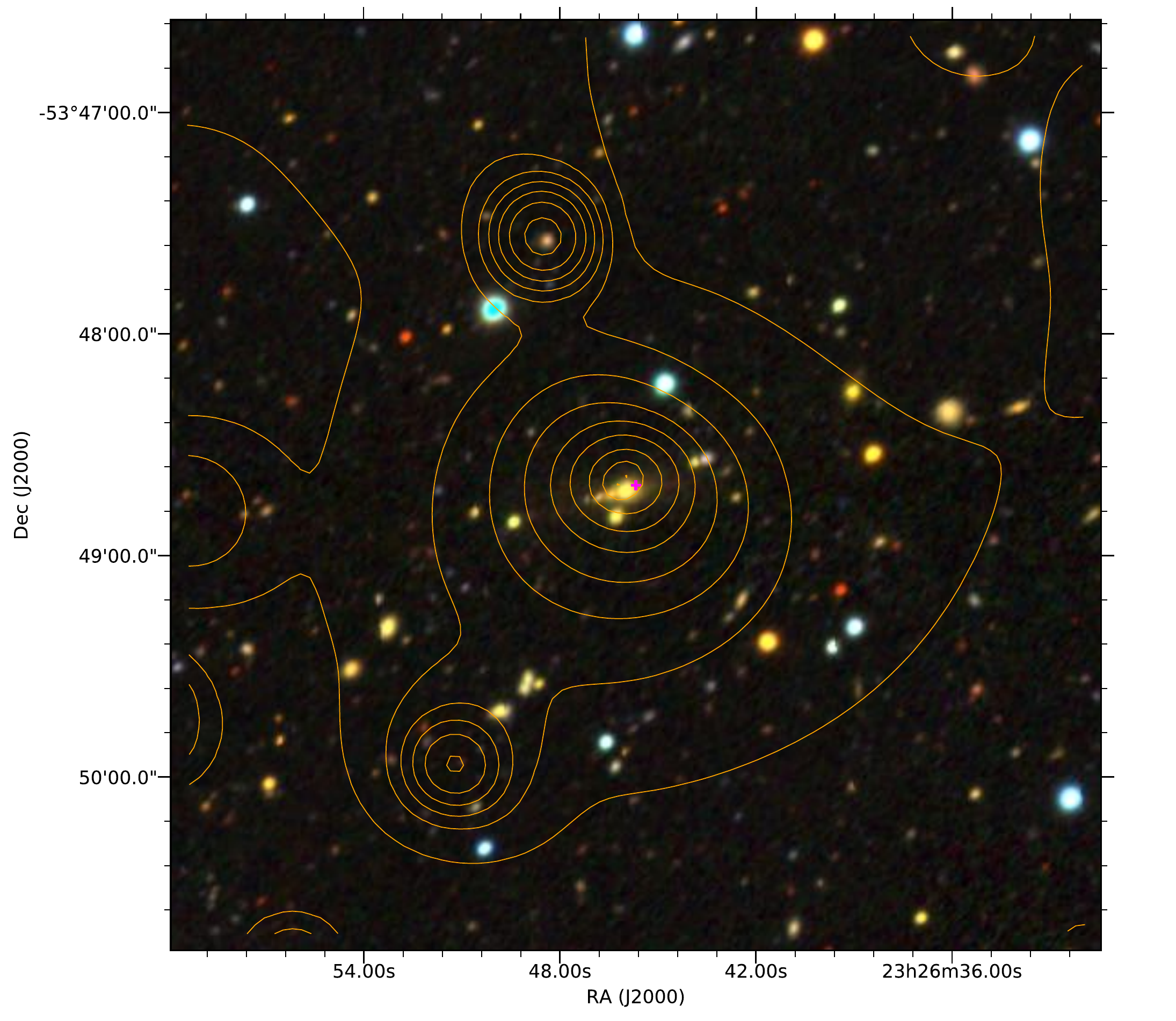}     
    \includegraphics[width=0.3\textwidth]{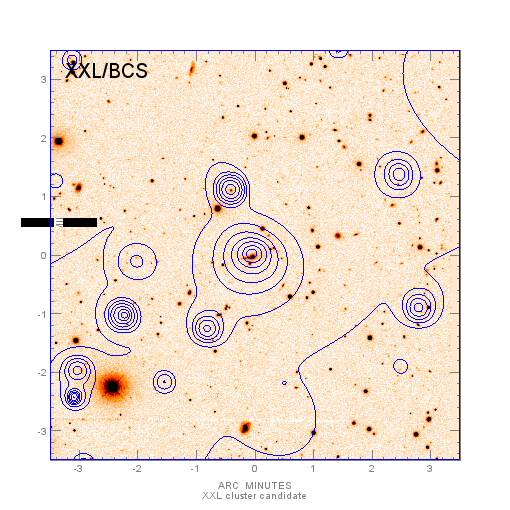}    
    \includegraphics[width=0.3\textwidth]{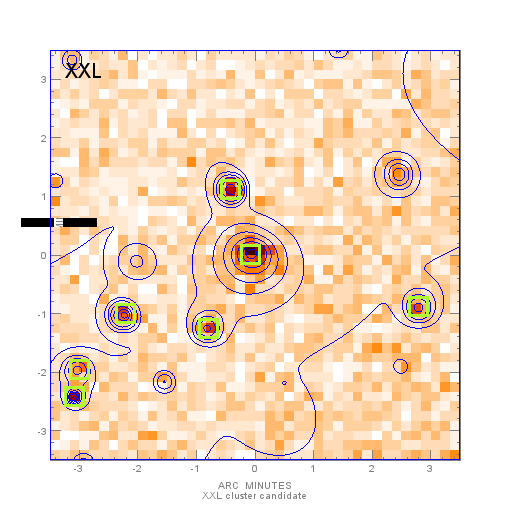}
    \caption{Cluster XLSSC 595. The optical spectrum of the BCG from AAT indicates probable AGN activity.}
    \label{fig:XLSSC595}
\end{figure*}

\begin{figure*}
    \centering
    \includegraphics[width=0.32\textwidth]{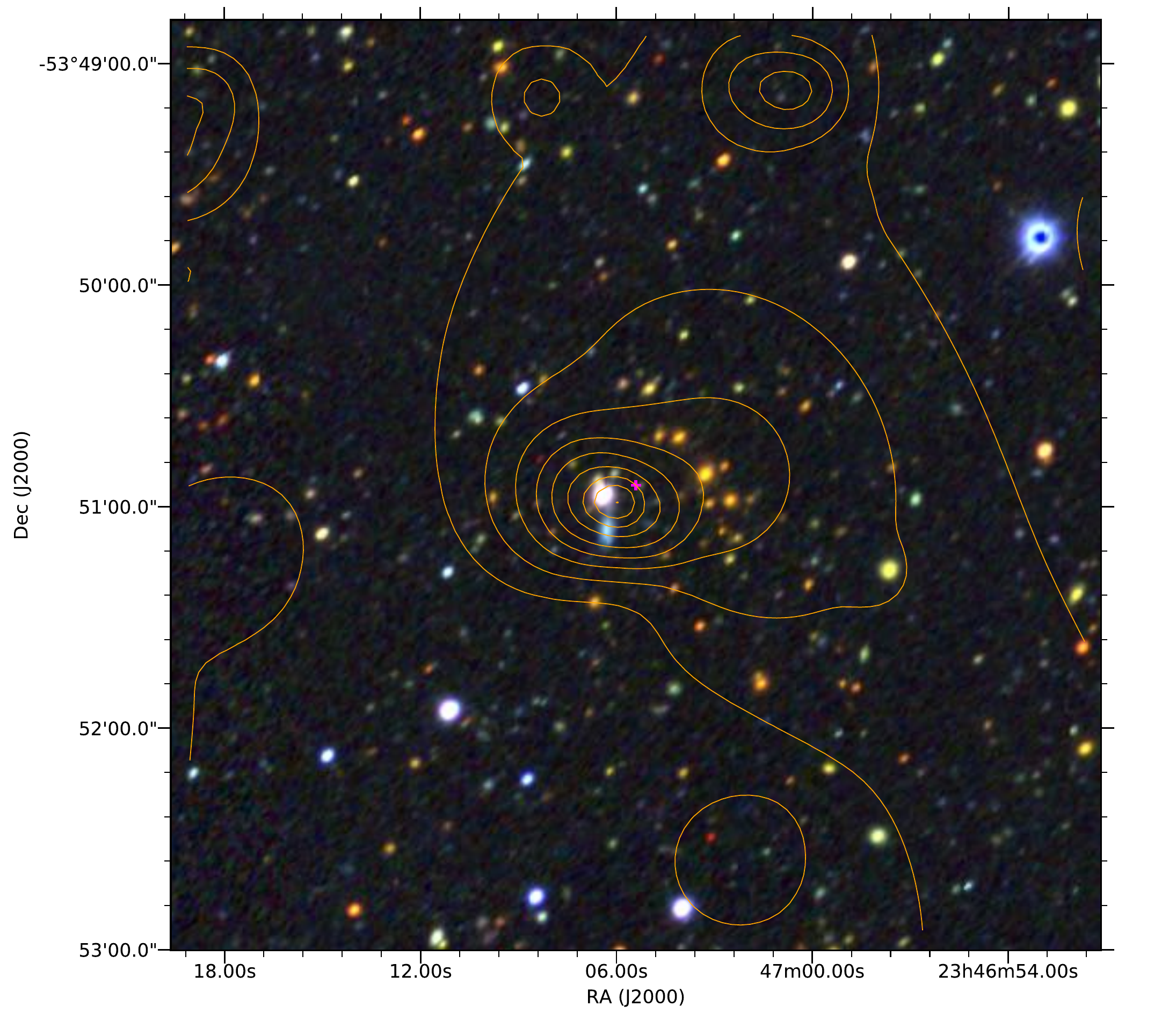}
    \includegraphics[width=0.3\textwidth]{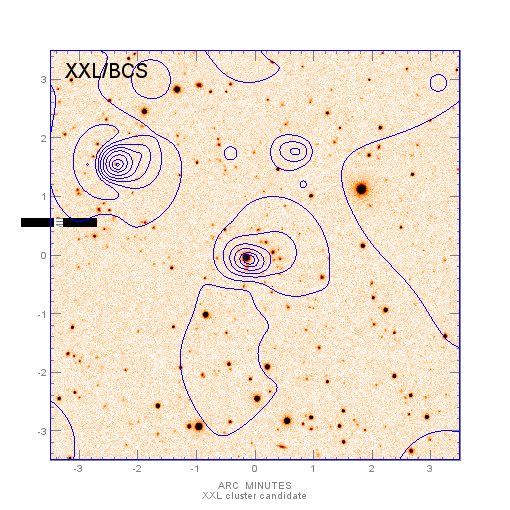}    
    \includegraphics[width=0.3\textwidth]{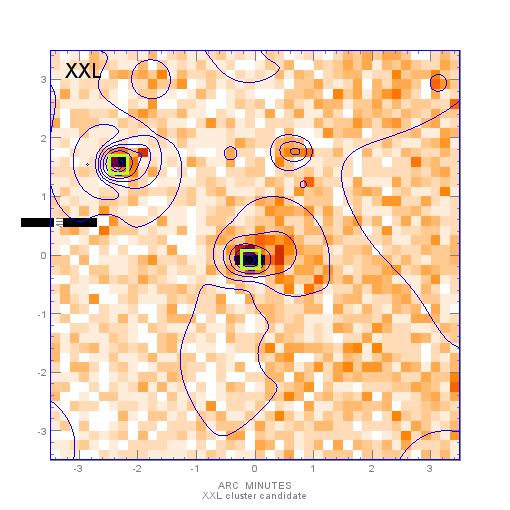}
    \caption{Cluster XLSSC 648 at $z_{spec} = 0.64$ contaminated by an X-ray bright star.}
    \label{fig:XLSSC648}
\end{figure*}

\begin{figure*}
    \centering
    \includegraphics[width=0.32\textwidth]{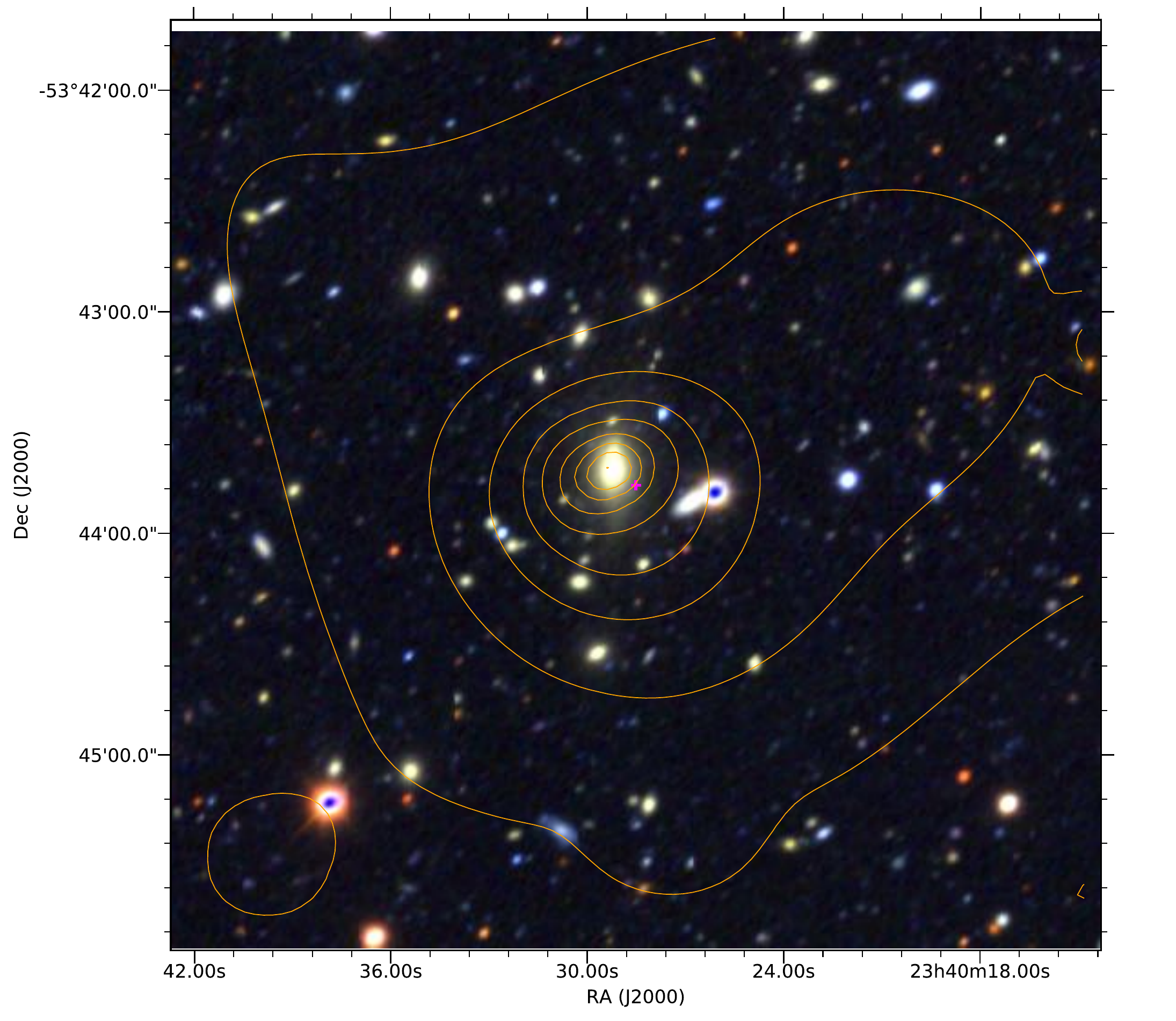}
    \includegraphics[width=0.3\textwidth]{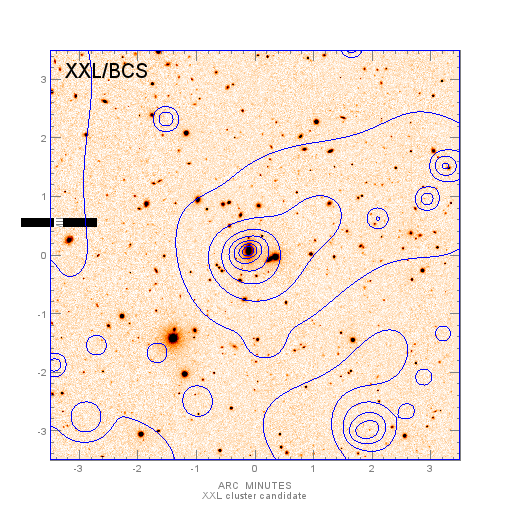}    
    \includegraphics[width=0.3\textwidth]{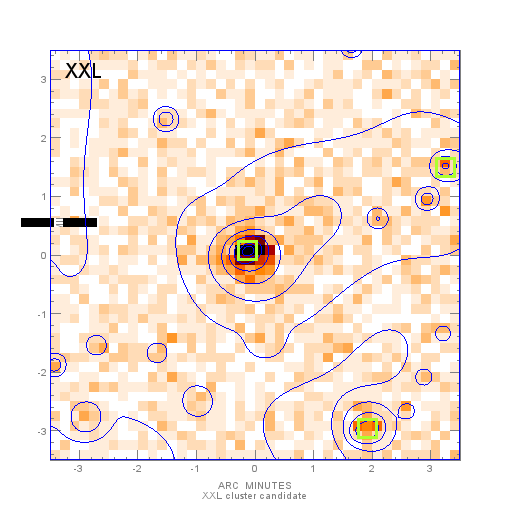}
    \caption{Cluster XLSSC 649 at $z_{spec}=0.19$ confirmed by NTT spectroscopic observations. The BCG hosts a broad line AGN.}
    \label{fig:XLSSC649}
\end{figure*}

\begin{figure*}
    \centering
    \includegraphics[width=0.32\textwidth]{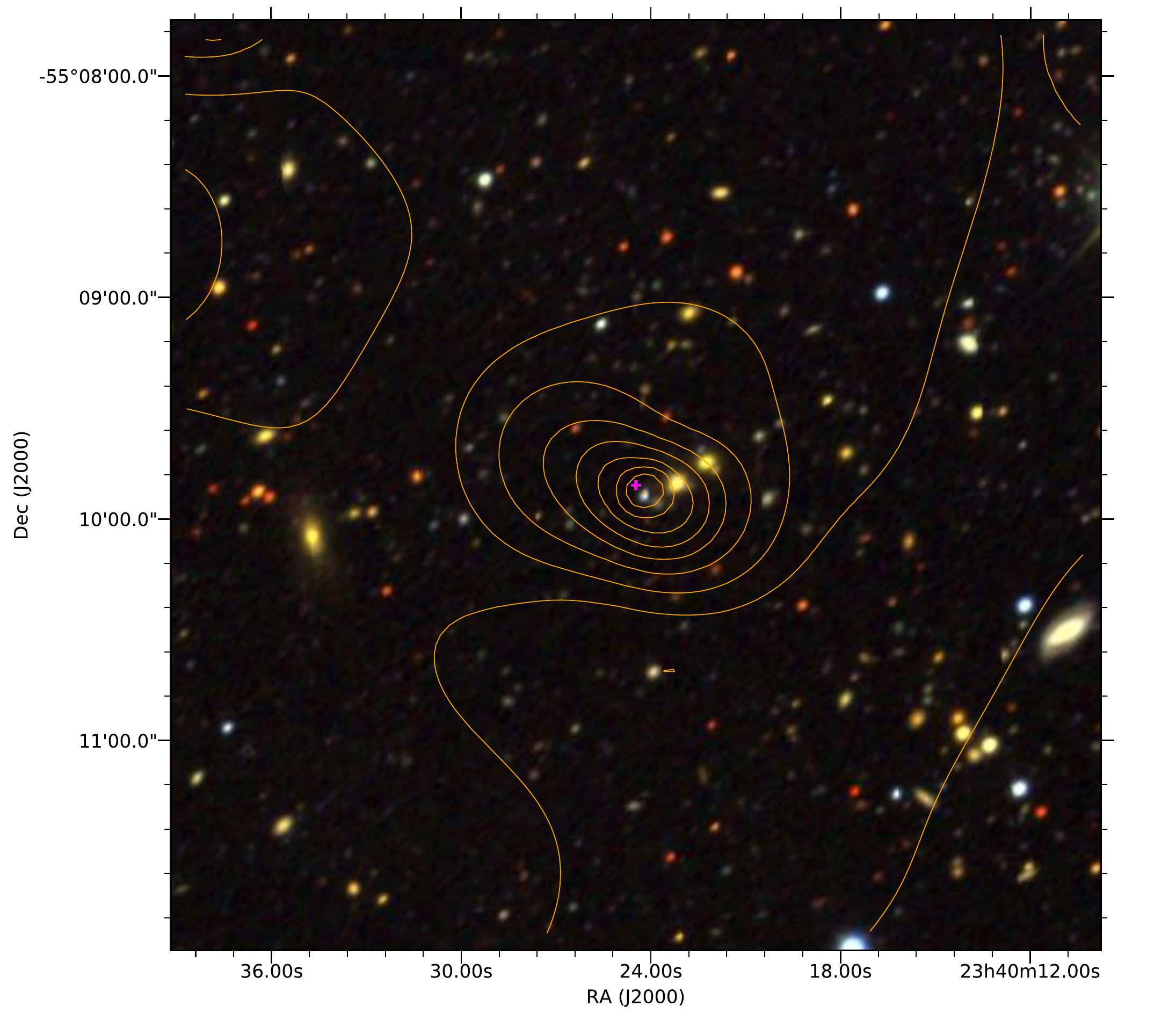}
    \includegraphics[width=0.3\textwidth]{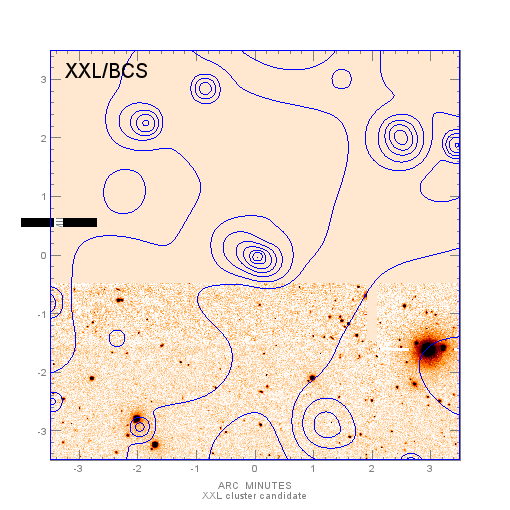}    
    \includegraphics[width=0.3\textwidth]{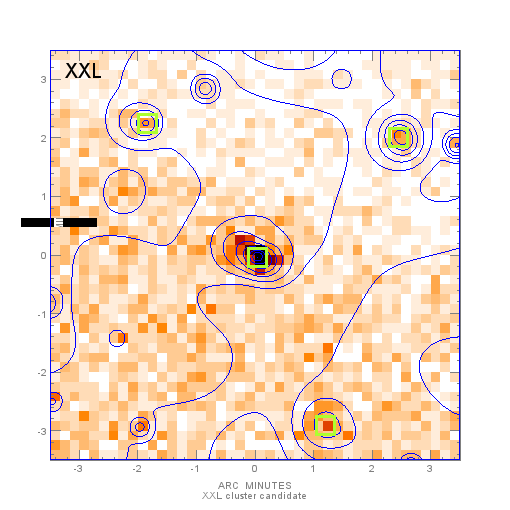}
    \caption{Cluster XLSSC 650 likely contaminated by a QSO. Two galaxies observed by NTT very close to the centre of the X-ray emission with $z_{spec}=0.29$.}
    \label{fig:XLSSC650}
\end{figure*}

\begin{figure*}
    \centering
    \includegraphics[width=0.32\textwidth]{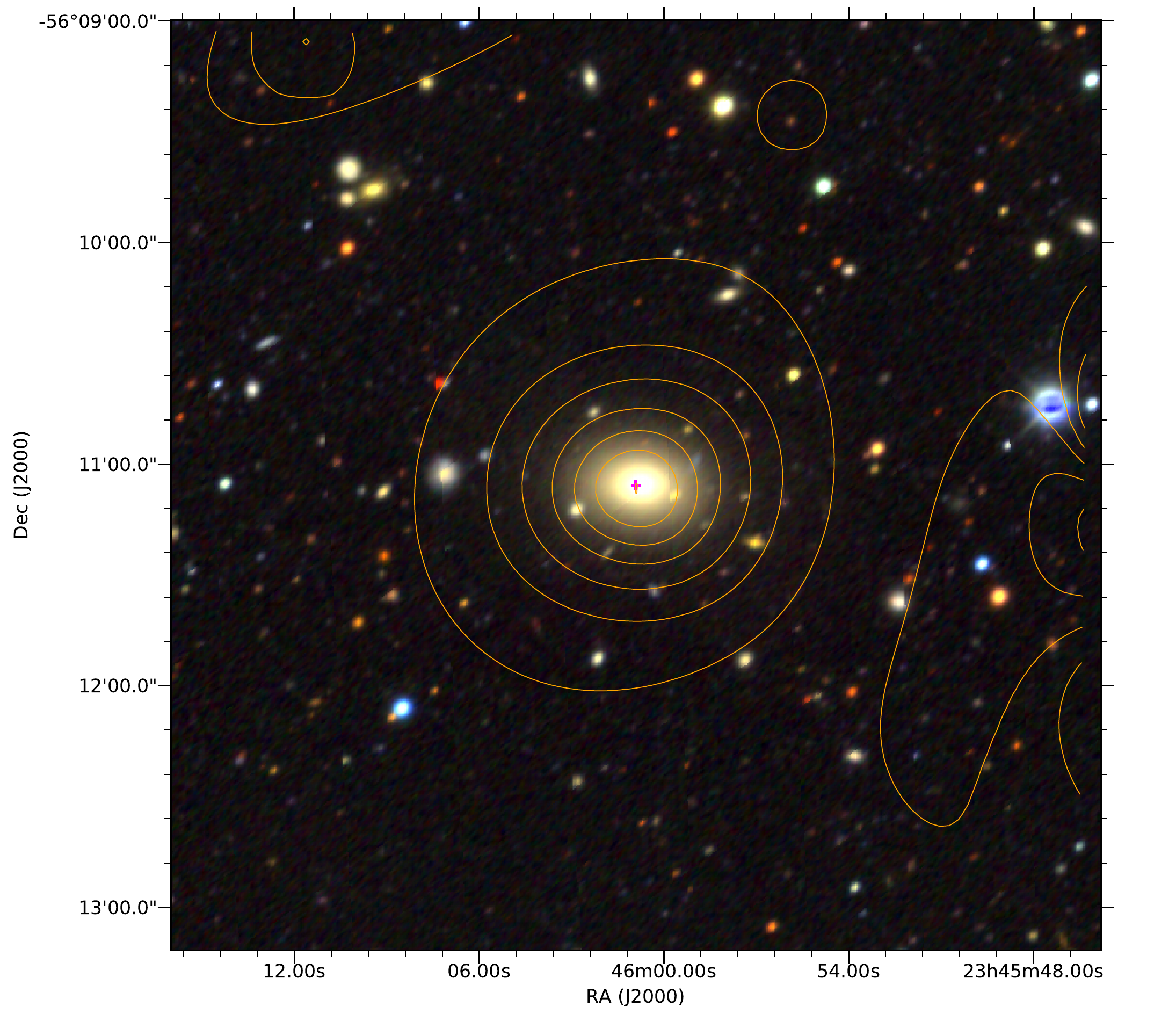}
    \includegraphics[width=0.3\textwidth]{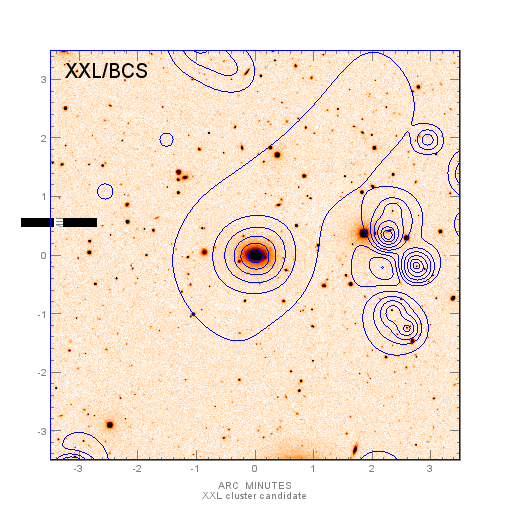}    
    \includegraphics[width=0.3\textwidth]{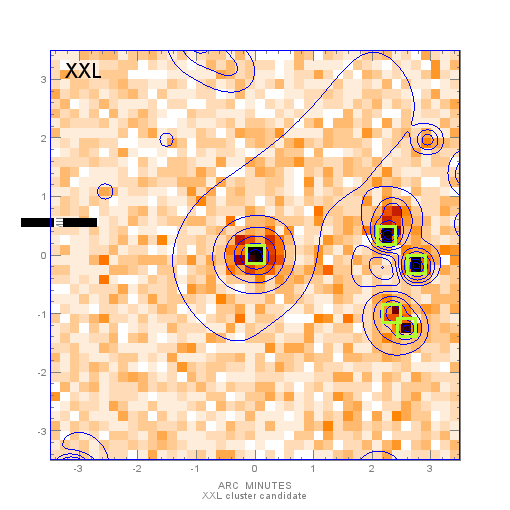}
    \caption{Fossil group XLSSC 651 at $z=0.102$.}
    \label{fig:XLSSC651}
\end{figure*}

\begin{figure*}
    \centering
    \includegraphics[width=0.3\textwidth]{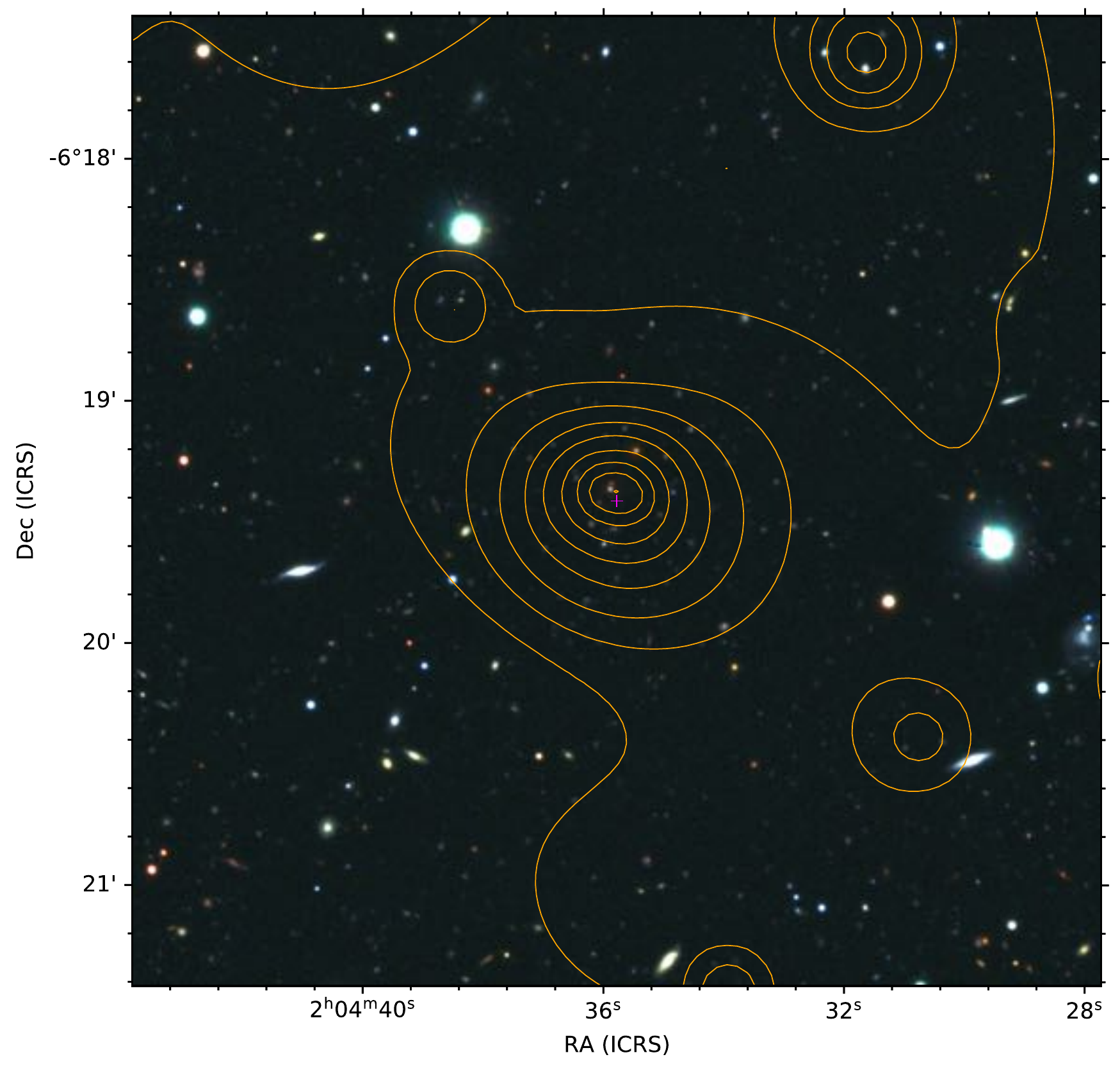}
    \includegraphics[width=0.3\textwidth]{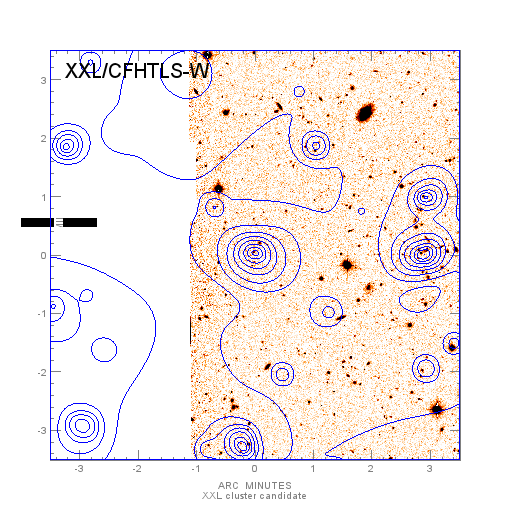}    
    \includegraphics[width=0.3\textwidth]{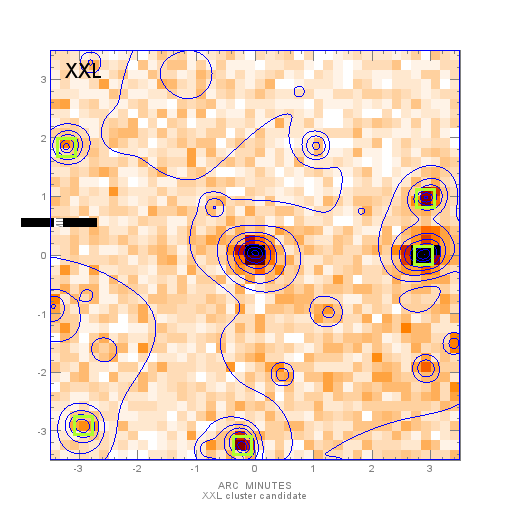}
    \caption{Cluster candidate XLSSU J020435.7-061922 with AGN at $z_{spec}=0.91$. Many HSC photo-zs within the field are found to be at the same redshift.}
    \label{fig:NT0095}
\end{figure*}

\begin{figure*}
    \centering
    \includegraphics[width=0.3\textwidth]{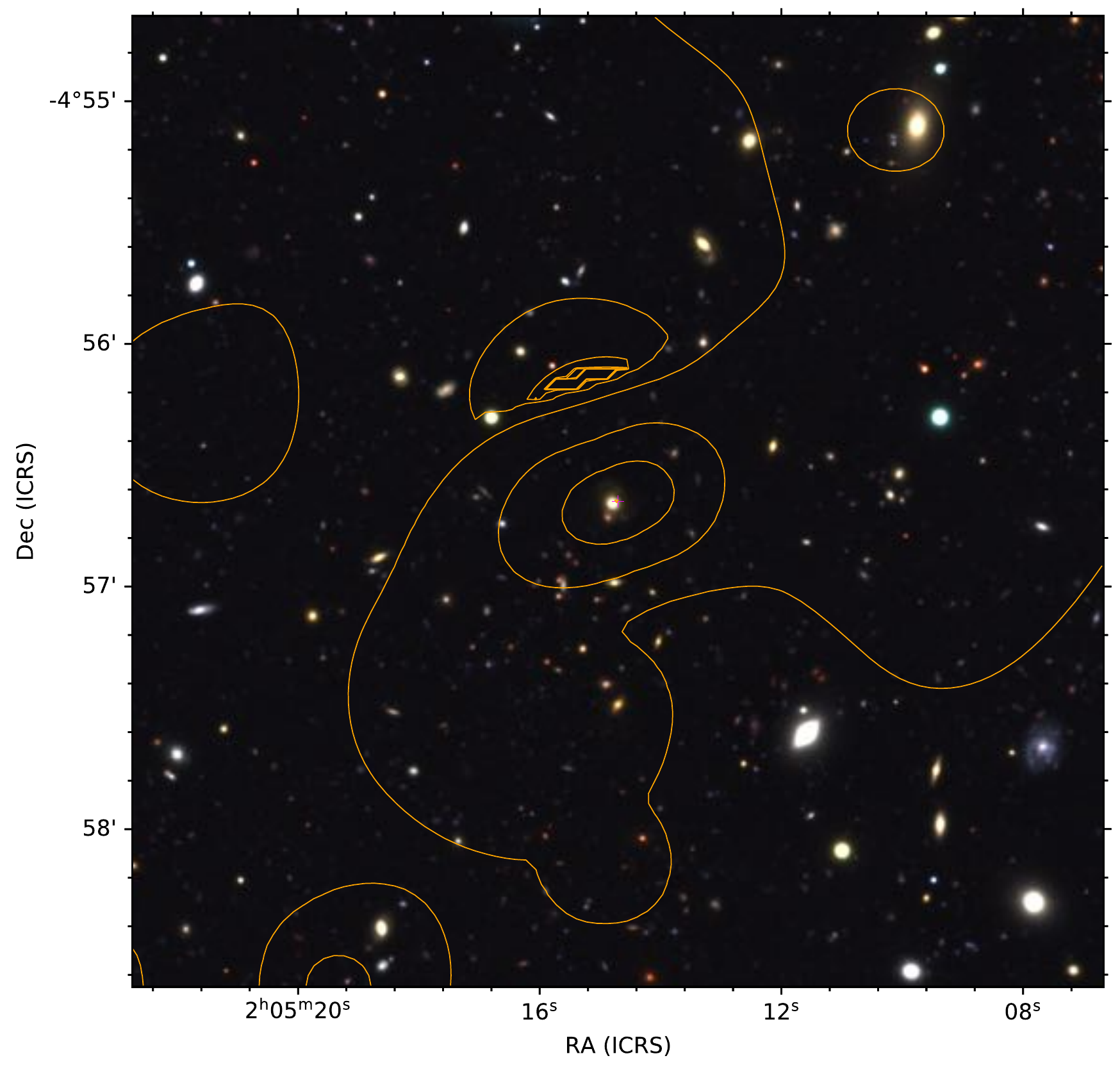}
    \includegraphics[width=0.3\textwidth]{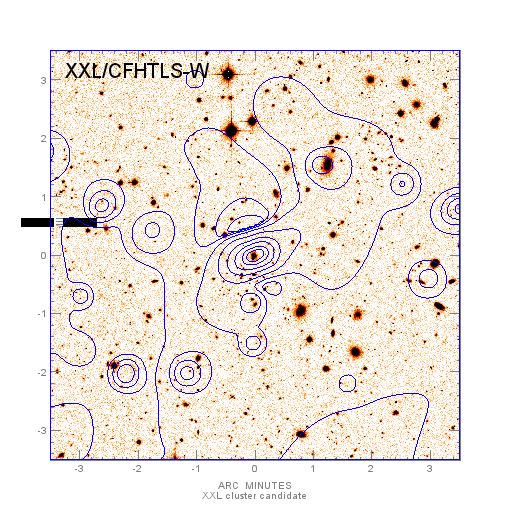}   
    \includegraphics[width=0.3\textwidth]{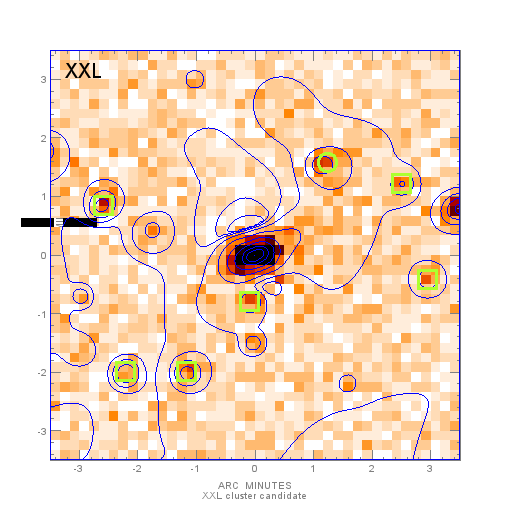}
    \caption{Cluster candidate XLSSU J020514.7-045638 with QSO at $z_{spec}=0.36$. Spectroscopic observations from the MISTRAL instrument contradict the QSO redshift, placing it at $z_{spec} = 0.31$.}
    \label{fig:NT0075}
\end{figure*}

\begin{figure*}
    \centering
    \includegraphics[width=0.3\textwidth]{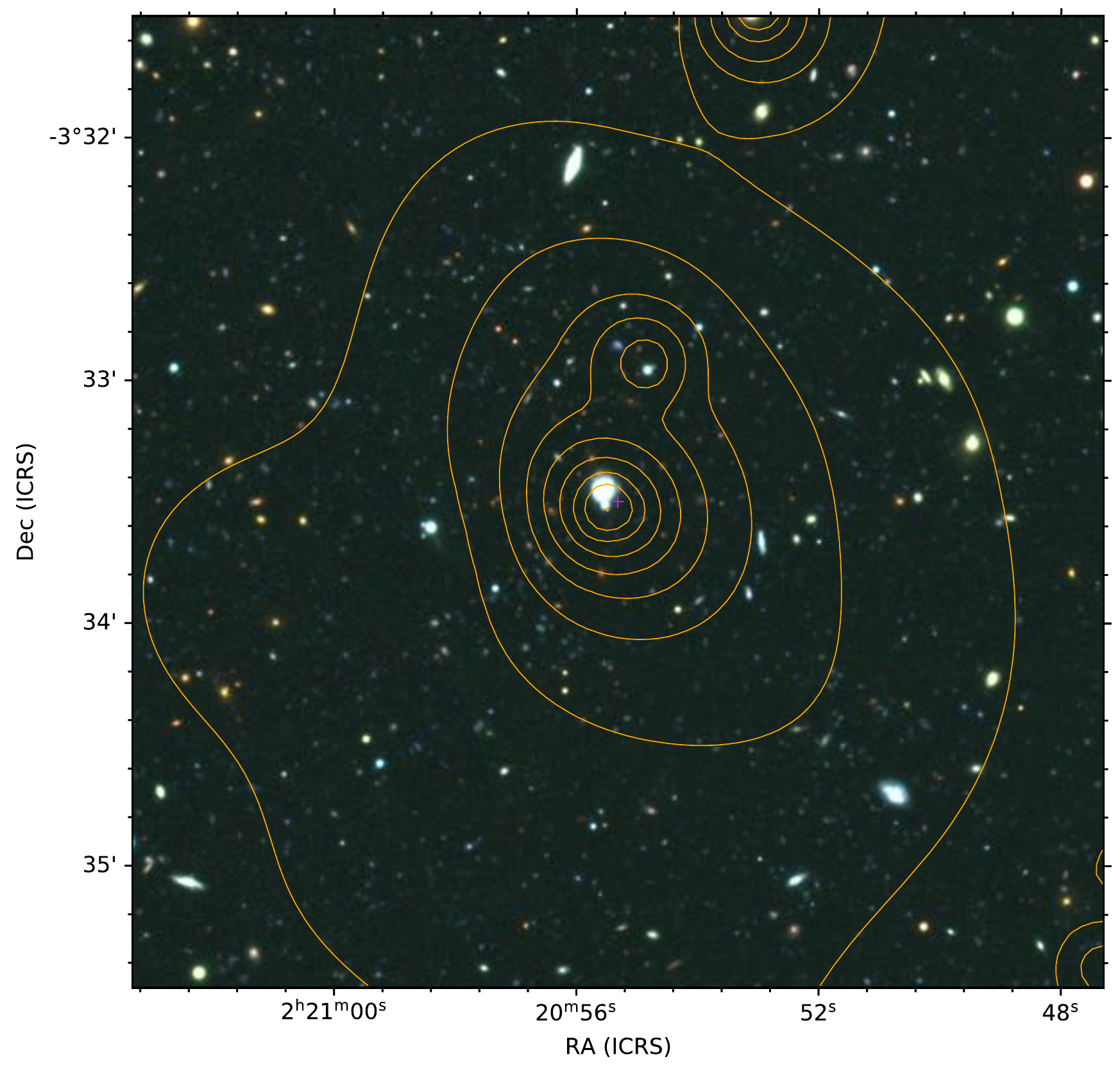} 
    \includegraphics[width=0.3\textwidth]{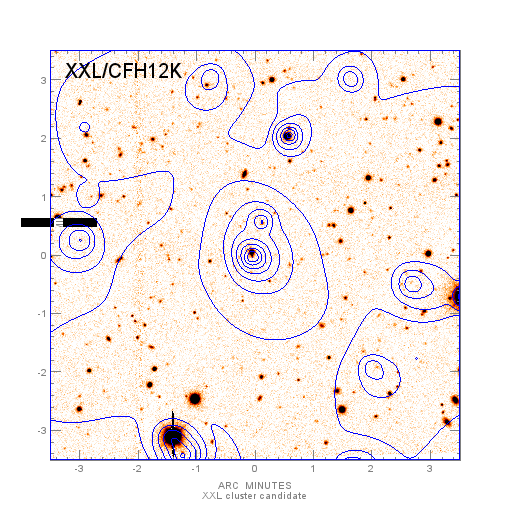}
    \includegraphics[width=0.3\textwidth]{figs/ac_images/XXLnTile-03_25_v4.3_rawraw.png}
    \caption{Cluster candidate XLSSU J022055.4-033332 (also known as ACT-CL J0220.9-0332) with $z_{spec}=1.03$ published in \cite{Hilton2018}. See Figure \ref{fig:HSTimageforNT005} for more details. MISTRAL observations of the central foreground object shows evidence of H$\alpha$, H$\beta$, and NII emission lines.}
    \label{fig:NT0005}
\end{figure*}

\begin{figure*}
    \centering
    \includegraphics[width=0.3\textwidth]{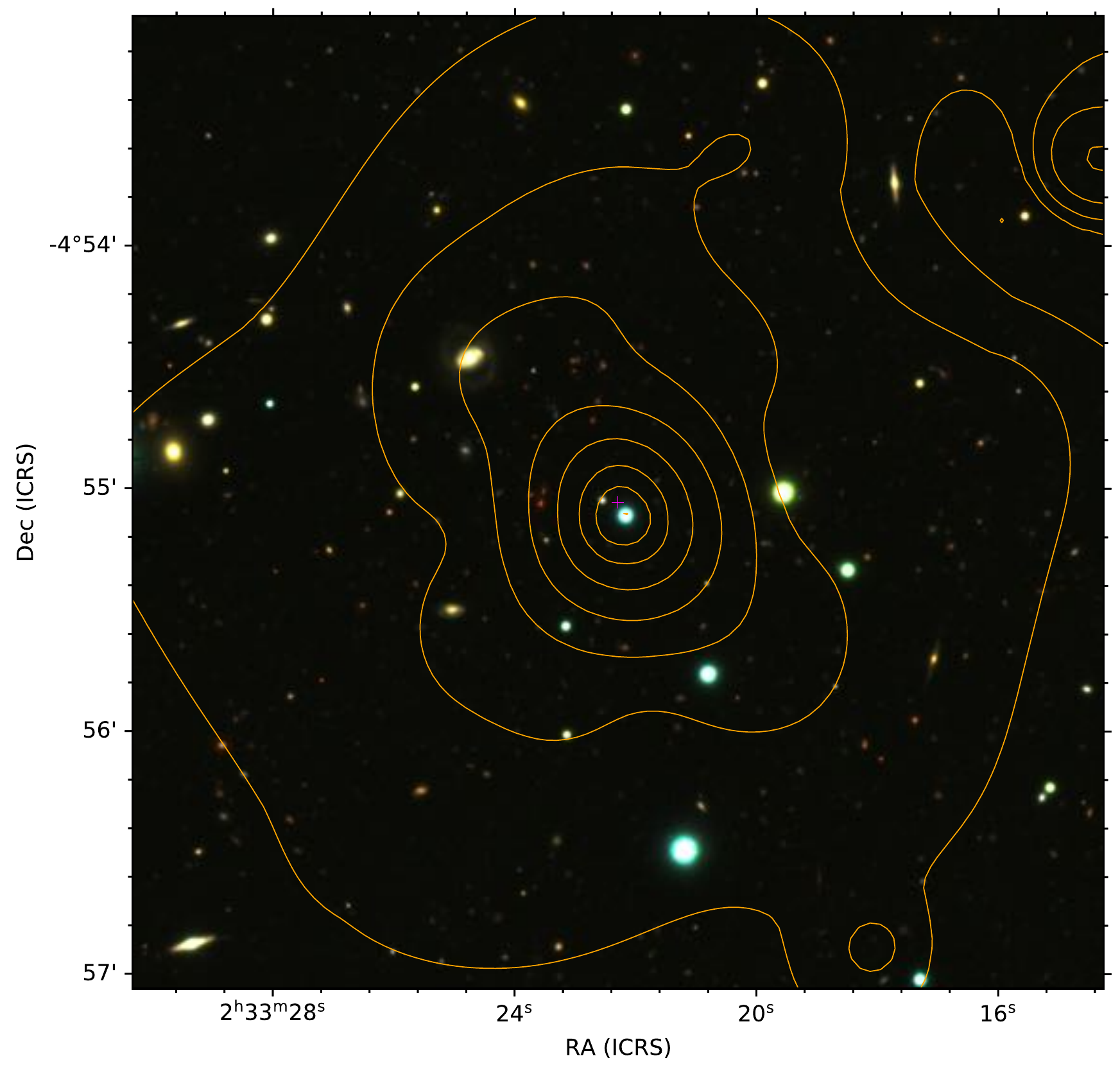}
    \includegraphics[width=0.3\textwidth]{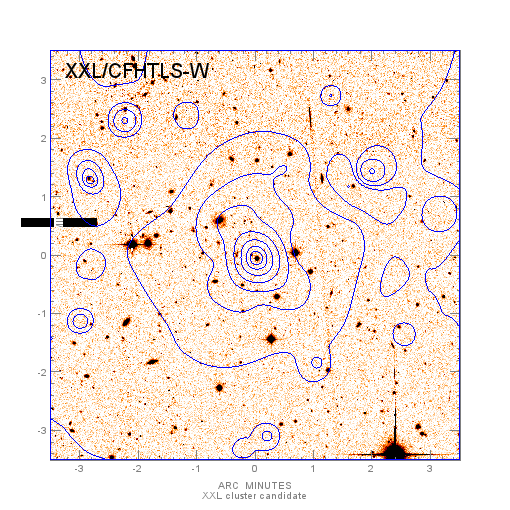}    
    \includegraphics[width=0.3\textwidth]{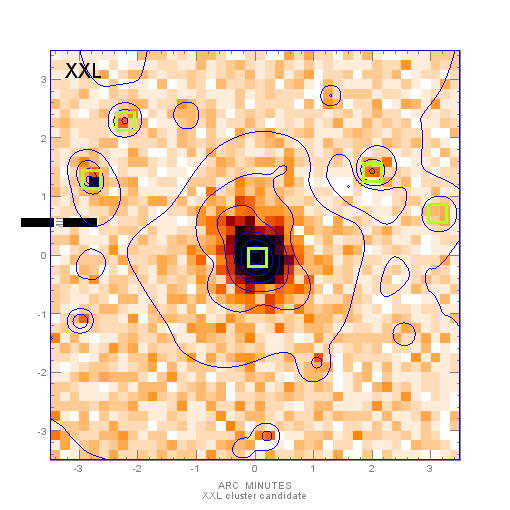}
    \caption{Cluster candidate XLSSU J023322.1-045506. The X-ray emission is centred on a QSO at z=0.78 (SDSS). The cluster is approximately at the same photometric redshift.}
    \label{fig:NT0031}
\end{figure*}

\begin{figure*}
    \centering
    \includegraphics[width=0.32\textwidth]{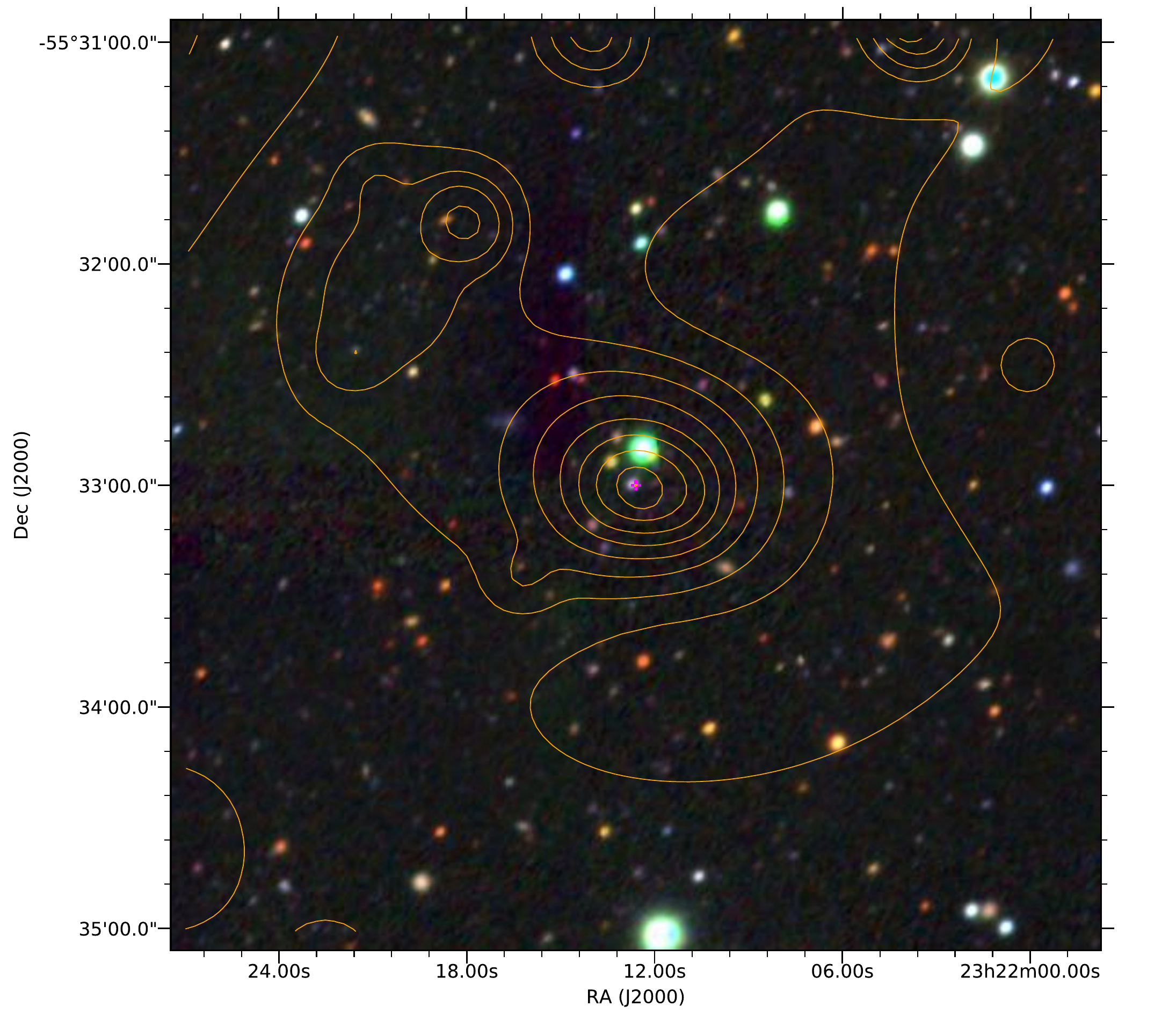}
    \includegraphics[width=0.3\textwidth]{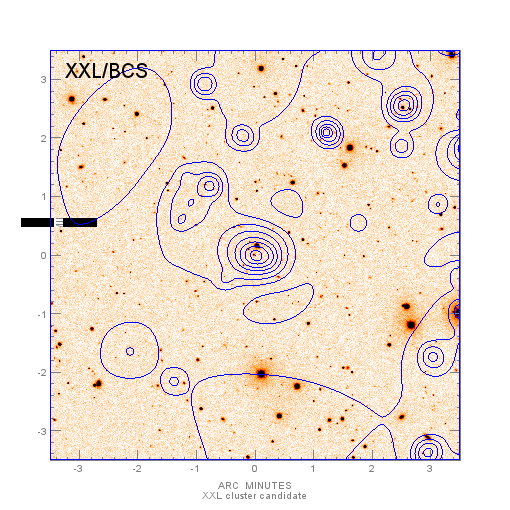}    
    \includegraphics[width=0.3\textwidth]{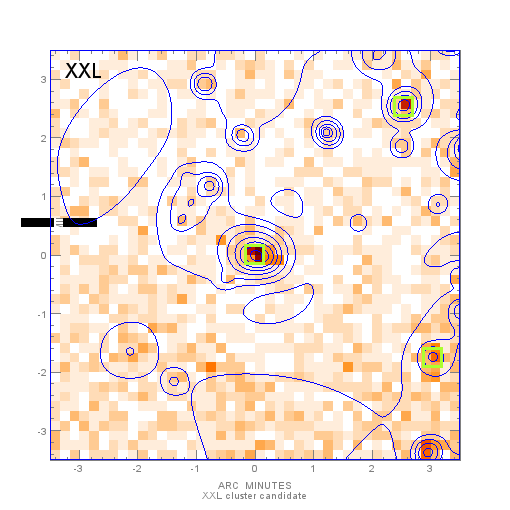}
    \caption{Cluster candidate XLSSU J232212.6-553259 with strong contamination from a QSO with photometric redshift of $z_{phot}=0.82$}
    \label{fig:ST0050}
\end{figure*}

\begin{figure*}
    \centering
    \includegraphics[width=0.32\textwidth]{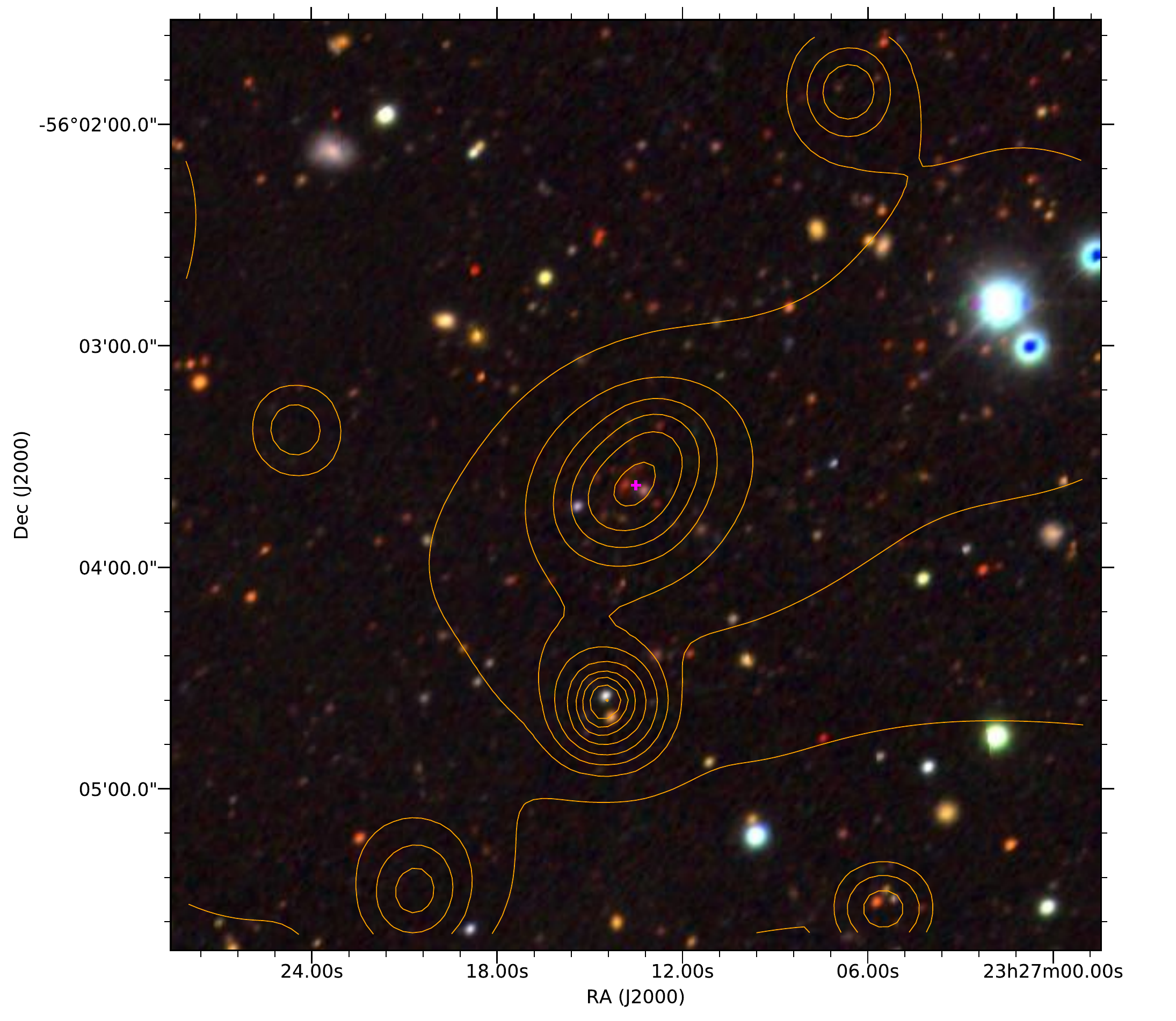}
    \includegraphics[width=0.3\textwidth]{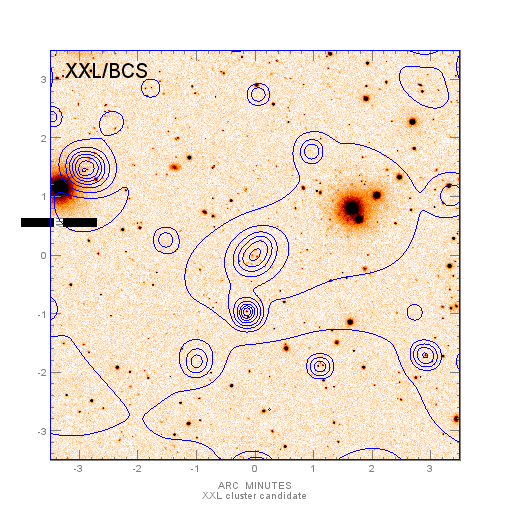}    
    \includegraphics[width=0.3\textwidth]{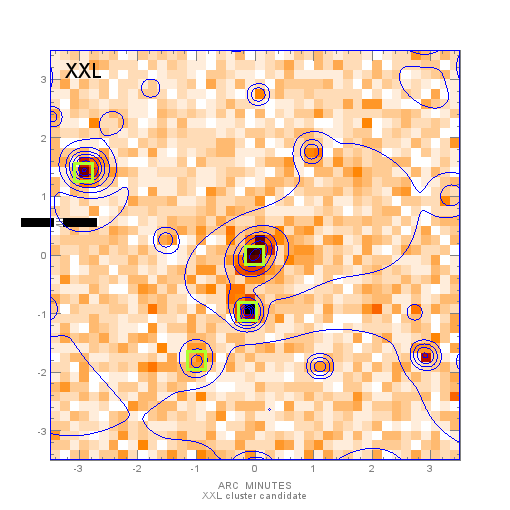}
    \caption{Cluster candidate XLSSU J232713.5-560337 (also known as XBCS J232713.7-560341) published in \cite{Suhada2012}.}
    \label{fig:ST0063}
\end{figure*}

\begin{figure*}
    \centering
    \includegraphics[width=0.32\textwidth]{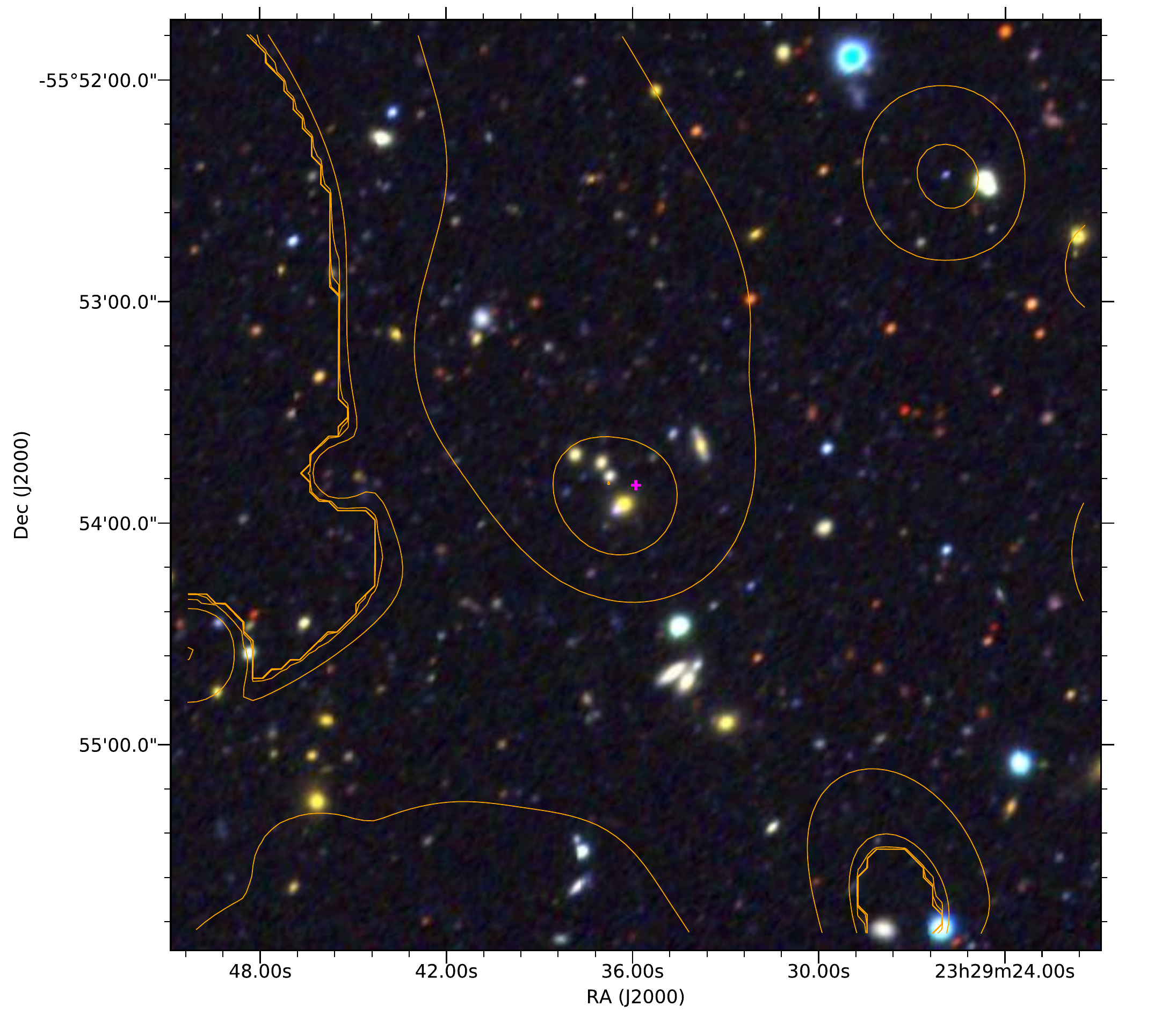}
    \includegraphics[width=0.3\textwidth]{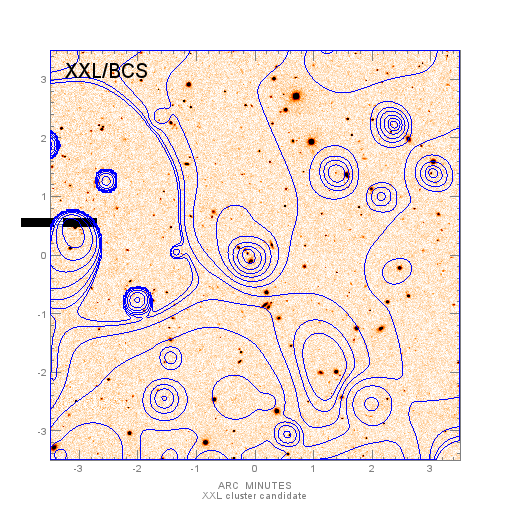}    
    \includegraphics[width=0.3\textwidth]{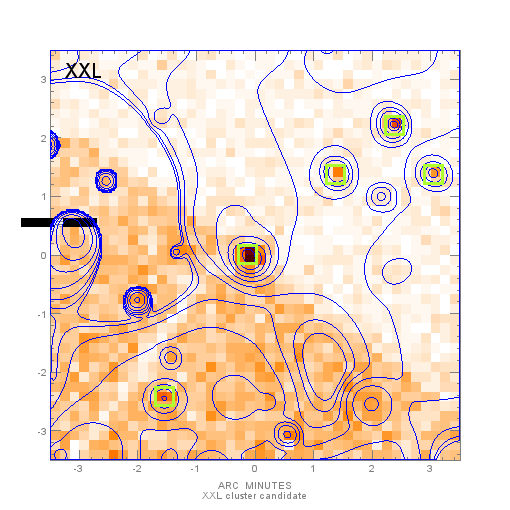}
    \caption{Cluster candidate XLSSU J232936.7-555349 showing two galaxies at the centre of the X-ray emission at $z=0.31$ and a QSO at $z=2.03$. Both galaxies host an AGN. Some additional galaxies nearby are possibly at the same redshift. This is a line-of-sight projection of AGN or a small group.}
    \label{fig:ST0062}
\end{figure*}

\begin{figure*}
    \centering
    \includegraphics[width=0.32\textwidth]{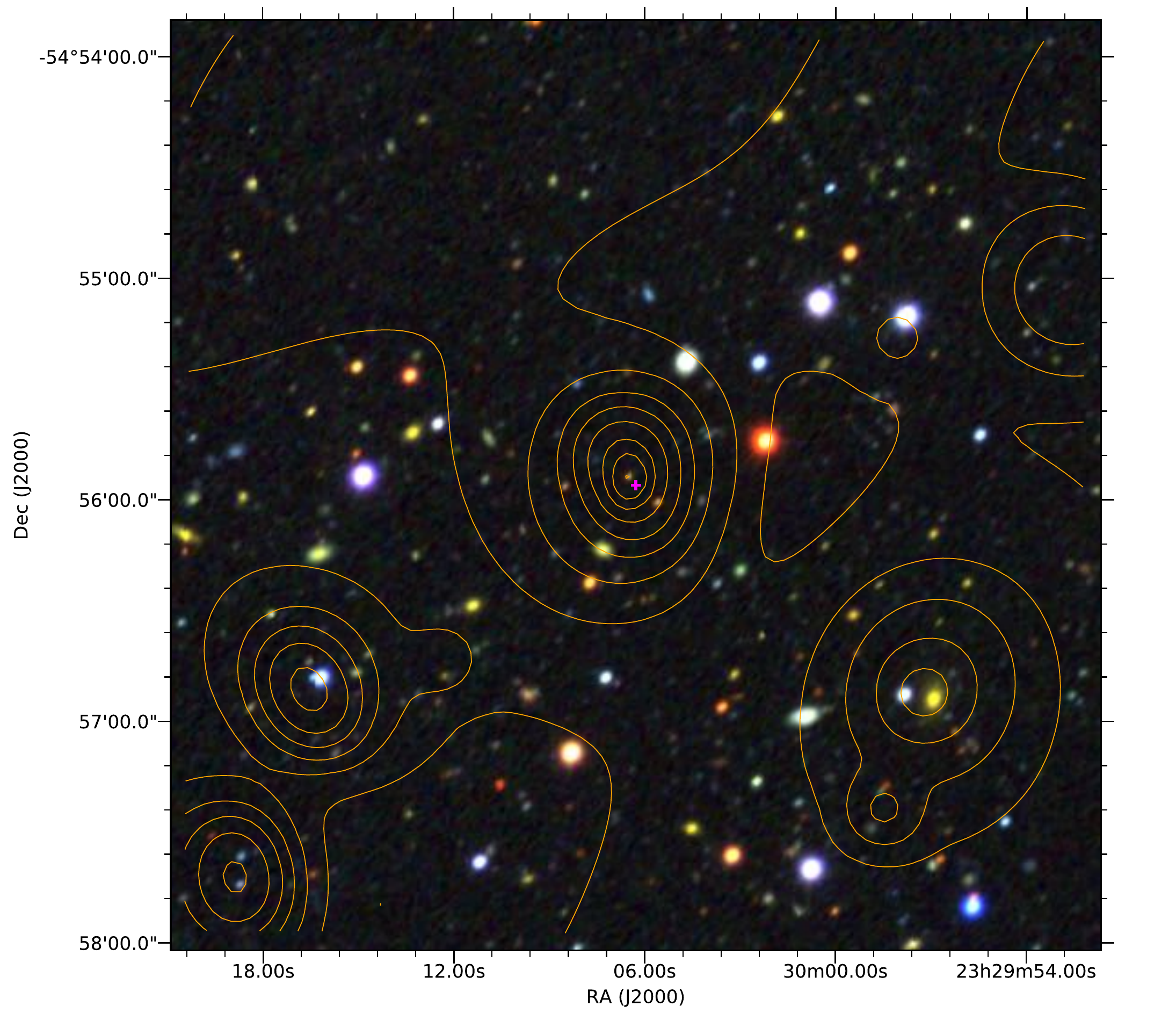}
    \includegraphics[width=0.3\textwidth]{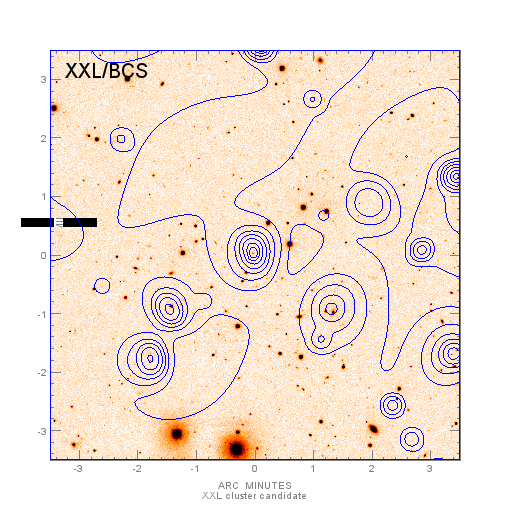}    
    \includegraphics[width=0.3\textwidth]{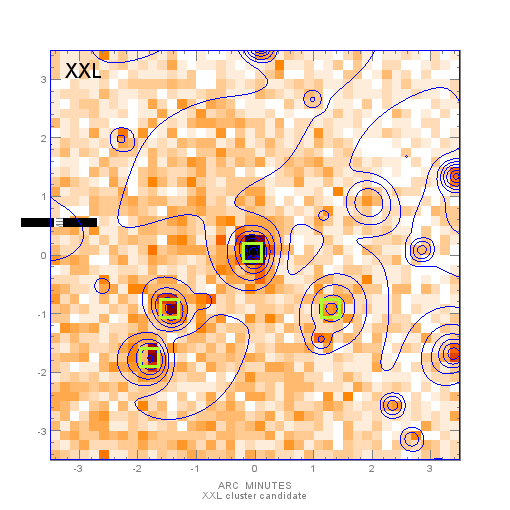}
    \caption{Cluster candidate XLSSU J233006.5-545553 with visible galaxies possibly at concordant redshift but with no spectroscopic information. There exists also the possibility of a high-redshift cluster. The origin of the X-ray emission is unclear. Nevertheless, this source falls within the type I/II AGN wedge based on WISE data.}
    \label{fig:ST0045}
\end{figure*}

\begin{figure*}
    \centering
    \includegraphics[width=0.32\textwidth]{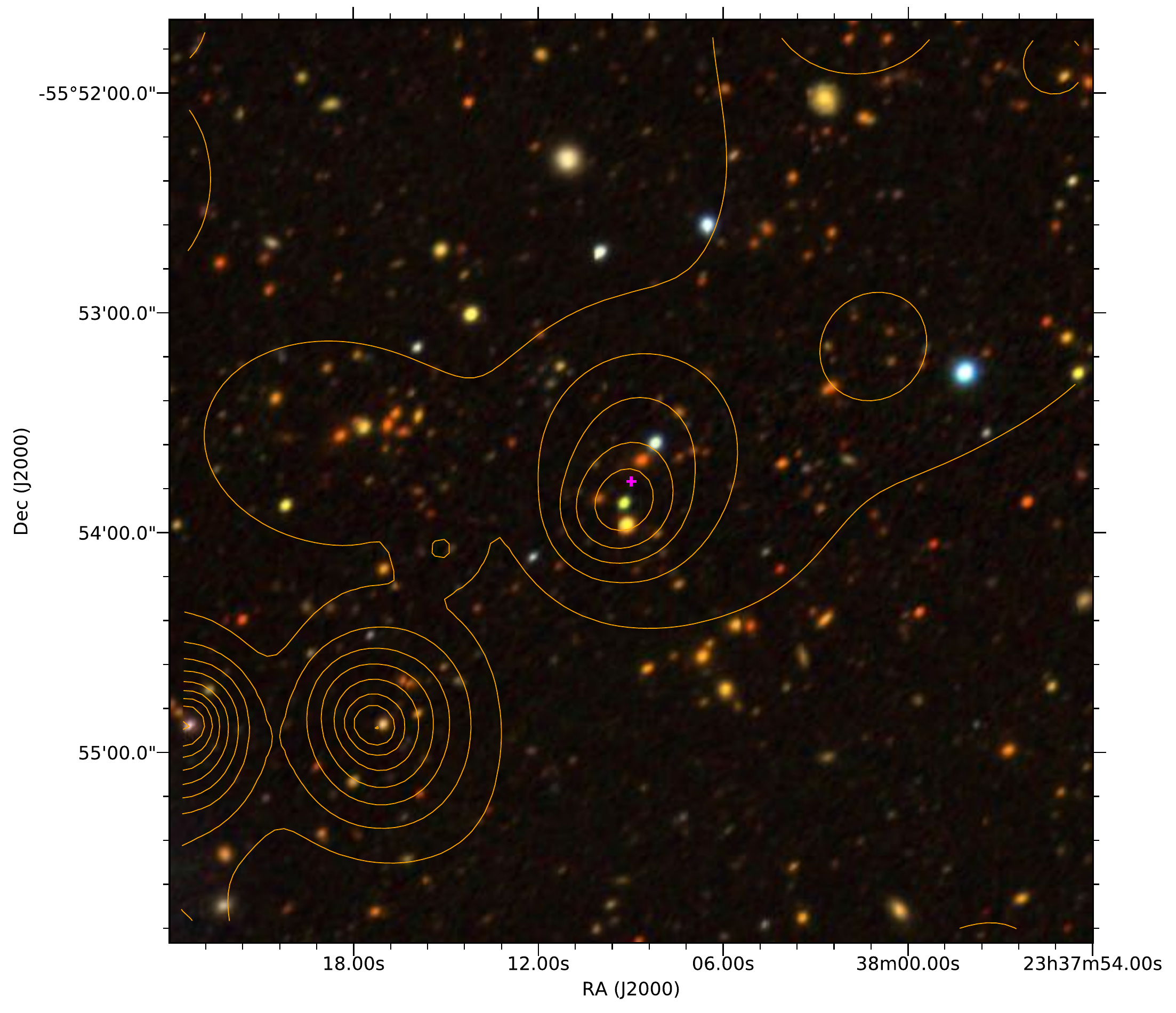}  
    \includegraphics[width=0.3\textwidth]{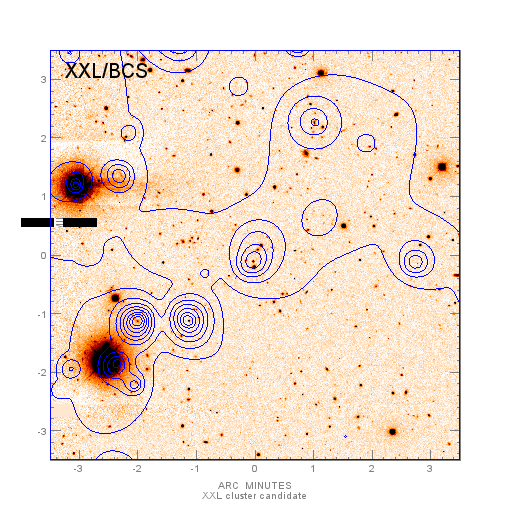}    
    \includegraphics[width=0.3\textwidth]{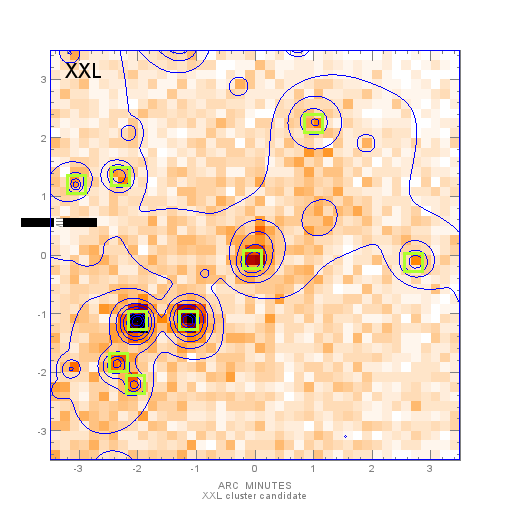}
    \caption{Cluster candidate XLSSU J233809.3-555350 with a QSO at the centre of the X-ray emission at $z=3.81$ and a possible BCG above the QSO. Cluster LCS-CL J233802-5553.3 with tentative spectroscopic redshift at z=0.6 is located 1 arcminute to the bottom left of the X-ray centre \citep{Bleem2015}}.
    \label{fig:ST0057}
\end{figure*}

\begin{figure*}
    \centering
    \includegraphics[width=0.32\textwidth]{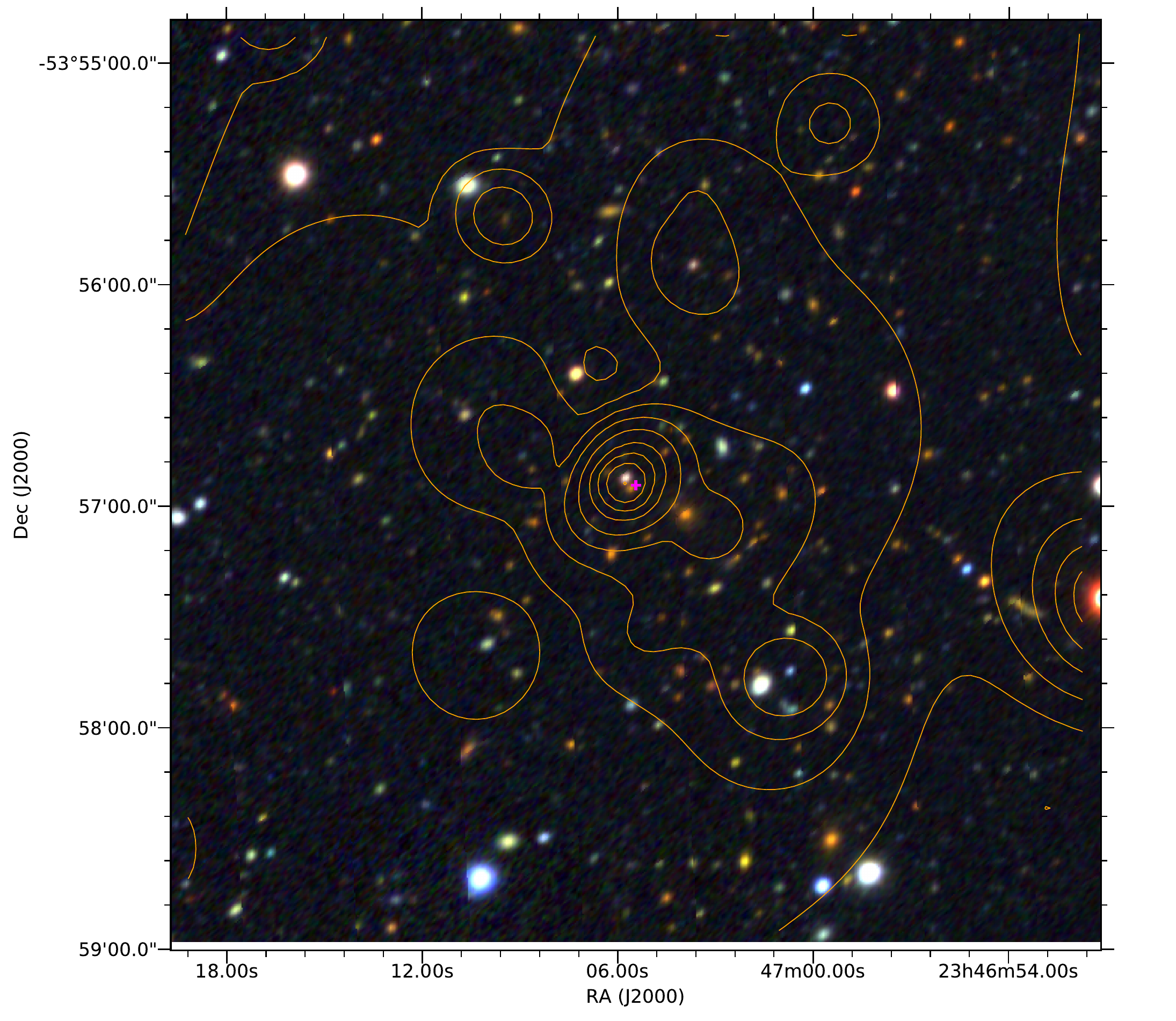}
    \includegraphics[width=0.3\textwidth]{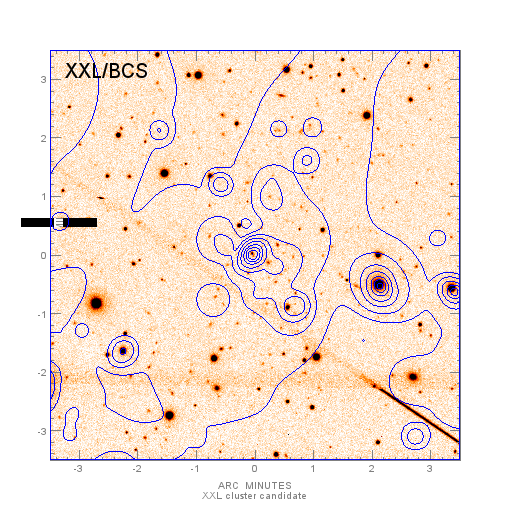}    
    \includegraphics[width=0.3\textwidth]{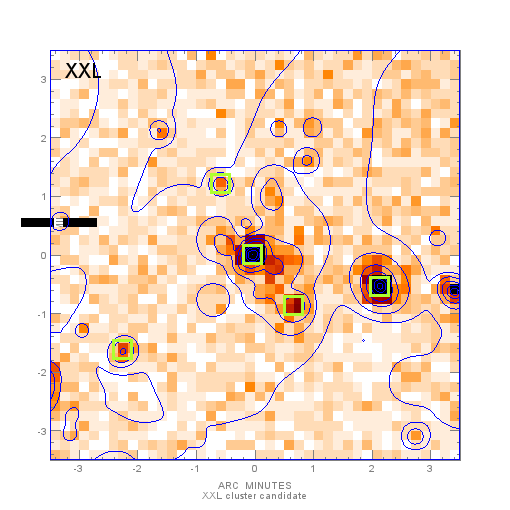}
    \caption{Cluster candidate XLSSU J234705.7-535653 without any spectroscopic information.}
    \label{fig:ST0028}
\end{figure*}

\begin{figure*}
    \centering
    \includegraphics[width=0.3\textwidth]{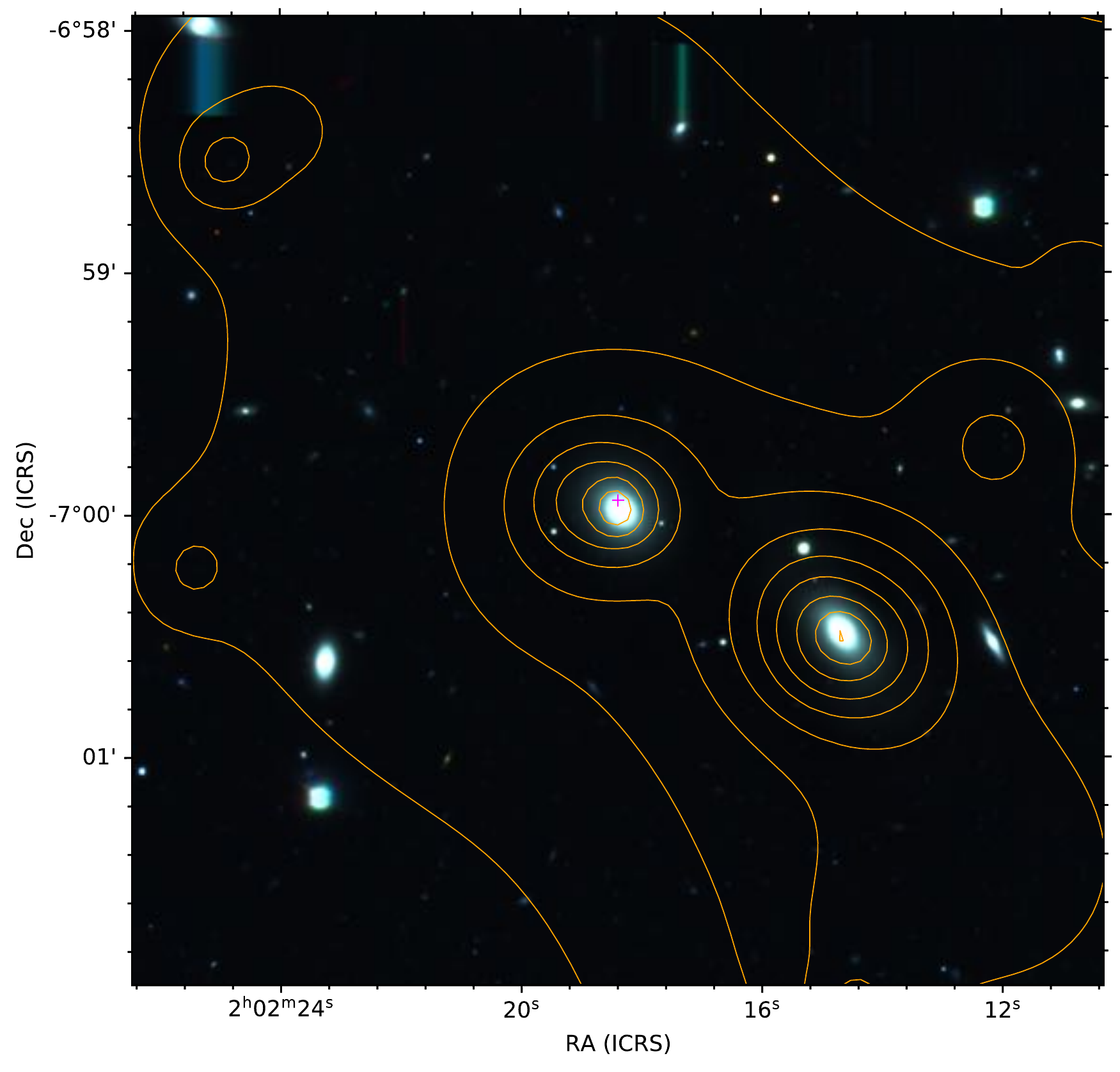}
    \includegraphics[width=0.3\textwidth]{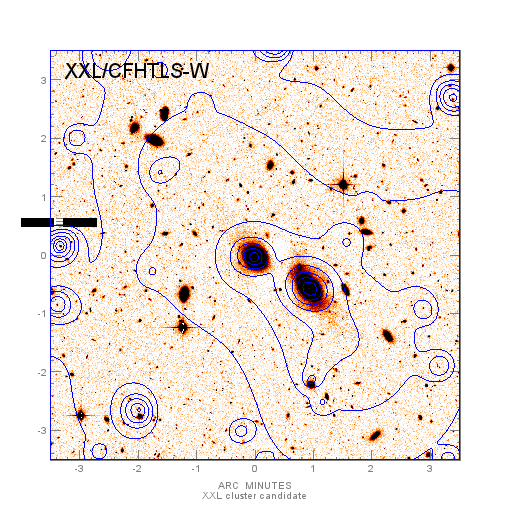}    
    \includegraphics[width=0.3\textwidth]{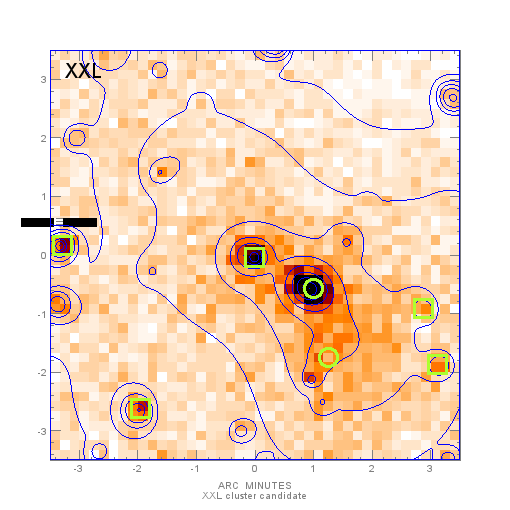}
    \caption{Active galaxy XLSSU J020218.3-065958 in a nearby group, known as XLSSC 113 at $z=0.05$. The optical spectrum (NED) shows no AGN activity.}
    \label{fig:NT0103}
\end{figure*}

\begin{figure*}
    \centering
    \includegraphics[width=0.3\textwidth]{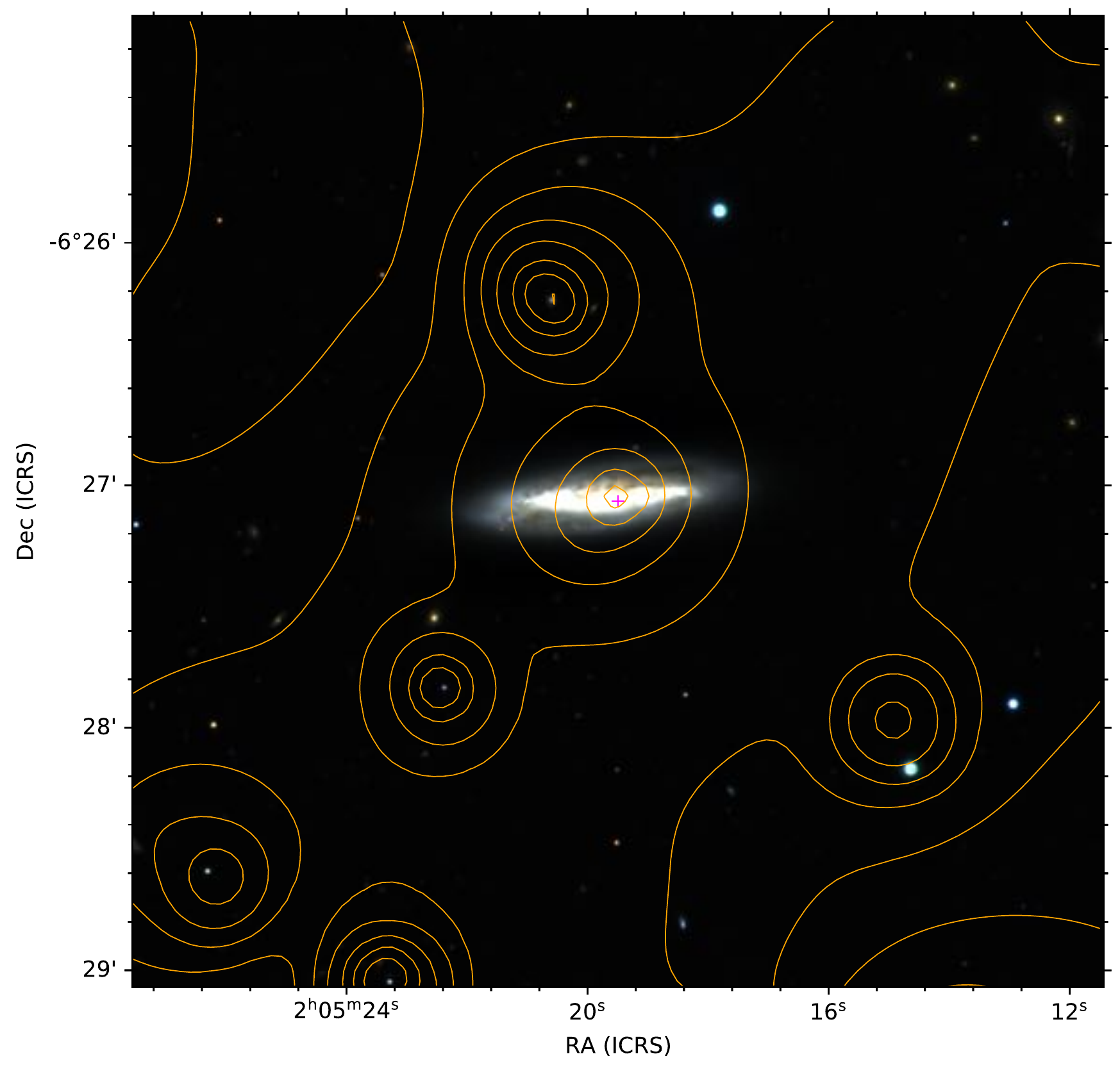}
    \includegraphics[width=0.3\textwidth]{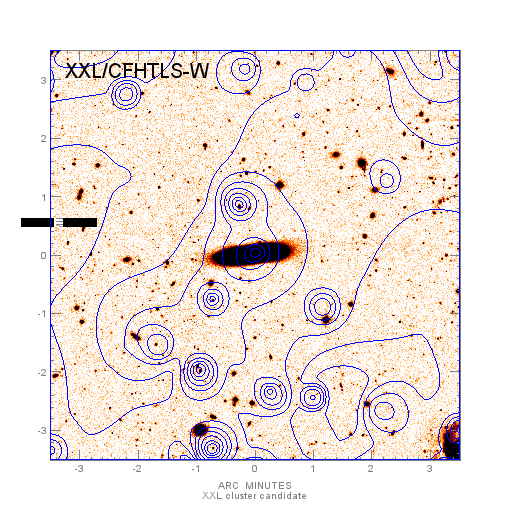}
    \includegraphics[width=0.3\textwidth]{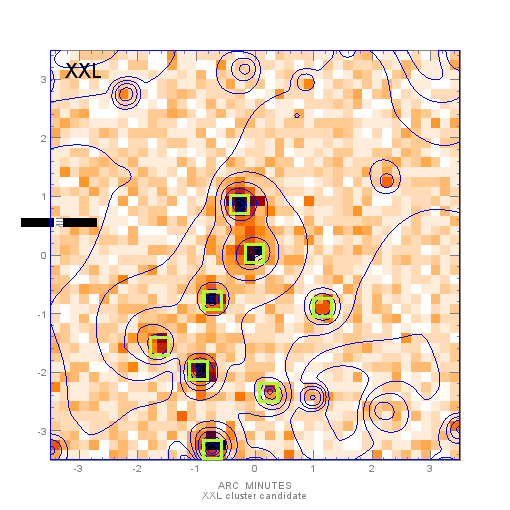}
    \caption{Active galaxy XLSSU J020519.5-062702 at $z = 0.01$.}
    \label{fig:NT0094}
\end{figure*}

\begin{figure*}
    \centering
    \includegraphics[width=0.3\textwidth]{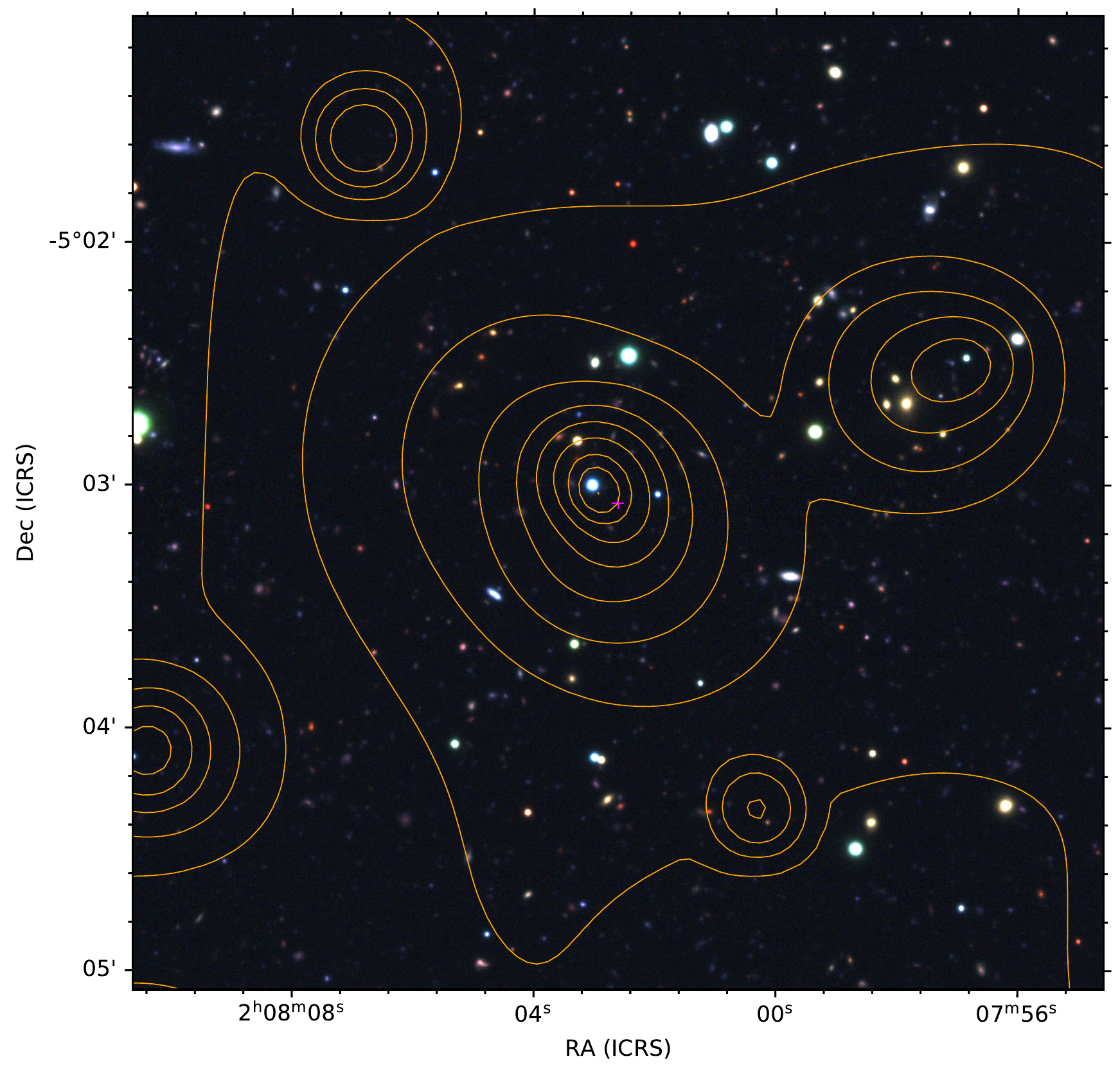}
    \includegraphics[width=0.3\textwidth]{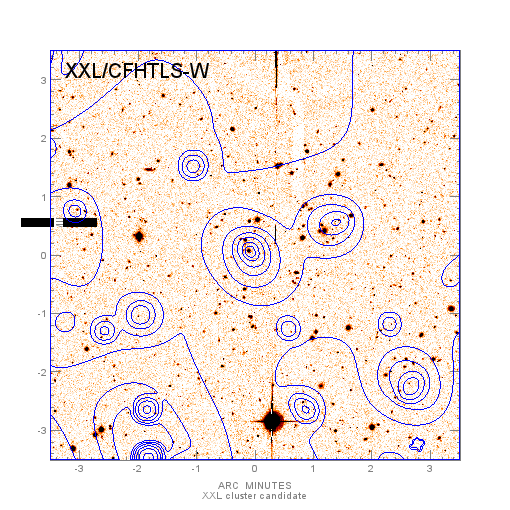}    
    \includegraphics[width=0.3\textwidth]{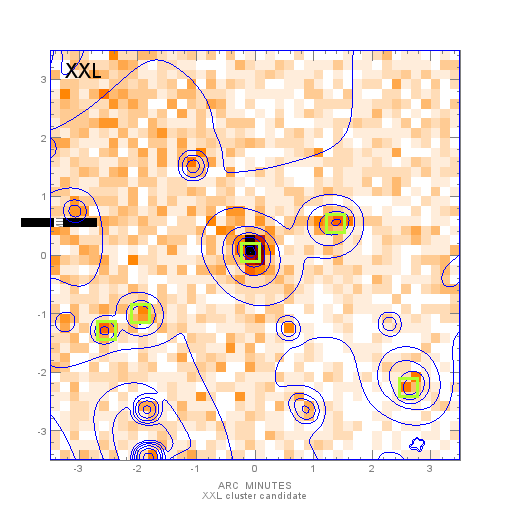}
    \caption{Active galaxy XLSSU J020802.9-050302 with a QSO at $z_{spec}=1.86$.}
    \label{fig:NT0074}
\end{figure*}

\begin{figure*}
    \centering
    \includegraphics[width=0.3\textwidth]{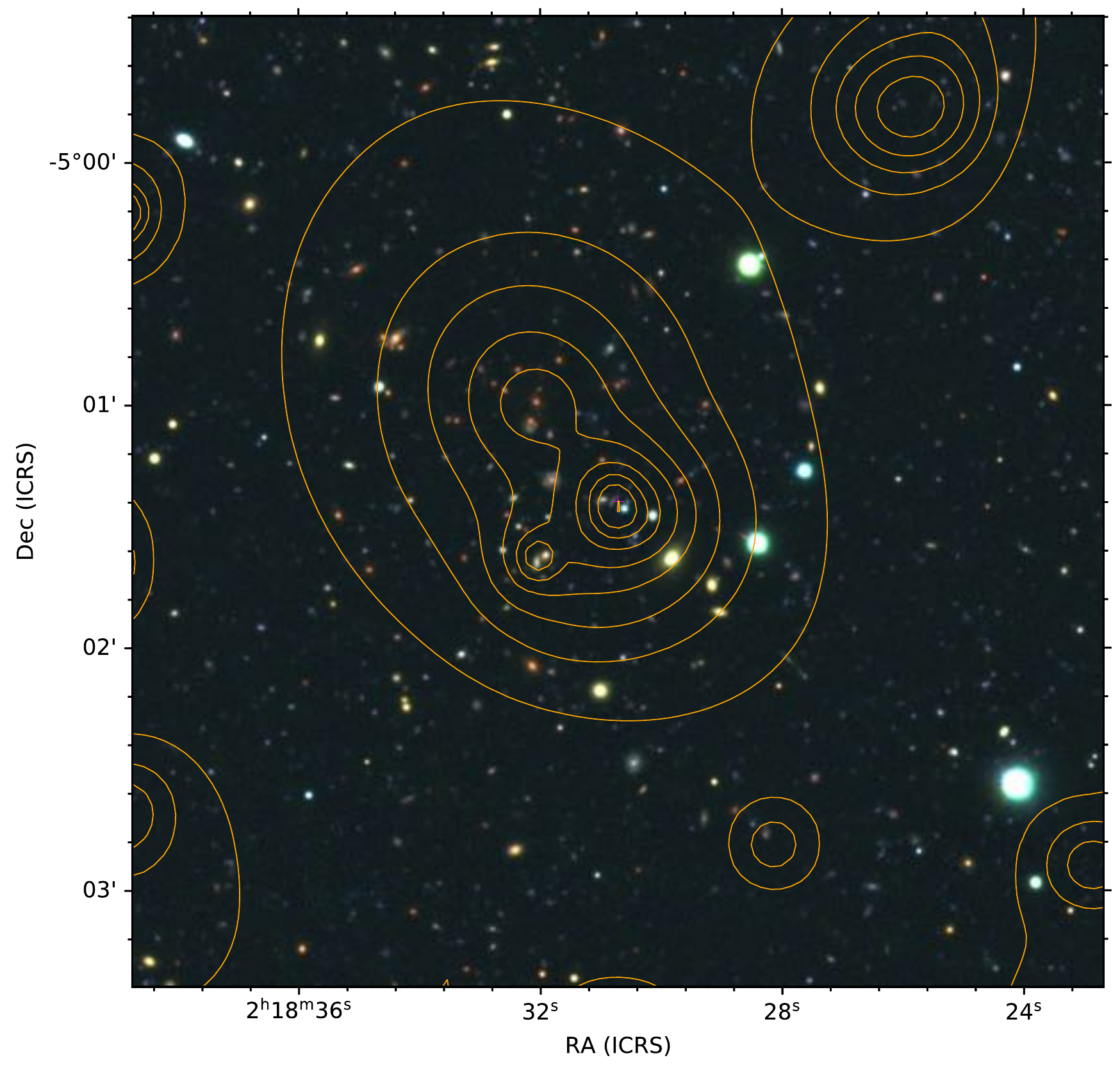}
    \includegraphics[width=0.3\textwidth]{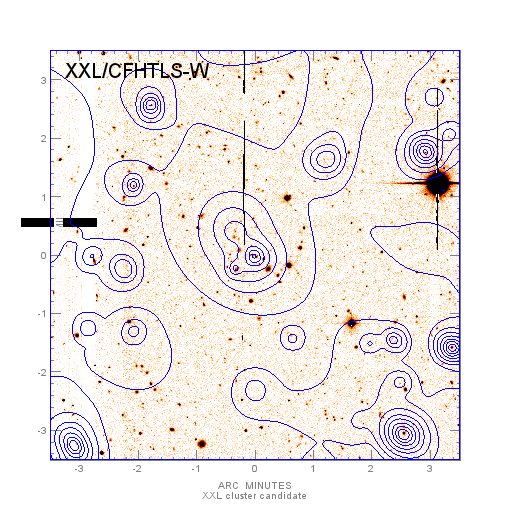}    
    \includegraphics[width=0.3\textwidth]{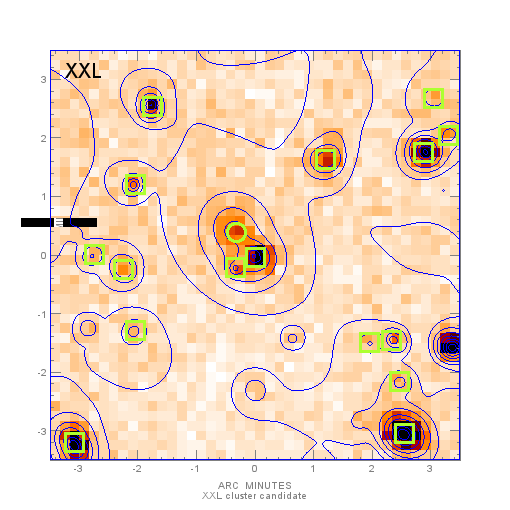}
    \caption{QSO XLSSU J021830.7-050126 with peaked X-ray emission observed at $z=3.00211$ (SDSS). XLSSC 64 is separately detected as a cluster by the pipeline.}
    \label{fig:NT0063}
\end{figure*}

\begin{figure*}
    \centering
    \includegraphics[width=0.3\textwidth]{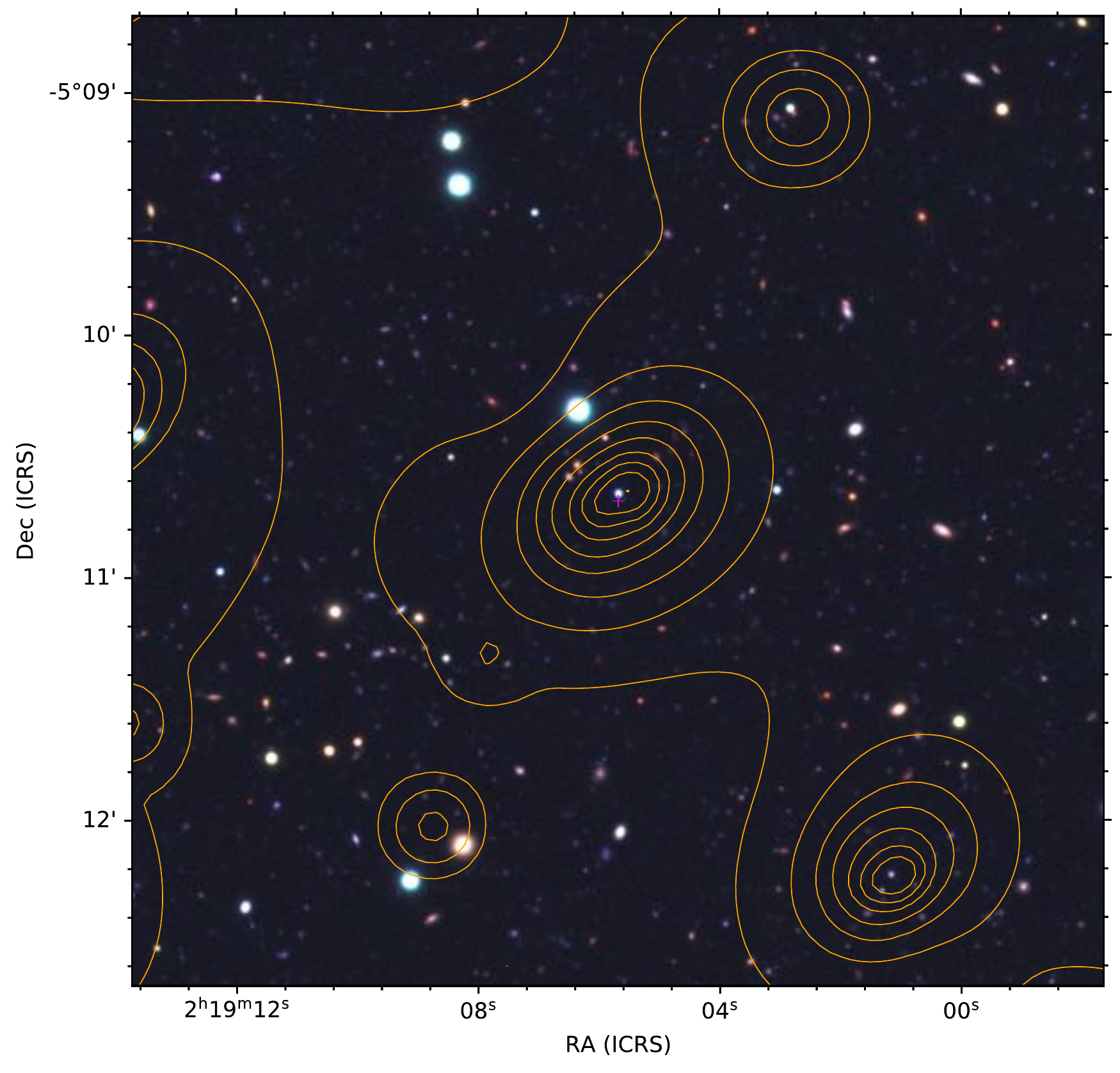}    
    \includegraphics[width=0.3\textwidth]{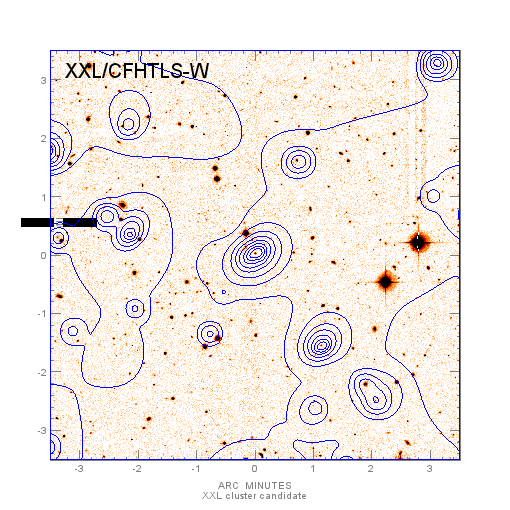}
    \includegraphics[width=0.3\textwidth]{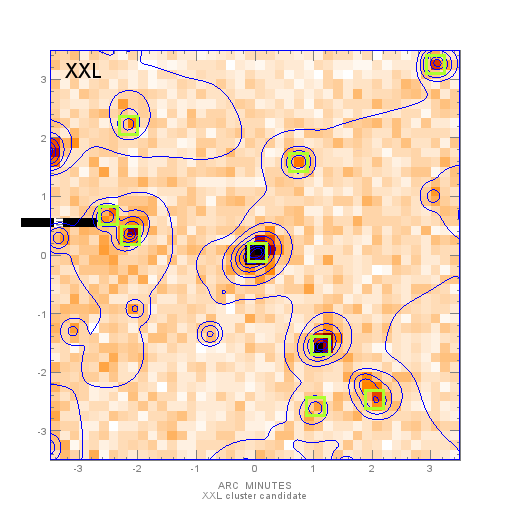}
    \caption{Active galaxy XLSSU J021905.5-051038. QSO at $z=1.66$. There is possibly a foreground cluster, but there is no further redshift information.}
    \label{fig:NT0064}
\end{figure*}

\begin{figure*}
    \centering
    \includegraphics[width=0.3\textwidth]{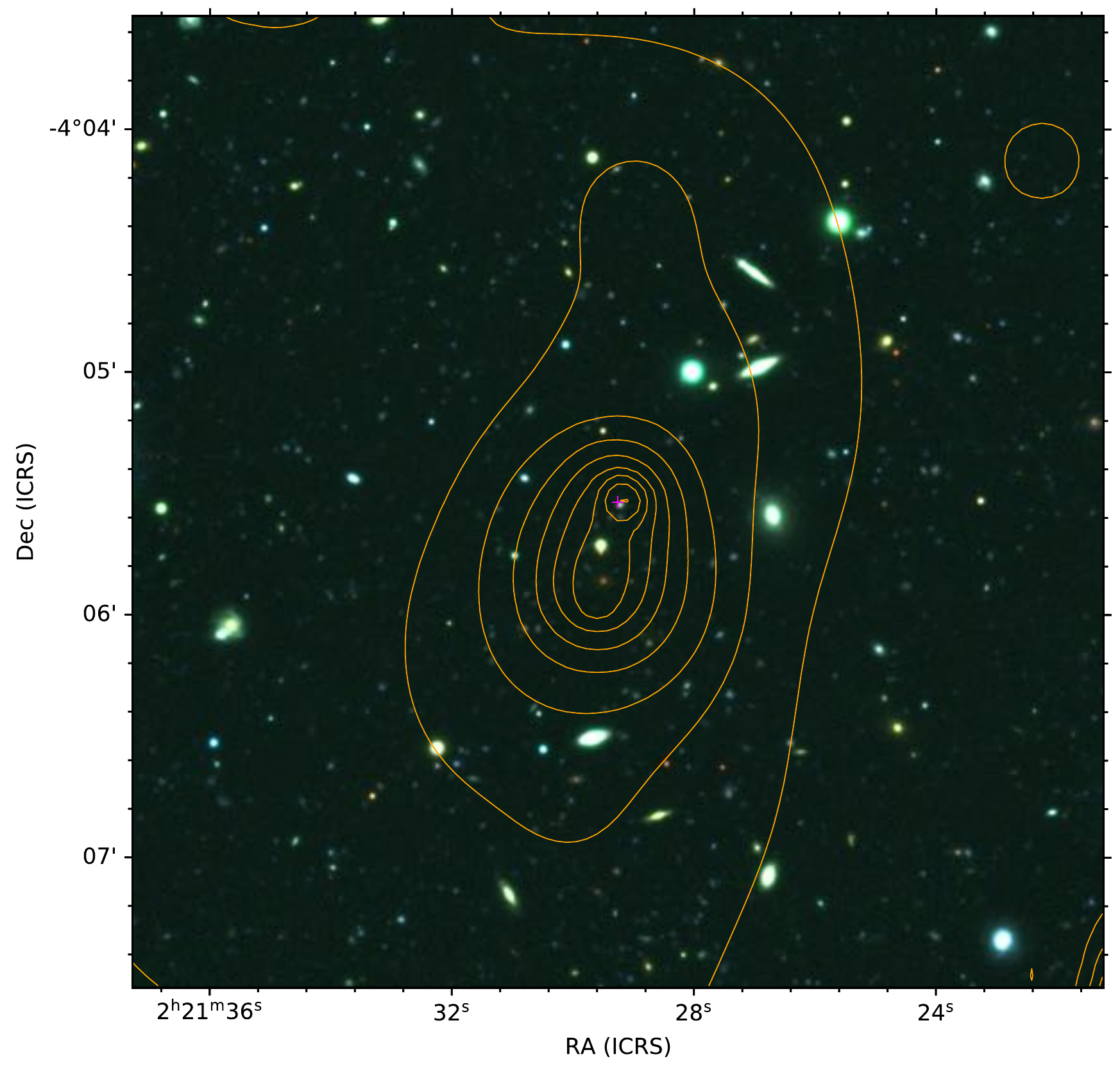}    
    \includegraphics[width=0.3\textwidth]{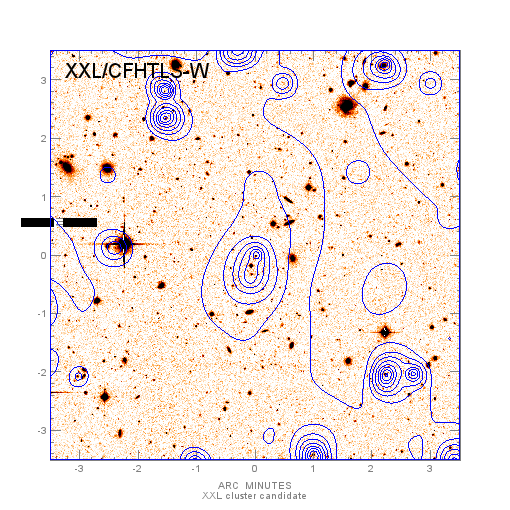}
    \includegraphics[width=0.3\textwidth]{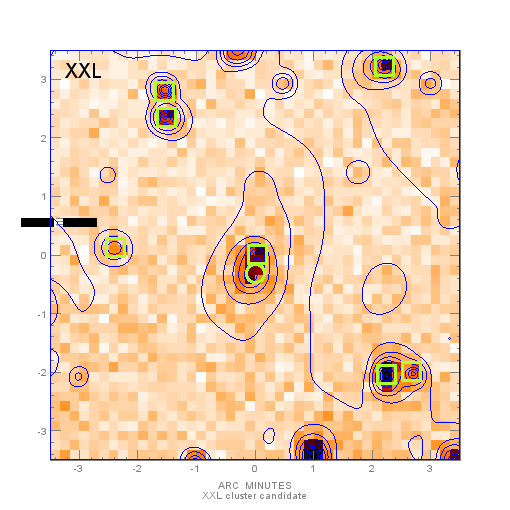}
    \caption{QSO XLSSU J022129.1-040531. XLSSC 34 at $z_{spec}=1.036$ is separately detected. However, the X-ray emission is centred on a possible AGN with uncertain redshift in SDSS (z=1.23). The AGN classification is dubious because of its low S/N spectrum.}
    \label{fig:NT0017}
\end{figure*}

\begin{figure*}
    \centering
    \includegraphics[width=0.3\textwidth]{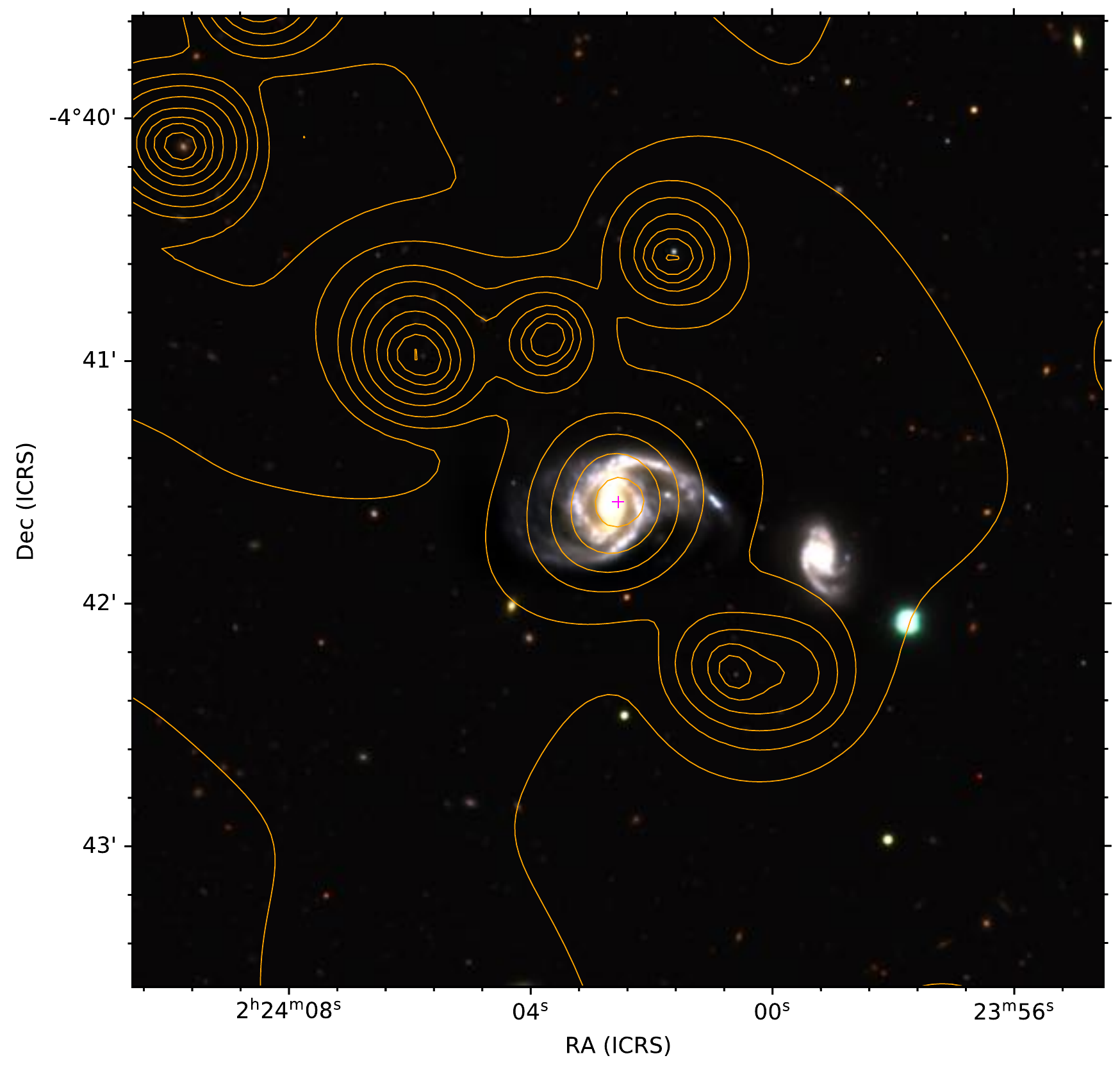}
    \includegraphics[width=0.3\textwidth]{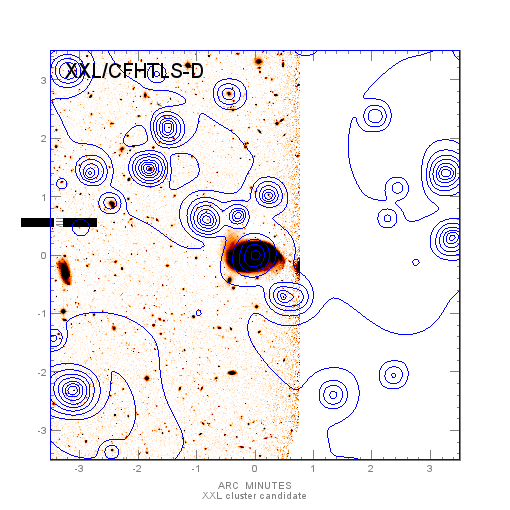}
     \includegraphics[width=0.3\textwidth]{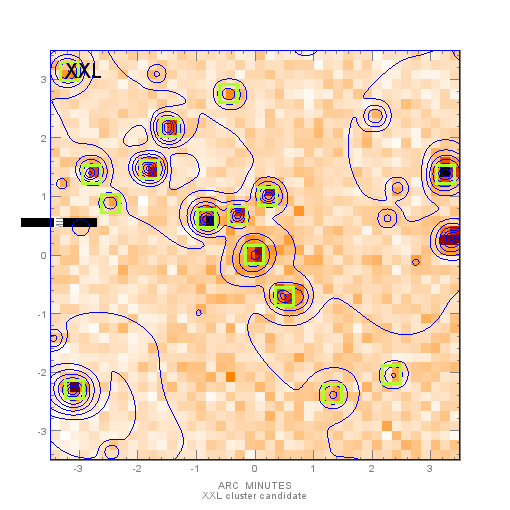}
    \caption{Active galaxy XLSSU J022402.5-044134 at $z=0.043$.}
    \label{fig:NT0045}
\end{figure*}

\begin{figure*}
    \centering
    \includegraphics[width=0.3\textwidth]{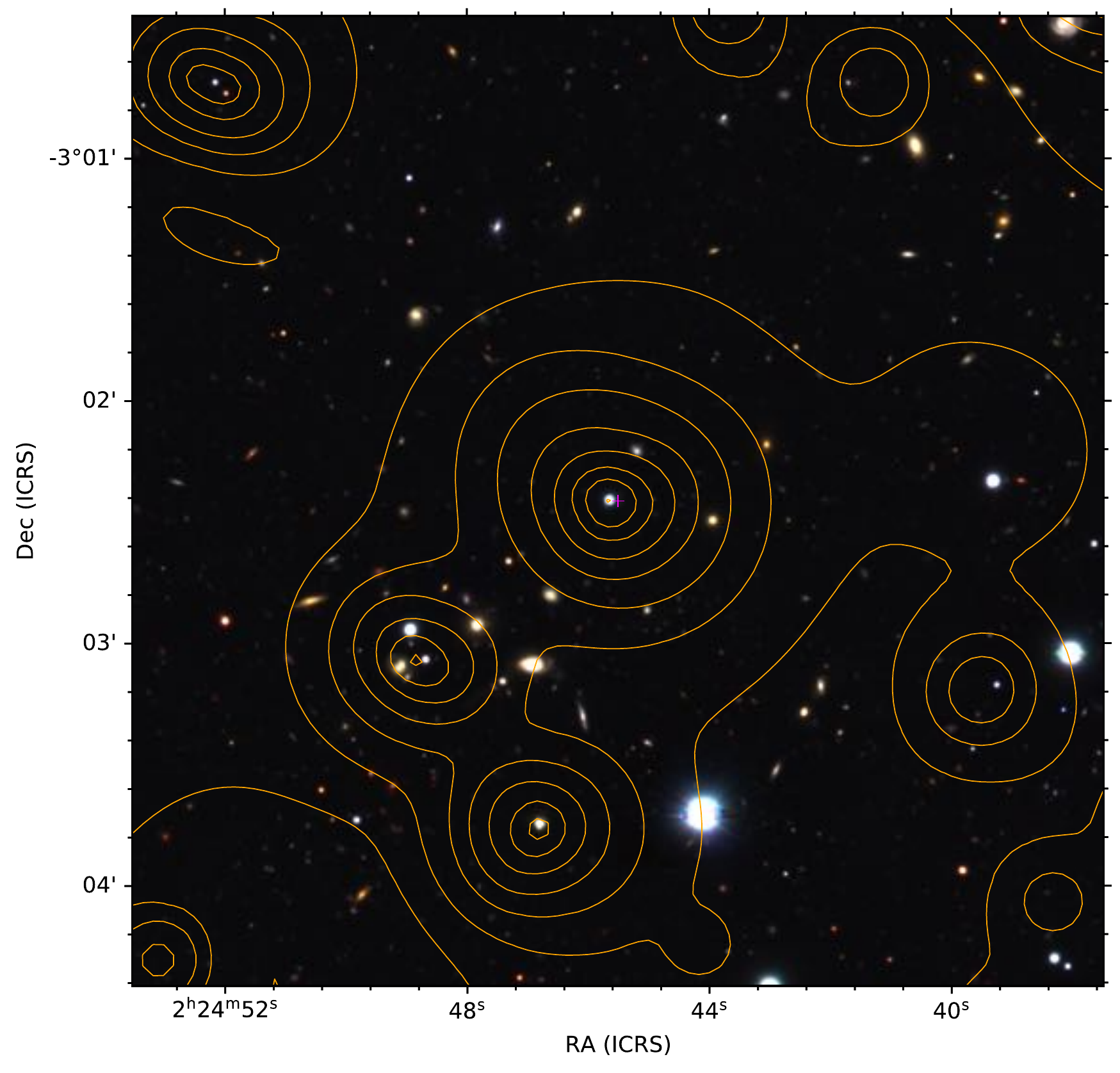}
    \includegraphics[width=0.3\textwidth]{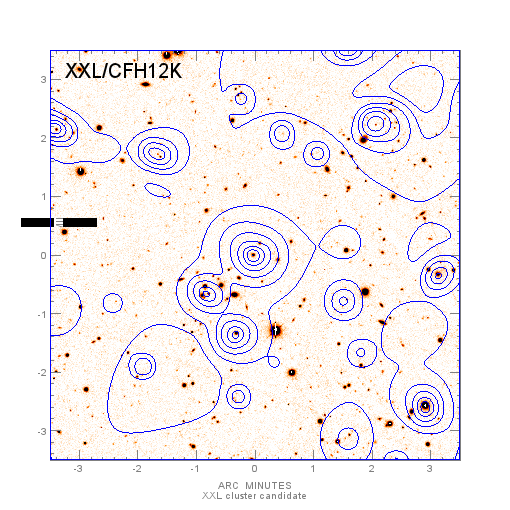}
    \includegraphics[width=0.3\textwidth]{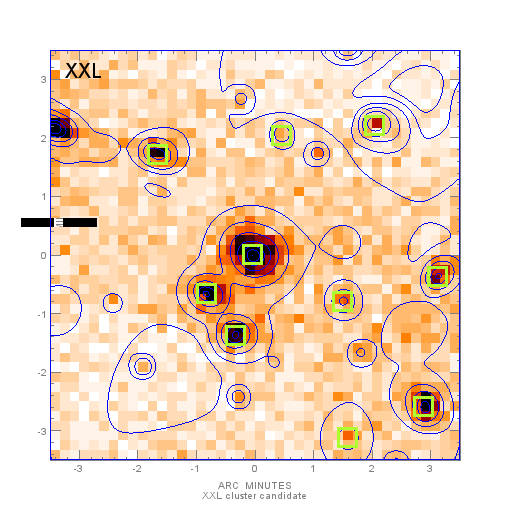}
    \caption{Active galaxy XLSSU J022445.6-030224 with a known QSO at $z=1.23$ and a cluster that is clearly visible to the south-east of the object.}
    \label{fig:NT0001}
\end{figure*}

\end{appendix}

\end{document}